# KRNC: New Foundations for Permissionless Byzantine Consensus and Global Monetary Stability

Clinton Ehrlich[1]    Anna Guzova[2]

February 27, 2020 Version 1.5

> It is not unlikely that a new security/trust protocol, with the magnitude of the influence from the invention of the public-key protocol, [will be] inspired by the study of animal communication networks.
>
> – Prof. Zhanshan Ma,
> *Chinese Academy of Sciences*, 2009

**Abstract** This paper applies biomimetic engineering to the problem of permissionless Byzantine consensus and achieves results that surpass the prior state of the art by four orders of magnitude. It introduces a biologically inspired asymmetric Sybil-resistance mechanism, Proof-of-Balance, which can replace symmetric Proof-of-Work and Proof-of-Stake weighting schemes.

The biomimetic mechanism is incorporated into a permissionless blockchain protocol, Key Retroactivity Network Consensus ("KRNC"), which delivers ~40,000 times the security and speed of today's decentralized ledgers. KRNC allows the fiat money that the public already owns to be upgraded with cryptographic inflation protection, eliminating the problems inherent in bootstrapping new currencies like Bitcoin and Ethereum.

The paper includes two independently significant contributions to the literature. First, it replaces the non-structural axioms invoked in prior work with a new formal method for reasoning about trust, liveness, and safety from first principles. Second, it formalizes two simple but powerful exploits — book-prize attacks and pseudo-transfer attacks — that undermine the security guarantees of all prior permissionless ledgers.

---

[1] Chief Computer Scientist, Krnc Inc.; Fmr. Visiting Researcher, MGIMO University, *clint@krnc.io*

[2] Lead Mathematician, Krnc Inc.; Fmr. Senior Applied Mathematician, AO UniCredit Bank

[3] Proof-of-Balance comprises the subject matter of the following published patent applications: US 16/261, 478; PCT/US2019/015732.

# Part I: Concept

## 1. Introduction

### 1.1 Summary

Reverse engineering biological systems has yielded rapid advancement in fields ranging from pharmacology and artificial intelligence to materials science and aerospace design. [1] This approach, known as *biomimicry*, allows humans to learn from and copy what are effectively "alien technologies" developed through billions of years of evolutionary optimization.

It has long been hoped that biomimicry could be the key to a major leap forward in trust-minimized computation. In 2009, the same year that Bitcoin was released, one of the world's few dual PhDs in computer science and biology predicted that adapting animal-communication techniques to fault-tolerant distributed systems could yield a breakthrough comparable to the invention of asymmetric encryption in the 1970s. [2] This paper vindicates that prediction: it adapts cue-authenticated biological signaling to construct the first asymmetric method of Sybil-resistance, which allows correct agents to verifiably retain control of a permissionless blockchain even if they are unable to match the adversary's budget for an attack. Adding biomimetic cost asymmetry to off-the-shelf consensus algorithms unlocks a roughly 40,000-fold increase in reliability, speed, and scalability over symmetric weighting methods, such as Proof-of-Work and Proof-of-Stake, which require correct agents to expend more resources than their faulty counterparts.

Those legacy technologies embody the "handicap principle," a theory that originated in biology to explain the evolution of seemingly wasteful traits, like the oversized tails of male peacocks. According to the handicap principle, the reliability of a signal depends on its verifiable cost to the signaler. [3] Inside the Bitcoin community, this theory has been elevated to the status of a supposedly universal natural law, which applies with equal weight to biology and computer science. [4] In Proof-of-Work, a handicap is imposed by forcing consensus participants to expend computing power. In Proof-of-Stake, a handicap is imposed by forcing participants to expend money. In both cases, the goal is to authenticate the results of consensus by auctioning control of the blockchain to the highest bidder. If voting power is assigned in proportion to verifiable handicaps, then a virtual network can be created on which the fraction of faulty replicas is guaranteed to fall below the security threshold of a specified consensus algorithm.

This approach predates Bitcoin by four years. It has been the foundation of permissionless Byzantine consensus since the publication of the first Sybil-resistant algorithms. [5] Unfortunately, it is deeply flawed. The present paper identifies three critical problems.

*A Confluence of Errors*

First, the traditional handicap principle is no longer good science. It reflects the state of biological signaling theory in the early 1990s. [6] Subsequent research, some of it by the same scientist who first formalized the handicap principle, has refuted the theory that a signal's reliability depends on its verifiable cost to the signaler. In reality, honest signals can be transmitted at zero cost, as long as the verifiable cost of a dishonest signal is sufficiently high. [7] There is thus no intrinsic reason that participants in permissionless consensus should be forced to waste money or computing power. The costs that Proof-of-Work and Proof-of-Stake systems impose on their users are a design flaw, not a feature.

Second, the assumption that Sybil-resistance is sufficient to guarantee consensus on a permissionless network is false. It conflates two distinct forms of statistical bias: sampling error and non-sampling error. The former relates to *which* elements of a population are included in a sample, the latter to *how* those elements are counted. [8] Sybil-resistance guarantees that the entities participating in consensus will be counted correctly, but it does not guarantee that those entities are an unbiased sample of the population that is axiomatically known to contain an honest majority or supermajority. The maximum fraction of corrupted entities within the protocol can be verified only if the set of protocol participants is large enough to ensure an accurate sample of the population whose minimum percentage of honest members is axiomatically known. Today's Sybil-resistant protocols do not satisfy the minimum threshold for statistical reliability, so they are vulnerable to "book prize" attacks, which exploit their reliance on non-probability sampling.

Third, axioms about control of a designated resource are incompatible with the concept of an adaptive adversary. Economic agents have varying resource endowments, so an adaptive adversary can alter how much of a given resource it controls simply by switching which agents it has corrupted. To establish the security of a resource-weighted protocol, it is therefore necessary to start with an axiom that is invariant in the face of adaptive corruption — such as the maximum combined value of all resources within the adversary's potential control — and to prove from that axiom that the adversary will be unable to acquire the fraction of the designated resource needed for an attack. Existing proofs of security for permissionless blockchains are tautological, because they start with this desired conclusion as their premise.

These three problems are related. Because computer scientists have mistakenly assumed that it is necessary to employ handicap-authenticated signaling, they have designed protocols that force all participants to waste money or computing power. Because today's protocols force all participants to waste money or computing power, most internet users decline to join, so the set of participating agents is not large enough to ensure a statistically unbiased sample. Because the set of protocol participants is not a reliable sample of the population, it is impossible to prove the existence of an honest majority within the protocol, so one has simply been assumed as an axiom.

It should be a red flag when the most rigorous proofs in a field all start with the same convenient-but-unreliable axiom. [9] The ultimate purpose of formal proofs of security is to provide information to end users about the real-world reliability of a given protocol. If the adversary model employed in a formal proof does not match the threats that are present in the real world, "the presented protocols may not be fully proven by the formalization." [10] A facially rigorous proof based on a flawed adversary model may be worse than no proof at all, because the illusion of security it provides can induce end users to leave themselves vulnerable to attack.

If society is going to employ permissionless ledgers to manage billions of dollars in value, the security of those systems should be derived from axioms that are known to be true with overwhelming probability. If proof of security cannot be obtained from reliable premises, then the public should be warned to treat permissionless ledgers as high-risk toys, not serious financial platforms for storing and exchanging value.

*Illusion of Security*

This pessimism may sound inconsistent with the track record of today's permissionless blockchains. It is not. Past results provide no credible assurances of future safety in this context, because — according to game-theoretic modeling — a rational adversary will delay its attack to inculcate a false sense of security among protocol participants, then "cash out" by executing a double-spending attack once a sufficiently large payoff is available. As Ponzi schemes famously illustrate, naïve induction is not a reliable method for assessing whether funds are safe.

In Proof-of-Work systems, the game-theoretic problem is that a free-entry condition exists, which prevents incentive compatibility between the agents currently in control of the protocol and the agents who will be in control of the protocol in the future. It has previously been hoped that Proof-of-Stake systems overcome this problem, because agents receive franchise value. [11] Our results demonstrate that the cure is worse than the disease: the franchise value available on a Proof-of-Stake ledger means that an adversary who pays the cost of an attack at the protocol's inception can execute a double-spending attack at any subsequent time, if it takes simple measures to avoid dilution.

This enables an adversary to profit through inter-temporal arbitrage; it can join a protocol early and purchase the "option" to attack, then once the option is "in the money" — i.e., once the ledger's market capitalization has expanded enough to make high-value double-spending attacks viable — the adversary can reap the reward for its patience.

The exploit we have just described severely decreases the externally verifiable security of ICO-launched Proof-of-Stake ledgers. The design of those protocols assumes that their security will automatically increase as their native assets appreciate. This assumption is in error. It is based on the conflation of time probability and ensemble probability, a mistake whose revelation rocked decision theory several years ago, but which has not yet been sufficiently digested in computer science. In an attempt to bridge the conceptual gap, we formalize a new class of "price adaptive" asynchronous adversary, which embodies the externally verifiable minimum cost of executing a successful attack.

The new formalism demonstrates that the "sleepy" or "dynamic" consensus achieved by longest-chain Proof-of-Stake algorithms does not fully replicate the beneficial attributes of Nakamoto consensus, as claimed in the prior literature. [12] [13] In Nakamoto consensus, late-joining agents are not forced to place their trust in the initial set of protocol participants, because control of the protocol becomes verifiably decentralized as the total hashing power on the network expands. In Proof-of-Stake algorithms, an initial "dishonest majority" can remain in control of the protocol indefinitely, because its voting power cannot be forcibly diluted by late-joining agents.

It has previously been argued that, even if such an attack is theoretically possible, it can be ruled out on a given blockchain once enough stake has migrated to new addresses. [14] The present paper disproves this claim. It demonstrates that the adversary can execute a *pseudo-transfer attack*, in which it shifts stake between its own cryptographic addresses to generate the illusion of decentralization. The movement of stake on a Proof-of-Stake ledger therefore does not assure security.

We demonstrate that, in formal terms, the externally verifiable security properties of such ledgers are indistinguishable from permissioned protocols with an initial, fixed set of participants. The difference is that, on permissionless ledgers, the identities of the agents are unknown. Proof-of-Stake protocols are effectively "trust maximized," because they require permanent trust in the honesty all past sets of participants. This trust may appear reasonable once a ledger achieves a large market capitalization, but the ledger's security guarantees may have actually been disabled by a "dishonest majority" in its infancy. Once that occurs, growth in its market capitalization does not provide additional security, since the adversary has already purchased enough stake to attack — and can preserve that power indefinitely by collecting its *pro rata* fraction of block rewards.

*A New Kind of Sybil-Resistance*

With KRNC, an initial "dishonest majority" cannot retain control of the protocol because its voting power will be verifiably diluted as new agents join and unlock their forked fiat money. Because the existence of an "honest majority" in the present does not require the consistent existence of an honest majority in the past, the security guarantees of KRNC are not path dependent— or, technically speaking, they become path dependent only after the cost of an attack stabilizes based on the ledger's mature market capitalization.

KRNC also surpasses past protocols because its Sybil-resistance mechanism, Proof-of-Balance, is *asymmetric*. Just as "a lock on a door increases the security of a house by more than the cost of the lock," Proof-of-Balance provides a degree of security that is exponentially greater than the costs it imposes on correct protocol participants. [11] The lock analogy is helpful in understanding why asymmetry is so important to efficiency and security. If locks were symmetric in the same sense as today's Proof-of-Work and Proof-of-Stake protocols, their purchase price would need to exceed the maximum "burglary budget" of all criminals. Homeowners would be locked in a constant financial battle with potential intruders, forced to waste their money on exorbitant locks, which still would not deliver any verifiable security.

Proof-of-Balance breaks this symmetry by harnessing the efficiency advantages of cue-authenticated signaling. Like a handicap, a cue is a trait that reliably communicates information because it is too costly to fake. The difference is the purpose and timing of the costs. A handicap is assumed for the purpose of signaling, so it is inherently wasteful and disadvantageous. A cue is a preexisting trait that is later adopted for signaling purposes at zero added cost. Cue-authenticated signaling eliminates the greatest barrier to blockchain adoption, because it makes it possible for agents to participate in consensus for free. Rather than incurring new costs, agents verify their pre-protocol sunk costs, which provides Sybil-resistance without the need to take anyone's money or waste their computing power.

Instead, Proof-of-Balance assigns cryptographic weights in proportion to each agent's existing "stake" in the fiat monetary system. The weight that is issued then serves two purposes. First, it allows weight owners to participate in consensus, just as if they had invested money in a Proof-of-Stake ICO. Second, the weight acts as a form of cryptographic "backing" for the original fiat money, which imbues it with inflation protection similar to Bitcoin. The goal is to eliminate the need for private cryptocurrencies by upgrading the money that everyone already owns.

The protocol can be analogized to a cryptographic return to the gold standard. However, it corrects the design flaw that caused the gold standard to collapse in the first place. Instead of entrusting central banks with cryptographic gold, KRNC distributes the new resource that backs the fiat monetary supply directly to the owners of fiat money. When they transact with one another, they transfer both their original fiat money and its cryptographic backing. This avoids any single point of failure that can default on the promise to maintain sufficient reserves.

*The Future of Consensus*

In a deep sense, KRNC is the logical progression of blockchain technology. For the participants in a blockchain protocol to reach consensus, it is necessary for them to agree on the protocol's initial state — i.e., the balances recorded in its genesis block. Because a monetary balance represents the quantity of goods or services that a given agent is entitled to receive from the other members of society, picking an immutable set of initial balances that the entire world will be willing to tolerate is a challenge that no prior protocol has been able to crack.

KRNC's solution is to import the abstract state of the existing global monetary ledger. This is the obvious Schelling point, because it is the distribution of balances that the world has already accepted. The results we obtain in this paper are an extension of that simple truth. In effect, the existing monetary system is the "longest chain" of value in society. KRNC is the first cryptographic protocol that allows that chain to be extended, rather than rejected and replaced.

## 1.3 Roadmap

In **Section 2**, we formalize the distinction between permissioned and permissionless Byzantine consensus. In our taxonomy, protocols are categorized according to the axioms their users must accept for liveness and safety to be common knowledge. Permissioned protocols require users to assume that an honest majority exists among the subset of the population who have received permission to participate in consensus, which makes them ideal for use by small groups of agents who know and trust one another. In a permissionless protocol, the entire population has the opportunity to participate in consensus, so users are required to assume that an honest majority exists among the subset of the population that chooses to participate in consensus.

In **Section 3,** we demonstrate that the security guarantees of permissionless protocols are statistically reliable if and only if a sufficient fraction of the total population is participating in the protocol to ensure an unbiased sample. Because the adversary can affect the composition of the set of agents that join a permissionless protocol, axiomatic knowledge that an honest majority exists in the population does not reliably translate into knowledge that an honest majority exists within the protocol itself. We employ a toy environment without Sybil attacks to establish the minimum-participation requirements needed to prevent book-prize attacks.

In **Section 4**, we extend the toy environment with a game-theoretic model of coordination in the presence of imperfect and incomplete information. We prove that the minimum sample size required for permissionless consensus is guaranteed if and only if the cost of participation is zero, because in that case the equilibrium strategy is for all agents to participate in consensus. However, when the cost is positive, no reliable security guarantees can be obtained, because the inherent risk of book-prize attacks deters correct agents from participating.

In **Section 5**, we prove that a protocol's vulnerability to book-prize attacks in the toy environment depends on the temporal structure of its sampling frame. If participation is limited to an initially small group of agents, the protocol will never achieve verifiable security, because those agents can take control and never relinquish it. The only verifiably secure solution is a dynamic system that causes the total quantity of voting power to expand as more agents join the protocol.

In **Section 6**, we replace the toy environment with a realistic model of resource-weighted Byzantine consensus, where the adversary can attempt Sybil and book-prize attacks. To apply the results from our toy model to the realistic model, we prove that the axioms employed by various resource-weighted protocols are special cases of a more fundamental "honest majority of capital" axiom. Defining a capital-normalized population of agents enables us to use our existing formulas to calculate how much capital a real-world protocol must sample for reliable security.

In **Section 7**, we demonstrate that the security of Proof-of-Stake protocols is limited by the temporal structure of their sampling frames. If stakes are sold to an initial set of agents in an ICO, an "option" to attack the ledger can be purchased for the price of buying a super-threshold fraction of the stake. To objectively quantify the seriousness of this risk, we define a new class of asynchronous adversary, which we refer to as "price adaptive." The new formalism confirms that Proof-of-Stake protocols are insecure and that a dynamic sampling frame is a necessary component of any solution.

In **Section 8**, we adapt our game-theoretic model to the realistic environment and derive the conditions that are necessary to achieve zero-cost Sybil resistance and protection against book-prize attacks. It is critical to select the staking resource that is already maximally favored by agents for its non-signaling value, because doing so allows sunk costs to be harnessed without the need for agents to assume new costs. We prove that, to employ this technique in the real world, a consensus protocol must assign cryptographic weights in proportion to ownership of money.

In **Section 9**, we demonstrate that Proof-of-Balance enables the incumbent monetary resource to be employed as the staking resource for a permissionless Byzantine consensus protocol. For ease of exposition, we first present the technology's security and performance advantages in a toy model where all balances are public, the fiat ledger is perfectly consistent, and a public-key infrastructure exists. We then remove these simplifying assumptions and iteratively specify the innovations that underpin the KRNC protocol, including synthetic consistency and retroactive weight assignments.

## Part II: Toy Environment

## 2. Requirements for Byzantine Consensus

### 2.1 Methodology

We begin with the foundational requirement of Byzantine consensus.

**Axiom 1**: Verifiable Byzantine consensus requires a constraint $k$ on the fraction of entities executing a consensus algorithm that are faulty.

This is our paper's equivalent of the "honest majority" axiom from Bitcoin or the quorum requirements of traditional BFT algorithms. The use of an abstract security threshold is merely a notational convenience to preserve the generality of our results across synchrony assumptions and algorithm families. The substantive innovation in Axiom 1 is that it does not assume that the security-threshold requirement will be satisfied; it merely specifies that the requirement exists. The core assumption in prior literature that the concentration of faulty agents or replicas would never exceed the security threshold rendered past proofs effectively tautological: they established that Byzantine consensus was guaranteed *if* the requirements for Byzantine consensus were met.

Proof that a system is reliable if a given condition is met, without an accompanying reason to believe that the condition will be satisfied when the system is deployed, does not meaningfully establish the system's real-world reliability. If distributed-systems engineering aims to deliver fault-tolerance assurances that meet the standards employed in the design of physical systems like bridges and airplanes, it must make concrete predictions based on verifiably accurate premises. The method advanced in this paper is an attempt to move the field in that direction.

### 2.2 Network and Protocol

We represent the population of agents with access to a computer network as a set $N$, whose elements are indexed by $n = 1, 2, …, |N|$. The population is divided into two sub-populations of $N_C$ correct agents and $N_F$ faulty agents, such that $N_C \cup N_F = N$. Faulty agents can engage in arbitrary behavior, including coordinated attacks, so we treat them as if they have been corrupted by an adversary, $\mathcal{A}$. The adversary embodies all the unpredictable factors that can interfere with protocol execution in the real world. It may be considered a formalization of Murphy's Law: anything that can go wrong, will go wrong, unless verifiably prevented by the protocol, $\mathcal{P}$, or specified as a limitation on the power of the adversary, $\mathcal{A}$.

All corrupted agents are faulty ($n \in N_F$) and all non-corrupted agents are correct ($n \in N_C$), unless otherwise specified. To track which agents $\mathcal{A}$ controls, we assign a *corruption status* of 0 to all correct agents and 1 to all faulty agents. Formally, the corruption status is an indicator $Y[n]$ associated with each agent $n$, where

$$Y[n] = \begin{cases} 0 \text{ if } n \in N_C \\ 1 \text{ if } n \in N_F \end{cases}$$

.

The mean corruption status $\overline{Y}$ within $N$ can be obtained using the formula

$$\overline{Y} = \frac{1}{N} \sum_{n=1}^{N} Y_n$$

,

which yields a decimal within the closed interval [0,1], such that $\overline{Y}$=0 if all agents on the network are correct and $\overline{Y}$=1 if all agents on the network are faulty. An "honest majority" of correct agents exists within $N$ only if $\overline{Y}$<0.5.

We define Byzantine consensus as atomic broadcast satisfying a list of required properties, which we assume to include liveness and safety, but which can be expanded to include additional properties, such as deterministic finality. The participants in a consensus protocol $\mathcal{P}$ running on the network are a collection $N_S$ drawn from $N$, such that $N_S \subseteq N$.[1] The faulty agents within the protocol comprise the intersection $N_S \cap N_F$ of all faulty agents and all protocol participants, which we write as $N_S^F$. Let the security threshold of the protocol's consensus algorithm be $k$, such that Byzantine consensus is guaranteed if and only if $N_S^F / N_S < k$.

Equivalently, the security threshold can be expressed in terms of the mean corruption status $\overline{y}N_S$ of protocol participants. Consensus within $N_S$ is guaranteed if and only if $\overline{y}N_S < k$. In contrast, when $\overline{y}N_S \geq k$, then the concentration of faulty agents within $N_S$ exceeds the capabilities of the protocol's consensus algorithm, and the adversary can therefore violate liveness or safety.

---

[1] Throughout this paper, re-indexing is employed when necessary for $\Sigma$ and $\Pi$ notation. Each subset $_{SUB}$ is treated as if it begins at the first element of its superset and ends at $|_{SUB}|$, the element of the superset indexed by the cardinality of the subset.

## 2.3 Revealed Trust

We model a protocol's trust assumptions as a set of objective properties, unrelated to the subjective intentions of the protocol's designer or users. Our approach is similar to prior formalizations of "access structures" and "adversary structures" in the literature on secure-multiparty computation. [15] It is also inspired by the concept of "revealed preference" in economics, which allows the preferences of consumers to be inferred objectively from their choices. [16] Similarly, we infer the objective trust assumptions of a consensus protocol based on the sets of agents that receive the power to successfully attack the protocol. This allows us to bootstrap an objective definition of trust from the existing, objective definition of Byzantine faults: a protocol trusts an agent if the corruption of that agent by the adversary would violate the protocol's liveness or safety guarantees.

We refer to a set of one or more agents who have the power to violate a consensus protocol's security guarantees as a *control set*, and we refer to the combination of all the control sets for a given protocol as its *control structure*. Control sets offer a formalized representation of Nick Szabo's concept of "security holes." [17] A control structure is, in a real sense, an objective map of the security holes that each protocol forces its users to tolerate.

Since we have an objective definition of trust, we can model a consensus protocol's *maximum user base* as the set of agents whose trust sets are compatible with the protocol's control structure. Ultimately, the value of a protocol depends on whether its control structure is well matched to the needs of its intended users. We will formalize two different paradigms, which can be used separately or in combination. *Permissioned consensus* allows a user to trust a consensus protocol if its control structure is verifiably dominated by agents whom the user knows and trusts. *Permissionless consensus* allows all agents in the population to trust a consensus protocol if the security of its control structure is validly derived from the axiom that a minimum fraction of all agents are trustworthy.

*Permissioned Consensus*

To model permissioned consensus, we assume that each agent $n$ believes that certain other, specific members of $N$ are correct. Let $T[n]$ be the output of a function that takes an agent $n$ and returns the set of agents within $N$ whom $n$ trusts. This collection of agents is the *trust set* of $n$. It is possible for different agents within $N$ to have different trust sets; in modal-logic notation, $\lozenge T[n] \neq T[n']$.

A protocol's *control structure* is the combinations of agents that have the power to violate its security guarantees. Formally, let the control structure be a family $F$ of sets over $N$, where $F$ contains every subset $f \subseteq N$ whose members can collectively violate liveness or safety, such that $(\bar{y}N_S \geq k) \equiv (N_S^F \in F)$. That is, the security threshold of a protocol's consensus algorithm is exceeded if and only if the set of faulty participants in the protocol belongs to the control structure $F$.

An agent $n$ can trust a protocol if every control set in the protocol's control structure contains at least one member of $n$'s trust set. Formally, let the set of agents who can trust a protocol be its *maximum user base*, $N_U$, which can be calculated using the formula

$$N_U = \{n \in N : T[n] \cap \forall f \in F \neq \emptyset\}$$

i.e., an agent $n$ is a member of the maximum user base of a protocol if each intersection of the agent's trust set and a control set within the protocol's control structure is not the empty set. In other words, an agent can trust a protocol if every possible combination of agents who can break the protocol by colluding contains at least one agent whom $n$ trusts not to participate in such an attack. If there is a combination of agents who can break the protocol, but none of those agents are in $n$'s trust set, then $n$ cannot trust the protocol.

Equivalently, an agent $n$'s ability to trust a consensus protocol can be defined based on the intersection of its trust set $T[n]$ and the population of protocol participants, $N_S$. The maximum user base $N_U$ of a protocol is then

$$N_U = \left\{ n \in N : \frac{|N_S| - |T[n] \cap N_S|}{|N_S|} < k \right\}$$

.

The left side of the inequality represents the number of agents within $N_S$ whom $n$ believes to be potentially faulty — i.e., who are not members of the trust set $T[n]$ — divided by the full number of agents within $N_S$. The number of agents participating in the protocol who do not belong to the trust set $T[n]$ must be small enough that, even if all of those agents were faulty, the adversary still would not control a super-$k$ share of voting power.

When trust is based on the specific identities of the members of $N_S$, the maximum user base $N_U$ depends on how much overlap exists between the trust sets of different members of $N$. For example, the Ripple protocol derives its security guarantees from the intersection of the trust sets of the members of $N_S$ [18] Unfortunately, this need for overlapping trust sets inherently limits the scalability of identity-based consensus.

*Permissionless Consensus*

Expanding a protocol's maximum user base $N_U$ to encompass all agents requires *permissionless consensus*, which allows an agent $n$ to trust a protocol without knowing the identifies of the participants in $N_S$. Indeed, permissionless consensus allows the agent $n$ to trust the protocol even if $n$ knows that none of the individual members of $N_S$ belong to $n$'s trust set, $T[n]$. Instead, the assumption is made that each agent $n$ knows that the set $N$ of agents with access to the network contains the "honest majority" (or supermajority, etc.) required for consensus, i.e. $\bar{Y} < k$. This enables $n$ to trust a protocol by verifying that the cardinality of the set $N_S$ of protocol participants is great enough to guarantee that the honest majority present on the network $N$ is retained within the protocol, i.e. $|N_S|$: $\bar{y}N_S < k$.

The existence of the requisite honest majority within $N$ is treated as common knowledge among the members of $N$. Each agent $n$ knows that an honest majority is present within $N$ because, as Micali observes, "society would not exist at all if there were not an honest majority out there." [19] This axiom can be expressed in our trust-set notation by defining a meta-agent, $\mathcal{N}$, who personifies the aggregated individual choices of each agent $n \in N$. Unlike the adversary $\mathcal{A}$, the meta-agent $\mathcal{N}$ does not control other agents or enhance their coordination abilities; rather, it is the individual behavior of the agents within $N$ that, when aggregated, produce a collective "choice" attributed to the meta-agent. Thus, if an agent $n$ believes that $\bar{Y} < k$, then the meta-agent necessarily belongs to that agent's trust set — i.e., $T[n] \ni \mathcal{N}$. The axiom that all agents with access to the network know that $\bar{Y} < k$ can likewise be expressed as all agents trusting the meta-agent — i.e., $\mathcal{N} \in \forall T[n \in N]$.

However, as Micali accurately warns, it is a misuse of this axiom to assume that an honest majority will exist within a "specialized group of people," who may not be a representative sample of the full population. [19] A meaningful permissionless security guarantee will exist if and only if the set of protocol participants, $N_S$, is large enough to verifiably capture the honest majority known to exist among $N$, the full set of agents with access to the network. If $N_S$ is not large enough, then agents who trust $\mathcal{N}$ will not have statistically reliable evidence that they should also trust the lesser meta-agent $\mathcal{N}_S$, and permissionless Byzantine consensus will therefore fail.

## 3. Book-Prize Attacks

How large is "large enough" for a set of consensus participants — i.e., what is the minimum cardinality $|N_S|_{MIN}$ required to ensure that $\bar{y}N_S < k$? We will first establish the strict deterministic minimum, then consider probabilistic alternatives.

### 3.1 Deterministic Threshold

*Formulas*

In the deterministic version of our model, it is assumed that the adversary $\mathcal{A}$ has control of every faulty agent $n \in N_F$ on the network. The adversary can therefore command all of those agents to join the protocol and become part of $N_S$. To provide deterministic protection against this attack, $N_S$ must be so large that it is guaranteed to contain the honest majority required for consensus, even if every faulty agent on the network joins the protocol, i.e., $\bar{y}N_S < k$ even if $N_F \subseteq N_S$. The minimum size of $N_S$ for a deterministic guarantee is provided by the formula

$$|N_S|_{MIN(D)} = |N_S^F|_{MAX} + |N_S^C|_{MIN} : \frac{|N_S^F|_{MAX}}{|N_S|} < k$$

,

where

$$|N_S^F|_{MAX} = (\bar{Y}_{MAX})(|N|_{MAX})$$

,

i.e., the cardinality of the minimum set of participants in the protocol for deterministic security is equal to the maximum cardinality of the intersection $N_S^F$ of all faulty agents and protocol participants, plus the minimum number of correct agents such that the ratio of the maximum number of faulty protocol participants to the number of protocol participants remains below the consensus algorithm's security threshold.

The maximum number of faulty protocol participants is the maximum cardinality of the intersection $N_S^F$, which is obtained by multiplying the maximum mean corruption status on the network by the maximum cardinality of the set of agents on the network.

*Application*

The population $N$ can be taken as a constant, which eliminates the need to specify its maximum cardinality. Security guarantees can instead be defined in terms of the minimum portion of $N$ that must be included within $N_S$ for consensus to be guaranteed. For example, if an agent $n$ knows that at least 90% of the population is correct, then at most 10% of the population can be faulty ($\bar{Y}_{MAX} = 0.1$), so if a protocol's consensus algorithm requires a simple majority of correct agents ($k = 0.5$), consensus is guaranteed if and only if more than 20% of the population participates in the protocol. If the agent knows that 99% of the population is correct, then any sample greater than 2% of the population is sufficient for deterministic permissionless consensus. If the percentage of the population known to be correct drops to 80%, then participation of more than 40% of the population is required.

We can apply the minimum participation requirements for permissionless consensus in our deterministic-participation model to protocols running on the internet by assuming a maximum cardinality of 4 billion for $N$, based on present estimates of the number of unique internet users. ($|N|_{MAX} \approx 4\text{x}10^9$). To exceed 2% participation, as required if 99% of the internet population is correct, a consensus protocol needs more than 80 million users. To exceed 20% participation, as required if 90% of the internet population is correct, the protocol needs more than 800 million users. To exceed 40% participation, as required if 80% of the internet population is correct, the protocol needs more than 1.6 billion users.

## 3.2 Probabilistic Threshold

*Formulas*

In the probabilistic version of our model, it is assumed that the adversary $\mathcal{A}$ can only control faulty agents once they join the protocol, i.e., if $n \in N_S^F$. The adversary therefore does not have the power to cause every faulty agent in the population to join the protocol. Instead, the set of participants in the protocol is a function of which agents in the population choose to participate. This enables us to obtain probabilistic security guarantees by adapting the well-known Horvitz-Thompson estimator from the field of statistical inference. [8] [20]

The minimum size of $N_S$ for a probabilistic guarantee is given by the formula

$$|N_S|_{MIN(P)} = E(|N_S^F|) + |N_S^C|_{MIN} : \frac{E(|N_S^F|)}{|N_S|} < k$$

where

$$E(|N_S^F|) = (\bar{Y}_{MAX} + Bias)(|N_S|)$$

The first term, $E(|N_S^F|)$, is the expected number of faulty agents within the protocol, i.e., the expected cardinality of the intersection of the set of faulty agents and the set of protocol participants. It is equal to the maximum mean corruption status within the population ($\bar{Y}_{MAX}$), plus the bias between that value and the mean corruption status within the protocol, multiplied by the cardinality $|N_S|$ of the set of protocol participants.

To formalize the *Bias* term, we must define the selection mechanism that dictates which members of $N$ are included within $N_S$. We assume that all members of $N$ know that joining the protocol requires paying a non-negative cost, $c$, in return for a reward, $r$. The agents within $N$ are rational but lazy, in the sense that an agent $n$ will join a protocol if and only if it perceives the value of its individualized reward $r_n$ as exceeding the value of its individualized cost, $c_n$. Thus, $r_n > c_n$ for all $n \in N_S$ and $c_n \leq r_n$ for all $n \notin N_S$. The probability that an agent $n$ perceives its reward as exceeding its cost is thus identical to the probability that the agent joins the protocol. To quantify its probability of joining, every agent $n$ is associated with a vector $j$ of $N$ indicators, where $j = (j_1, j_2, \ldots, j_N)$. Let the $n$-th indicator take the value $j_n$, where $j_n = 1$ if $n$ joins the protocol ($n \in N_S$) and $j_n = 0$ if not ($n \notin N_S$). The probability $\rho_n$ that an agent $n$ will participate in the protocol is equal to the expected value $E(j_n)$. We refer to this as the agent's *participation propensity*.

The agents on the network who choose to join the protocol form the set of participants, $N_S$, where

$$N_S = \sum_{n=1}^{N} j_n$$

If every agent in the population has the same probability of joining the protocol — i.e., if $\rho_n$ takes the same value for all $n$ — then the mean corruption status within the protocol will closely approximate the mean corruption status within the population. When formalized as

$$\bar{y}_{N_s} = \frac{1}{N_s} \sum_{n=1}^{N} j_n Y_n$$

this yields an expected value of

$$E\left(\bar{y}_{N_s}\right) \approx \bar{Y}_N^* = \frac{1}{N\bar{\rho}} \sum_{n=1}^{N} \rho_n Y_n$$

where $N\bar{\rho}$ is the mean participation propensity of agents within $N$.

Based on the expected value $E\left(\bar{y}_{N_s}\right)$, the *Bias* term can be calculated according to the formula

$$Bias = E\left(\bar{y}_{N_s}\right) - \bar{Y}_N \approx \bar{Y}_N^* - \bar{Y}_N = \frac{C(\rho, Y)}{\bar{\rho}}$$

in which

$$C(\rho, Y) = \frac{1}{N} \sum_{n=1}^{N} (\rho_n - \bar{\rho})(Y_n - \bar{Y})$$

is the covariance between corruption status and participation propensity, and $\bar{\rho}$ is the mean participation propensity. Note that the *Bias* term can never be negative because, as we prove in the next section, the mean participation propensity of faulty agents will always be greater than or equal to that of correct agents.

The amount of bias depends on two factors. First, the lower the covariance between corruption status and participation propensity, the smaller the sample size required. Second, the greater the mean participation propensity, the larger the sample size achieved. These factors are related: the maximum potential covariance between corruption status and participation propensity, $C(\rho, Y)_{MAX}$, remains at its peak until the number of participants in the protocol is equal to the maximum number of faulty agents in the population, i.e., $|N_S| = |N_S^F|_{MAX}$, because until that threshold is reached it remains possible that every agent who has joined the protocol is faulty. Once the potential supply of faulty agents has been exhausted, every additional agent who joins the protocol increases $|N_S^C|_{MIN}$, thereby lowering $C(\rho, Y)_{MAX}$ until it assumes a value of 0 once $|N_S| = |N|_{MAX}$.

*Application*

By applying these formulas, we can compare the strength of different protocols' probabilistic security guarantees.

First, we can quantify how much increasing participation improves security by measuring how much it reduces bias. [20] Let $\bar{\rho}[\mathcal{P}_A]$ be the mean of participation propensity for Protocol A, and $\bar{\rho}[\mathcal{P}_B]$ be the mean participation propensity for Protocol B, where the latter is greater by a positive constant $\Delta$, such that $\bar{\rho}[\mathcal{P}_A] + \Delta = \bar{\rho}[\mathcal{P}_B]$. The bias of Protocol A, $Bias[\mathcal{P}_A]$, will be larger than the bias $Bias[\mathcal{P}_B]$ of Protocol B, per the formula $Bias[\mathcal{P}_A] = Bias[\mathcal{P}_B] + \frac{\Delta}{\bar{\rho}[\mathcal{P}_B]}$. The lower the baseline mean participation propensity, the greater the reduction in bias achieved for a given $\Delta$. For example, if the mean participation propensity is increased from 0.1 to 0.2, bias is reduced by 50%. If it is increased from 0.5 to 0.6, bias is reduced by 17%. In both cases, $\Delta = 0.1$, but the bias reduction in the former is ~3-times greater due to the insufficiency of the baseline sample size.

Second, we can calculate the maximum possible bias for a given protocol by making worst-case assumptions. When this value is obtained for multiple protocols, their reliability in the worst-possible conditions can be compared. To start, we decompose the bias estimator, using the formula

$$Bias \approx \frac{C(\rho, Y)}{\bar{\rho}} = \frac{J(\rho, Y)S(\rho)S(Y)}{\bar{\rho}}$$

where $J(\rho, Y)$ is the correlation coefficient between corruption status and protocol participation, $S(\rho)$ is the standard deviation of response probabilities, and $S(Y)$ is the standard deviation of corruption status.

For a given mean participation propensity $\bar{\rho}$ there is a maximum value that cannot be exceeded by the standard deviation $S(\rho)$ of participation propensity. Formally,

$$S(\rho) \leq S(\rho)_{max} = \sqrt{\bar{\rho}(1 - \rho)}$$

The worst-case scenario for bias occurs when $S(\rho) = S(\rho)_{max}$ and the correlation coefficient $R(\rho, Y)$ takes its maximum value, $R(\rho, Y) = 1$. We can thus calculate the maximum bias for a given protocol using the formula

$$Bias_{max} = S(Y)\sqrt{\frac{1}{\bar{\rho}} - 1}$$

However, identifying an acceptable amount of potential bias to tolerate in a trust-minimized protocol is difficult, which in turn makes it hard to objectively define a minimum acceptable ratio between $N_S$ and $N$. To avoid the need for subjective judgments, we take an apophatic approach and instead identify an objectively insufficient value.

It is indisputably necessary for the strength of a trust-minimized protocol's security guarantees to exceed the degree of assurance mandated for political polls and opinion surveys. Based in part on the $Bias_{max}$ formula just described, mean response propensities below 0.2 are considered unacceptable for polls and surveys due to the severe risk of bias from even modest correlations between inclusion probability and the variable of interest. [20] Under that standard, $N_S$ must include at least 1/5 of $N$. If we once again assume $|N|_{MAX}$=4 billion and $k$=0.5, then permissionless Byzantine consensus requires at least 800 million participants.

### 3.3 Sybil resistance is not enough

So far, we have only considered what is required for Byzantine consensus in a toy environment where the adversary cannot perform Sybil attacks, but the results we have obtained are already a cause for concern. The toy environment was intentionally constructed to be more forgiving than the real world. [21] Yet, as we have just demonstrated, the adversary can still violate liveness and safety by biasing the composition of the set of protocol participants.

We refer to this exploit as a *book-prize attack*, because the strategy was once used to alter the winner of the Book of the Year prize in the Netherlands. [8] The honor is supposed to be bestowed on the top example of Dutch literature published in a given year. The winner of the prize is determined through direct voting, in which any citizen of the Netherlands can participate. In 2005, there were 92,000 participants, more than double the 40,000 readers who participated the prior year. [22] Yet more than 70% of the votes were cast in favor of a new translation of the Bible — which had previously been ruled ineligible, since it was not a work of Dutch literature.

The voting process was not compromised by pseudo-spoofing or a Sybil attack. A strict "one-person, one-vote" condition was enforced using an external identity register. However, the outside group affiliated with the Bible translation was still able to bias the results: it simply recruited its members to participate. Because the mean participation propensity of Dutch readers was too low to ensure a representative sample, the outside group was able to dictate the outcome of voting.

A book-prize attack exploits the same vulnerability to compromise permissionless consensus protocols. If the number of agents participating in a consensus protocol is too low to ensure a representative sample of the agents with access to the underlying permissionless network, then the mean corruption status within the protocol may be much greater than axiomatically assumed. If the bias is sufficient to raise the protocol's mean corruption status above the security threshold of its consensus algorithm, then the adversary can violate liveness and safety.

The distinguishing feature of a book-prize attack is that, whether by adding faulty agents or excluding correct ones, the adversary raises the mean corruption status of participants in a permissionless protocol above the mean corruption status of agents who have access to the underlying permissionless network. This risk has historically been overlooked in the distributed-systems literature due to a subtle error in the way that resource-weighted consensus has traditionally been modeled.

We illustrate this error in **Figure 3.1**, which is based on the three-level diagram from the first paper on using Proof-of-Work to achieve Sybil-resistant Byzantine consensus on a permissionless network. [5]

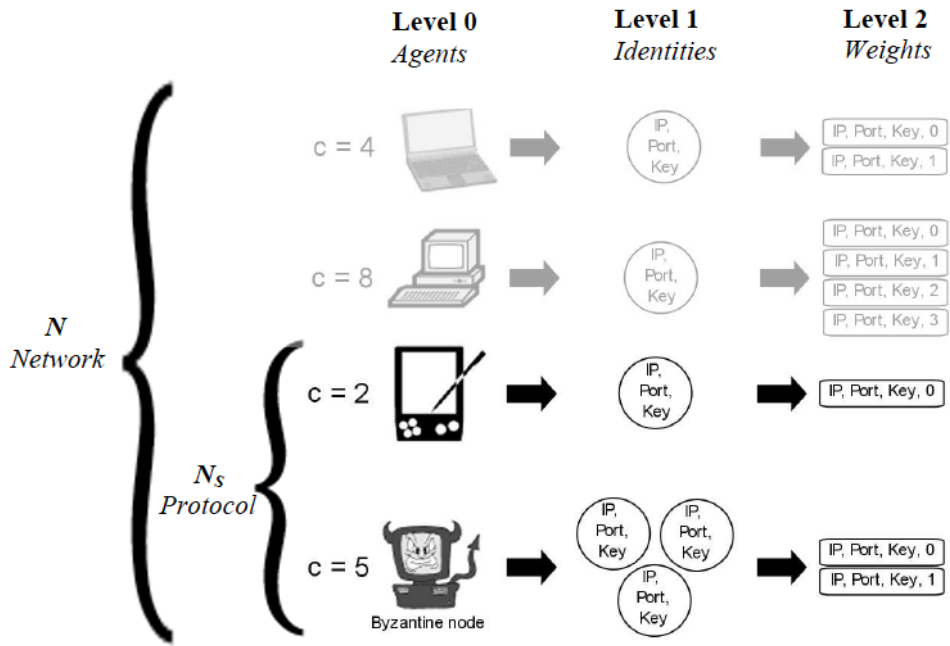

**Figure 3.1:** Consensus is guaranteed at Level 0 and Level 2 of the network, but fails at every level of the protocol.

In the original version of the diagram, there was no distinction between the collection of agents with access to the permissionless network and the subset of those agents that join the consensus protocol. The proof that Byzantine consensus could be achieved on the permissionless network if $k$=1/3 was based on counting the voting power of the entire population now labeled $N$: The adversary controls 1/4 agents at Level 0, so consensus is possible; it performs a Sybil attack at Level 1, acquiring 3/6 votes, which makes consensus impossible; weighing votes in proportion to computing power restores the ability to reach consensus at Level 2, where the adversary receives 2/9 of voting weight, an approximation of its 5/19 share of all computing power.

The assumption that the paper flagged as being potentially unrealistic was the relative parity of computing power between the agents. But there is a more fundamental problem. Given that the protocol is permissionless, all of the agents with access to the underlying network cannot automatically be relied on to participate in consensus. Instead, only the subset of agents who exercise the option to join the protocol will actually cast votes. The fact that the mean corruption status on the network is low enough for Byzantine consensus does not establish that the mean corruption status of protocol participants is also low enough for consensus. For that deduction to be valid, the number of agents participating in the consensus protocol must verifiably exceed the minimum participation threshold. If the participation propensity on the network is too low for a protocol to reach the minimum participation threshold, then a permissionless protocol is vulnerable to book-prize attacks.

**Figure 3.1** illustrates how an adversary can exploit that vulnerability to successfully attack a Sybil-resistant protocol. The adversary controls only 1/4 of the agents with access to the network, but 1/2 of the agents on the network have failed to join the protocol. Within the protocol itself, the adversary therefore controls 1/2 of the voting power at level 0, which makes Byzantine consensus impossible. At Level 1, the adversary executes a Sybil attack, so it becomes even more powerful, controlling 3/4 identities registered in the consensus protocol. At Level 2, Byzantine consensus remains impossible, even after weights are assigned using Proof-of-Work; the adversary controls 5/7 of the computing power among the set of consensus-protocol participants, and it receives 2/3 of voting weight.

This illustrates how book-prize attacks can defeat countermeasures designed to prevent pseudo-spoofing. A real-world permissionless protocol must verifiably resist both Sybil attacks and book-prize attacks, because the adversary can employ both in combination.

*Performance Implications*

The failure of existing protocols to deliver verifiable protection against book-prize attacks undercuts not only their security, but also their performance. The lower the maximum fraction of the participants in a consensus protocol who may be faulty, the more quickly transactions in the protocol can be confirmed. This is true in two senses.

First, sharding may be employed to assign responsibility for designated data to a subset of all protocol participants. The gains in efficiency from this approach are offset by a reduction in security, since confirmations are received only from the fraction of agents participating in the shard. This may have seemed like a reasonable tradeoff before book-prize attacks were formalized. However, now that the maximum mean corruption status of all protocol participants is in doubt, the mean corruption status within individual shards is highly questionable.

Second, and more important, the maximum mean corruption status of protocol participants dictates the confirmation times required by algorithms based on Nakamoto consensus. Even if we prohibit Sybil attacks by assuming complete resource-parity among agents, the risk of book-prize attacks means that transaction confirmation in today's state-of-the-art protocols actually takes a minimum of hours or days, not minutes.[2]

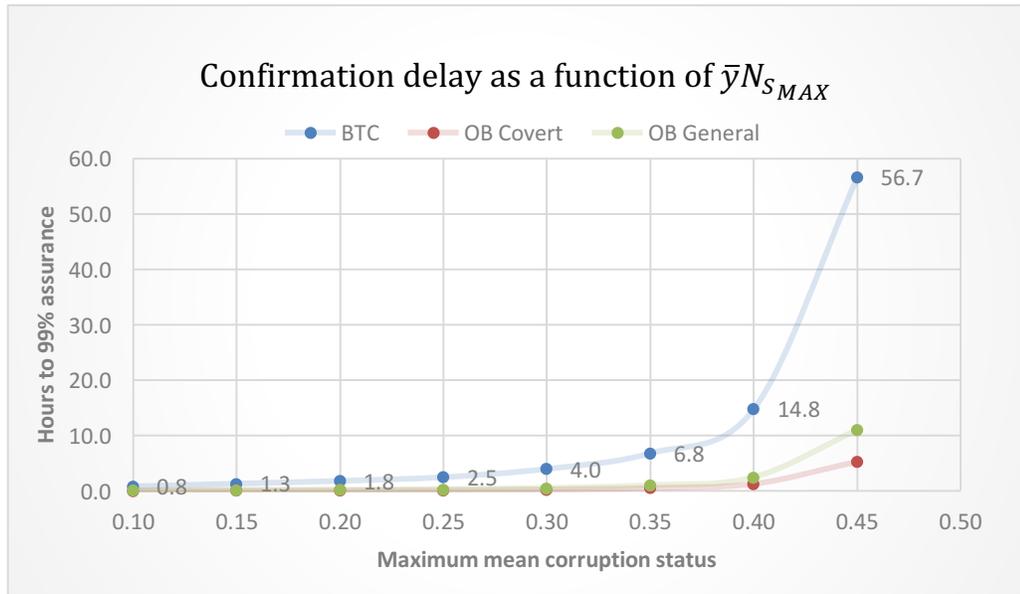

**Figure 3.2:** The risk of book-prize attacks delays transaction confirmation.

This problem is illustrated in **Figure 3.2**, which shows how both Proof-of-Work and Proof-of-Stake ledgers employing longest-chain algorithms require unacceptably long periods of time to achieve 99% assurance of finality due to the risk of book-prize attacks. In the legend, BTC is Bitcoin, and OB Covert and OB General are two variations of the Ouroboros protocol. To preempt any accusation of bias, the consequences of book-prize attacks are plotted according to the confirmation speeds claimed by the IOHK team in the Ouroboros paper. [12]

---

[2] Because maximum mean corruption status is the critical determinant of both reliability and performance in longest-chain algorithms, we employ it as our core metric of a protocol's security and speed.

## 4. Cost of Participation

The problem with existing Sybil-resistance methods is not merely that they are ineffective against book-prize attacks, but that they make protocols *more* vulnerable to those attacks. In an attempt to prevent Sybil attacks, such systems impose a verifiable cost for voting; this increases the risk of book-prize attacks, because it decreases the number of agents who choose to participate in consensus.

As we prove in this section, the cost tolerance of faulty agents weakly dominates the cost tolerance of correct agents. Imposing costs on all agents therefore produces an inherent risk of adversarial bias in the composition of the set of protocol participants. The only way to verifiably guarantee permissionless consensus is therefore to reduce the cost of participation to zero. For now, we will defer the question of how this can be accomplished. The results in this section merely establish that the task is necessary.

We start by specifying a game-theoretic representation of our Sybil-free environment with probabilistic participation, which **Figure 4.1** shows in extensive form. We then use that model to prove the theorem that permissionless Byzantine consensus requires zero-cost participation. First, we prove that consensus can be guaranteed among the set of agents with access to the network if and only if the equilibrium strategy is for both faulty and correct agents to join the consensus protocol. Second, we prove that this outcome — known in signaling theory as a "pooling equilibrium" — can be guaranteed if and only if the cost of joining the protocol is zero.

### 4.1 Extensive-Form Game

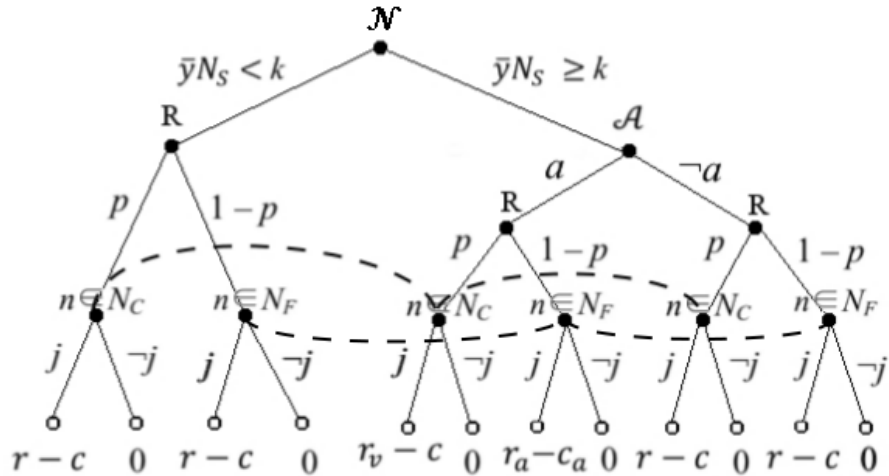

**Figure 4.1:** Sybil-Free Permissionless Consensus

*Players*

There are $|N|+3$ players in the game: $|N|$ agents with access to the network, the adversary $\mathcal{A}$, a meta-agent $\mathcal{N}$ that represents the aggregated individual decisions of every agent $n \in N$, and a stochastic function $R$ that represents the probability that such an agent is faulty ($n \in N_F$) or correct ($n \in N_C$). The goal of every agent $n$ is to maximize its individual payoff; the goal of $\mathcal{A}$ is to maximize the net payoff of all faulty agents. No goal is specified for $\mathcal{N}$, because its moves are determined by the individual goals of the agents that comprise $N$.

*Nodes and Payoffs*

At the root node, the meta-agent $\mathcal{N}$ chooses whether the set of agents who join the protocol will have a mean corruption status $\bar{y}_{N_s}$ that exceeds the security threshold of the protocol's underlying consensus algorithm. Unlike the adversary, $\mathcal{N}$ has no power to control or communicate with agents, so its decision node does not permit the members of $N$ to make a coordinated choice about whether to join the protocol.

Instead, $\mathcal{N}$'s choice is dictated by the criteria already specified in the probabilistic participation model: an agent $n$ will choose to join the protocol if and only if it expects its reward to outweigh the cost of participation. The reward and cost are specified at the bottom of the game tree, which introduces an element of reflexivity: every agent $n$ must make its choice based on its belief about $\mathcal{N}$'s choice, but to infer $\mathcal{N}$'s choice $n$ must take into account its own choice, because $n$ is itself one of the agents whose aggregated individual choices comprise $\mathcal{N}$'s decision.

If $\mathcal{N}$ decides that the mean corruption status of the agents who join the protocol will fall below the security threshold of its consensus algorithm ($\bar{y}N_S < k$), then consensus is guaranteed, so the adversary has no power to attack. If $\mathcal{N}$ decides that the mean corruption status of the agents who join the protocol will exceed the security threshold, ($\bar{y}N_S \geq k$), then the adversary can choose whether it will attack ($a$) or comply with the protocol ($\neg a$). If the adversary decides not to attack ($\neg a$), then the resulting subgame is isomorphic to the case in which consensus is guaranteed ($\bar{y}N_S < k$).

After the ability of the adversary to attack has been determined, the stochastic function $R$ dictates whether each agent $n$ within $N$ is faulty or correct, according to the binomial distribution $\rho \oplus 1\text{-}\rho$. The location of the three stochastic nodes within the game tree is not significant; an equivalent extensive-form representation can be constructed by placing a single stochastic node at the root and modifying the information sets. The version of the game shown in **Figure 4.1** begins at the meta-agent's decision node as a matter of convenience. The structure of the resulting game tree pairs together payouts for faulty and correct agents in the same protocol execution, which simplifies the notation required to prove the Bayesian Nash equilibrium.

In the left subgame — i.e., the branch of the tree in which $\bar{y}N_S < k$ — the action set and payoffs are identical for correct and faulty agents, because the adversary $\mathcal{A}$ has no power to attack. An agent $n$ can either stake ($j$) and receive a payoff of the standard reward minus the standard cost ($r$–c) or decline to join ($\neg j$) and receive a payoff of zero.

In the right subgame — i.e. the branch of the tree in which $\bar{y}N_S \geq k$, b $\mathcal{A}$ has declined to attack ($\neg a$) — the actions and payoffs are the same as in the left subgame. By voluntarily choosing not to attack, the adversary effectively replicates the condition in which it is unable to attack.

In the center subgame — i.e., the branch of the tree in which $\bar{y}N_S \geq k$, and the adversary $\mathcal{A}$ has chosen to attack ($a$) — correct and faulty agents both receive a payoff of zero if they decline to participation, but they receive different payoffs if they join. Correct agents who join pay the standard cost ($c$) while receiving a diminished reward $r_v$ that reflects the fact that the ledger has been victimized in an attack. Faulty agents who join the protocol participate in the attack, so they receive a larger reward that reflects their share of the profits generated by the attack ($r_a$), while paying a cost ($c_a$) that includes both the standard cost and any punishment for the attack.

*Imperfect Information*

Each agent $n$ knows whether it is faulty or correct, so the decision nodes for faulty and correct agents do not belong to the same information set. However, when an agent must choose between joining the protocol ($j$) and declining to participate ($\neg j$), it does not know whether $\bar{y}N_S < k$ or whether the adversary $\mathcal{A}$ has chosen to attack ($a$). As a result, the agent is unable to determine which subgame play has reached when its decision node arrives.

The limitation applies to both correct and faulty agents in the probabilistic-participation model, because the adversary cannot induce faulty agents to join the protocol. All three of the decision nodes for correct agents therefore belong to a single information set, and all three of the decision nodes for faulty agents also belong to a single information yet. Because none of the agents know which subgame has materialized, both correct and faulty agents will attempt to maximize their utility by choosing whichever action ($j \oplus \neg j$) delivers the greater expected payoff when averaged across the probability distribution for the three subgames.

An agent's expected payoff from joining the consensus protocol is determined by its expected reward, minus its expected cost. We denote the expected reward by $\lambda_Y$, where the subscript $Y$ indicates type-dependence, such that all correct agents expect to receive a reward of $\lambda_C$ and all faulty agents expect to receive a reward of $\lambda_F$. The values of the expected rewards are given by the formulas

$$\lambda_C = (r)(\rho_L + \rho_R) + (r_v)(\rho_C)$$

and

$$\lambda_F = (r)(\rho_L + \rho_R) + (r_a)(\rho_C)$$

where $\rho_L$ is the probability of the left subgame, $\rho_R$ is the probability of the right subgame, and $\rho_C$ is the probability of the center subgame.

We denote the expected cost with $\psi_Y$, which is also type dependent. The cost paid by correct agents is identical in all three subgames, so $\psi_C = c$. The same is not true for faulty agents, whose expected costs are given by the formula

$$\psi_F = (c)(\rho_L + \rho_R) + (c_a)(\rho_C)$$

These definitions allow us to extend the conditions previously specified in our probabilistic-participation model. The axiom that an agent will join the protocol if and only if the expected reward exceeds the expected cost can be formalized as a pair of type-dependent decision rules.

For correct agents,

$$n \in N_S^C \equiv \lambda_C > \psi_C$$

For faulty agents,

$$n \in N_S^F \equiv \lambda_F > \psi_F$$

*Incomplete Information*

It is common knowledge that the binomial distribution $(\rho \oplus 1 - \rho)$ employed by the stochastic function $R$ to select a faulty or correct agent at random from the set of agents with access to the network is constrained by the mean corruption status, $\overline{Y}$, such that $\overline{Y}_{MAX} \geq 1 - \rho \geq \overline{Y}_{MIN}$.

The amount of the standard cost $c$ is known to all players. The value of $r$ is not known to the players, but it is common knowledge that $r > 0$. This reflects the fact that, although the value of a reward denominated in a new cryptographic asset is highly uncertain, negative or zero values can be excluded.[3] Each player estimates the value of $r$ according to its own probability distribution. Those distributions are private information, so players are uncertain how much value other players attach to receiving $r$. However, to exclude trivial solution concepts in which only correct agents participate, we require that the average estimated value of $r$ by faulty agents be at least equal to the average estimated value of $r$ by correct agents.

---

[3] If a value of zero for receiving the reward cannot be excluded, then even if the cost of participation is zero, a net payoff of zero for joining the protocol cannot be ruled out. Participation will weakly dominate non-participation, because it guarantees the best possible payoff. However, participation will not strongly dominate, because the scenario in which both choices yield a zero payout cannot be excluded.

It is common knowledge that the successful execution of an attack decreases the value of a ledger's cryptocurrency ($r_v < r$). The realized amount of the decrease is not known to the players, but $r_v$ is known to be positive. It is common knowledge that $r_v < r_a$, because $r_a$ is composed of $r_v$ plus the faulty node's share of the attack profits. However, the value of $r_a$ is known only to the adversary, $\mathcal{A}$. This reflects the fact that the profits available from attacking a cryptocurrency ledger may be higher than the quantity of wealth that can be stolen in a double-spending attack, since there are agents (competitors, terrorists, etc.) who may assign a value to harming the ledger that exceeds its extant market capitalization.

It is common knowledge that $c_a \geq c$, because $c_a$ is composed of $c$ plus any punishment inflicted on the adversary. The concept of punishment is broad enough to encompass both criminal penalties and protocol-enforced penalties such as slashing. However, the amount of the difference, if any, between $c_a$ and $c$ is unknown, and players' beliefs about this value are private information. This reflects real-world uncertainty about the effectiveness of these deterrents at safeguarding distributed ledgers.

*Equilibrium Conditions*

We refer to a Bayesian Nash equilibrium in which one or more agents join the protocol as a *signaling equilibrium*, and one in which no agents join the protocol as a *non-signaling equilibrium*. Formally, a signaling equilibrium exists in the event that

$$(\lambda_C > \psi_C) \vee (\lambda_F > \psi_F)$$

and a non-signaling equilibrium exists when

$$(\lambda_C \leq \psi_C) \wedge (\lambda_F \leq \psi_F)$$

By definition, permissionless Byzantine consensus requires a signaling equilibrium, because in a non-signaling equilibrium no agents broadcast or receive messages.

A *pooling* equilibrium exists when the dominant strategy for faulty and correct agents is identical, such that an agent $n$'s decision to join the protocol provides no information about whether the agent is faulty or correct. A *separating* equilibrium exists when an agent's type affects its dominant strategy, such that knowledge of $n$'s decision to participate in the protocol conveys information about whether $n$ is faulty or correct. If that information indicates that $n$ is faulty — i.e. if the fact that an agent joined the protocol increases the probability that it has been corrupted by the adversary — then a separating equilibrium is *adversarially biased*.

Formally, a pooling equilibrium exists if
$$\{(\lambda_C \leq \psi_C) \wedge (\lambda_F \leq \psi_F)\} \oplus \{(\lambda_C > \psi_C) \wedge (\lambda_F > \psi_F)\}$$

A separating equilibrium exists if
$$\{(\lambda_C \leq \psi_C) \wedge (\lambda_F > \psi_F)\} \oplus \{(\lambda_C > \psi_C) \wedge (\lambda_F \leq \psi_F)\}$$

and the separating equilibrium is adversarially biased if
$$(\lambda_C \leq \psi_C) \wedge (\lambda_F > \psi_F)$$

The game-theoretic representation of the toy environment with probabilistic participation is now complete, and we are ready to prove this section's core result.

## 4.2 Zero-Cost Theorem

**Theorem 1**: *Permissionless Byzantine consensus is a verifiable Bayesian Nash equilibrium if and only if the cost of participation is zero.*

**Proof:** This theorem is a straightforward consequence of the two lemmas we prove below. Once we identify the potential Bayesian Nash equilibria in the extensive-form game, we analyze how the realized equilibrium changes depending on whether $c$ is positive or zero. Our model indicates that if the cost of participation in a protocol is positive then there is a risk that the set of participants will be adversarially biased, and that risk makes it impossible to recruit correct agents since those agents cannot guarantee that the protocol's mean corruption status will remain below its consensus algorithm's security threshold.

**Lemma 1**: *Any separating equilibrium between correct and faulty agents is adversarially biased.*

**Proof**: In the left and right sub-subgames — i.e., when $\bar{y}N_S < k$ or when $\mathcal{A}$ chooses $\neg a$ — the payoffs for faulty and correct agents are identical: $j$ yields $r$–$c$ and $\neg j$ yields zero for any agent $n$. Therefore, a separating equilibrium can only arise from the center subgame, where $\bar{y}N_S \geq k$ and the adversary chooses $\mathcal{A}$.

For a separating equilibrium to emerge in the center subgame, there must be a type-dependent payoff differential. There is the potential for such a differential, because the payoffs specified for faulty and correct agents are not identical. Both types receive 0 for $\neg j$, but a type-dependent payoff differential exists for $j$: faulty agents receive $r_a$–$c_a$ while correct agents receive $r_v$–$c$. A pooling equilibrium will exist if $r_a$–$c_a$ and $r_v$–$c$ are both $> 0$ or both $< 0$, because the dominant strategy for faulty and correct agents will be the same. If $r_a$–$c_a > 0$ and $r_v$–$c < 0$, a separating equilibrium will exist in which $j$ is the dominant strategy for faulty agents and $\neg j$ is the dominant strategy for correct agents. If $r_a$–$c_a < 0$ and $r_v$–$c > 0$, a subgame separating equilibrium will exist in which $j$ is the dominant strategy for correct agents and $\neg j$ is the dominant strategy for faulty agents.

However, the latter subgame separating equilibrium is never realized, because it is blocked by the adversary $\mathcal{A}$ at its decision node. If $r_v - c > 0$, then $r - c > 0$, because $r > r_v$. Play will therefore never reach the center subgame, because the payoff to $\mathcal{A}$ from $\neg a$ will weakly dominate $a$: if $n \in N_F$ plays $\neg j$, then $\neg a$ and $a$ are both worth 0 to $\mathcal{A}$, but if $n \in N_F$ plays $j$, then $\neg a$ returns the higher payoff, because $r - c > 0$ and $r_a - c_a < 0$. The dominant strategy for $\mathcal{A}$ whenever $r_a - c_a < 0$ and $r_v - c > 0$ is to prevent a subgame separating equilibrium by playing $\neg a$, since the pooling equilibrium within the right sub-subgame will maximize the payoff for faulty agents.

A realized subgame separating equilibrium exists if and only if $r_a - c_a > r - c$ and $r_a - c_a > 0$ while $r_v - c < 0$. In that circumstance, the inequalities required for an adversarially biased separating equilibrium in the center subgame are present, because $j$ for $r_a - c_a$ is the dominant move for faulty agents and $\neg j$ for 0 is the dominant move for correct replicas. The separating equilibrium is realized, because the weakly dominant strategy for $\mathcal{A}$ is to force the center subgame by playing $a$: the payoff available if $n \in N_F$ chooses $j$ is greater than in the right subgame, and it is equal if $n \in N_F$ chooses $\neg j$. □

**Lemma 2**: *A pooling equilibrium between correct and faulty agents is guaranteed if and only if there is no cost to joining the protocol.*

**Proof**: For correct agents' dominant strategy to be $j$, it is necessary but not sufficient that $r > c$. If $c \geq r$, then $c > r_v$, because $c \geq r > r_v$, and the payoff to correct agents from $j$ is therefore $\leq 0$ in all subgames. In that scenario, $j$ is weakly dominated by $\neg j$, because $\neg j$ guarantees a 0 payoff, while the payoff from $j$ may be $< 0$.

Even if $r > c$, the dominant strategy for correct agents will still be still be $\neg j$ if $r_v < c : \lambda_C \leq \psi_C$. If $c > 0$, the possibility that $r - c \leq 0$ cannot be excluded, because the value of $r$ is uncertain. However, if $c = 0$ then the possibility that $r - c < 0$ can be excluded, because $r$ is known to be positive, such that $r - 0 > 0$ for all potential values of $r$. Likewise, if $c = 0$, then it is guaranteed that $r_v - c > 0$, because $r_v$ is known to be positive, such that $r_v - 0 > 0$ for all potential values of $r_v$. The dominant strategy for correct replicas is therefore guaranteed to be $j$, even with imperfect information, because $j$ yields a larger payoff than $\neg j$ in all three subgames.

Per Lemma 1, if the dominant strategy for correct agents is $j$, then the dominant strategy for faulty agents is also $j$, because no realized separating equilibrium can be biased in favor of correct agents. □

## 5. Inter-Temporal Trust

We have so far modeled permissionless Byzantine consensus without a rigorous formalism for representing how trust in a protocol evolves as time passes. Yet time interacts with our results in subtle and important ways. In this section, we add temporal restrictions to our formal notation, and prove that a protocol's security and utility depend significantly on the temporal structure of its sampling frame.

### 5.1 Temporal Parts

We divide time into a sequence $S$ of discrete slots. The set of all slots is $S$. A slot $s \in S$ is the smallest unit of time that can be specified when restricting an entity's temporal parts. A protocol's initial slot is $s_{GEN}$, which we index with 0, so that the ensuing slots are $s_1, s_2, \ldots, s_\infty$. In our formal notation, we set out temporal restrictions in superscript using angle brackets. For example, $n$ refers by default to agent $n$ in all slots of time, whereas $n^{\langle s \rangle}$ refers to agent $n$ in only slot $s$, and $n^{\langle s-s' \rangle}$ refers to agent $n$ in the closed interval $[s, s']$ from slot $s$ to a subsequent slot, $s'$.

This extension of our notation makes it possible to express inter-temporal trust assumptions. The trust set of agent $n$ during slot $s$ is written $T[n^{\langle s \rangle}]$. Note that this temporal restriction applies to the time when the agent holds its belief. If we wish to restrict the temporal parts of the agent who is being trusted, then the angle brackets are added to the elements of the trust set. For example, $n'^{\langle s \rangle} \in T[n]$ indicates that, in all slots, agent $n$ trusts that agent $n'$ was correct during slot $s$.

The same notation can be employed to specify temporal parts of control sets and control structures. A control set during a specific slot is written $f^{\langle s \rangle}$, which denotes one or more agents that have the collective power to violate the protocol's security guarantees during slot $s$. A protocol's control structure for a specific slot is $F^{\langle s \rangle}$, which denotes all the control sets that have the power to violate the protocol's security guarantees during slot $s$.

These definitions make it possible to formalize the temporal parts of a protocol's maximum user base, $N_U$. Let $N_U^{\langle s \rangle}$ be a protocol's maximum user base during slot $s$, which can be calculated according to the formula

$$N_U^{\langle s \rangle} = \left\{ n \in N : T[n^{\langle s \rangle}] \cap \forall f F^{\langle s \rangle} \in F^{\langle s \rangle} \neq \emptyset \right\}$$

,

i.e., the protocol's maximum user base during slot $s$ consists of the members of $N$ whose trust sets during slot $s$ intersect with all of the protocol's control sets during slot $s$. Put differently, a user can only trust a protocol during a slot if it believes that every set of agents capable of violating the protocol's security guarantees during that slot is verifiably correct.

## 5.2 Temporal Sampling Frames

With our temporal notation formalized, we are ready to examine the practical implications of temporal restrictions.

*Fixed Permissioned Frame*

We begin with the simple example of a permissioned protocol, where the set of participants in consensus, $N_S$, is a fixed group of identifiable agents. This is consistent with pre-Bitcoin consensus algorithms, such as pBFT. [23] All of the members of $N_S$ generate asymmetric key pairs before the protocol is initiated. Their public keys are embedded in the protocol's genesis block, so that a quorum of the corresponding private keys is sufficient to achieve consensus in any subsequent slot.

If we assume that the agents retain exclusive control of their private keys, then all temporal parts of the protocol's control structure are identical. The same identifiable agents who control the protocol in the genesis block will necessarily retain control of the protocol in all future slots. In formal terms, $F^{\langle s_{GEN} \rangle} \equiv F^{\langle \forall s \in S \rangle}$.

The advantage to this arrangement is that the fixed membership of $N_S$ has "skin in the game." Every agent $n^{\langle s \rangle} \in N_S^{\langle s \rangle}$ who controls a private key during slot $s$ knows that it will retain the same voting power in any future slot $s'$. If there is an economic value to voting power, such as the ability to earn block rewards or transaction fees by updating a cryptographic monetary ledger, the consensus participants will have an incentive to act honestly in order to make the protocol a success. If the adversary $\mathcal{A}$ offered a quorum of consensus participants a bribe in the hopes of persuading them to destroy the protocol in the present, it would be economically rational for the quorum members to accept $\mathcal{A}$'s offer if and only if the present value of the bribe exceeds the time-discounted value of their future earnings as transaction validators.

The downside to the arrangement is that, because the protocol's control structure cannot change, its maximum potential user base cannot expand. If the composition of the total population remains constant, then the set of agents who can trust the protocol at its inception will be identical to the set of agents who can trust the protocol in all subsequent slots, i.e., $N_U^{\langle s_{GEN} \rangle} \equiv N_U^{\langle \forall s \in S \rangle}$. It is therefore impossible to "bootstrap" the protocol by starting with an initially small group of consensus participants. Those agents will remain permanently in control of the protocol, so the limits of their trustworthiness are the limits of the protocol's scalability.

*Disposable Permissionless Frame*

The situation is very different for a permissionless protocol with free entry. In our toy environment, the set of agents present in the protocol during a given slot can update the ledger by executing a new round of pBFT. Agents with access to the network join and leave the consensus protocol whenever they want, so the composition and size of $N_S$ always has the potential to change. The set of agents who control the protocol in one round may lose some of their members in the next round and be supplanted by a different coalition of agents. Even if they retain all their members, an even larger group of agents may suddenly join and seize power.

The agents who are present for a given round of consensus are always collectively in control of the protocol. Formally, for all $s$, $F^{\langle s \rangle} \subseteq N_S^{\langle s \rangle}$, i.e., the protocol's control structure in any round $s$ is a subset of the set of agents participating in that round. The downside to this is that a given round of consensus will be vulnerable to book-prize attacks unless the set of active protocol participants is large enough to satisfy the minimum-participation threshold. Formally, $\mathcal{N}_S^{\langle s \rangle} \in T[\forall n^{\langle s \rangle} \in N^{\langle s \rangle}] \equiv \left| N_S^{\langle s \rangle} \right| \geq |N_S|_{MIN}$, where $\mathcal{N}_S^{\langle s \rangle}$ is the meta-agent comprised of protocol participants during slot $s$. Maintaining the required degree of participation in every slot may be difficult or impossible.

The advantage is that new agents can start trusting the protocol without trusting the past sets of protocol participants. If the adversary $\mathcal{A}$ is able to successfully corrupt a super-threshold fraction of consensus participants during slot $s$, so that $\bar{y}N_S \geq k$, then $\mathcal{A}$ is only guaranteed the ability to attack during that slot — not during subsequent slots. Because new agent may join the consensus protocol, an influx of correct agents may lower the protocol's mean corruption status so that $\bar{y}N_S < k$ in slot $s'$. In our temporal notation, $\mathcal{A} \in F^{\langle s \rangle} \not\equiv \mathcal{A} \in F^{\langle s' \rangle}$. The security guarantees of such a protocol are therefore "ergodic" in the sense that the risk of failure in a given slot is path independent, i.e., agents can join in any order. [24] [25]

However, the free-entry condition used to achieve ergodicity has a downside. It creates a principal-agent problem: the members of $N_S^{\langle s \rangle}$ cannot guarantee to themselves during slot $s$ that they will remain in control of the protocol during a future slot $s'$, so they may have an incentive to take actions in the present — like defecting to the adversary — which provide immediate rewards. The harm caused by damaging the protocol so that it cannot achieve its full potential by time $s'$ may not be a sufficient incentive to defer defection by the members of $F^{\langle s \rangle}$, because each agent $n \in f \in F^{\langle s \rangle}$ will discount those future rewards by the probability that $n \in f \in F^{\langle s' \rangle}$. In economic terms, the present consensus participants do not receive franchise value in the protocol, so it is rational for them to defect against the future protocol "owners" as soon as they receive a sufficiently attractive opportunity. [11]

*Fixed Permissionless Frame*

One solution to the problems just described is to create a permissionless protocol without an ongoing free-entry condition. Let us imagine that an open invitation is issued for all agents with access to the network to sign up for the protocol by registering their public keys in advance. All registered keys are embedded in the genesis block, so that every agent who signed up in advance is eligible to participate in consensus. An agent with a registered key can either vote directly in pBFT or delegate its vote to one of the agents in its trust set. Every consensus decision is thereby backed by a quorum of registered keys.

If the cardinality of the set of agents who register their keys, $|N_{REG}|$, meets the minimum participation threshold $|N_S|_{MIN}$, then this arrangement offers the proverbial "best of both worlds" between the permissioned and permissionless paradigms. It has the same fixed control structure $F$ and incentive compatibility as the permissioned protocol, because the members of $N_{REG}$ will verifiably remain in control of the protocol in all future slots. Yet it achieves the maximum user base of permissionless consensus, because every agent with access to the network can trust the results of consensus given that $|N_{REG}| \geq |N_S|_{MIN}$ guarantees $\bar{y}N_S < k$.

However, if the set of agents who register their keys falls *below* the minimum participation threshold — i.e., if $|N_{REG}| < |N_S|_{MIN}$ — then the protocol is strictly defective. The set of keys embedded in the genesis block is not large enough to exclude the possibility that the mean corruption status of protocol participants exceeds the consensus algorithm's security threshold, so users cannot rationally trust the results of consensus without knowing the identities of the agents whose keys are registered. The result is equivalent to a permissioned protocol in which, rather than carefully recruiting trustworthy agents, keys are arbitrarily distributed to whomever shows up. Per Lemma 1, this arbitrary set of agents cannot rationally be trusted, because it may represent the adversary $\mathcal{A}$ executing a book-prize attack.

*Closed Permissionless Frame*

In practice, it may be impossible to arrange for a sufficient fraction of agents with access to the network to register their public keys before the creation of a protocol's genesis block. It is therefore desirable to design a mechanism that preserves incentive compatibility in every slot while also permitting the control structure $F$ to evolve. Ideally, such a mechanism would permit the set of consensus participants to expand over time, such that $|N_{REG}| < k \not\equiv \left|N_S^{(s)}\right| < |N_S|_{MIN}$.

The mechanism employed for this purpose in today's Proof-of-Stake protocols can be represented in our toy environment as a special case of delegation. An agent $n \in N_{REG}$ can irreversibly delegate an arbitrary fraction of its voting power to a different agent, $n'$, who may in turn irreversibly delegate arbitrary fractions of its voting power to other agents. We refer to this form of delegation as a *transfer* of voting power.

In theory, transfers enable the control structure $F$ to evolve, because an agent $n$ who transfers voting power to $n'$ loses control of that voting power. If a series of transfers resulted in voting power being evenly distributed among a sufficiently large number of agents by slot $s'$, such that $\left| N_S^{\langle s' \rangle} \right| \geq |N_S|_{MIN}$, then $\bar{y} N_S^{\langle s' \rangle} < k$ is guaranteed, and every agent $n^{\langle s' \rangle}$ would become a member of the protocol's maximum user base, $N_U^{\langle s' \rangle}$.

There are two significant shortcomings to this approach.

First, there is nothing to force the members of $N_{REG}$ to transfer their voting power, so the evolution of the control structure $F$ by slot $s'$ is possible but in no way guaranteed. The coordination problem from our extensive-form game therefore remains in any slot $s$ prior to $s'$. In the earlier slot, an agent $n$ has no means of predicting the protocol's mean corruption status in the future slot, so it cannot place its future trust in the protocol. Formally, $n^{\langle s \rangle} \notin N_U^{\langle s' \rangle}$.

Second, there is no way for agents to verify that the control structure $F$ has actually evolved. This is not a problem in our toy environment, where Sybil attacks are prohibited, but it a decisive flaw in the real world.

In the toy environment, every agent has access to a single private key, so a transfer of voting power from the public key associated with $n$ to the public key associated with $n'$ guarantees an actual change in the distribution of cryptographic voting power. This is what enables agents who could not initially trust the protocol to later become members of its maximum user base: even if they cannot trust $F^{\langle s \rangle}$, they can verify that subsequent transfers have redistributed voting power in a manner that makes $F^{\langle s' \rangle}$ trustworthy.

In the real world, each agent $n$ can generate an unlimited number of private keys with corresponding public keys. The members of any control set $f$ within $N_{REG}$ or $N_S^{\langle s \rangle}$ can therefore execute a special form of pseudo-spoofing, which we refer to as a *pseudo-transfer attack*. Specifically, every agent $n \in f^{\langle s \rangle}$ generates an arbitrarily large number of new key pairs and executes a series of transfers before slot $s'$ that redistribute all of its voting power among the new keys. To the outside world, these transfers make it appear that the control structure $F$ has evolved, such that $F^{\langle s \rangle} \not\equiv F^{\langle s' \rangle}$. Yet the evolution of $F$ is an illusion: the members of the control set $f^{\langle s \rangle}$ have simply transferred their voting power to themselves. Agents executing a pseudo-transfer attack can maintain this charade indefinitely with future transfers, invisibly preserving their power to successfully attack the protocol at the time of their choosing.

*Dynamic Permissionless Frame*

Pseudo-transfer attacks can be neutralized if, rather than relying on the members of $N_{REG}$ to transfer their voting power, new voting power can be issued to late-joining agents. The control structure $F$ will thereby verifiably evolve, even if the members of $N_{REG}$ retain all of their original voting power. To envision a simple scenario in which this is possible, let us imagine that a world government has established a global online voting system. This provides an approachable preview of the more complex mechanisms that the KRNC protocol employs to bootstrap consensus from the world's online-banking system.

In our hypothetical, the world government has used its coercive power to ensure that every agent $n \in N$ has one and only one account on the online voting system. Every account has a number, which is effectively a pseudonym because the identity of the agent associated with each account number is known only to the world government. Agents log in to the world government's online voting portal using their account number and a password known only to them. Once logged in to its online-voting account, an agent $n$ can broadcast arbitrary messages to the other members of $N$, which contain unforgeable proof of the agent's account number.

The list of account numbers acts as a sampling frame, so that one and only one unit of cryptographic voting power is issued to every agent. An agent who wishes to join the permissionless consensus protocol generates an asymmetric key pair, uses its existing password to log in to its online-voting account, and then broadcasts a message containing its public key. This message is treated as a registration of the public key in the permissionless consensus protocol, so one unit of voting power is assigned to the public key. If another public key is broadcast from the same account number, no voting power is assigned to the later key, because the voting power associated with the account has already been claimed by its owner.

Public keys registered before the initiation of the consensus protocol are embedded directly in the protocol's genesis block, but they do not limit the total amount of voting power in the protocol. When new public keys are registered from unused account numbers after the protocol has been initiated, they are simply appended to the list of public keys eligible to participate in consensus, which can continue growing until every agent has claimed its voting power. Even if the list of keys embedded in the genesis block is too short to initially guarantee security — i.e., if $|N_{REG}| < |N_S|_{MIN}$ — the protocol's ultimate security is guaranteed as long as the list of registered keys eventually satisfies the minimum participation threshold.

# Part III: Realistic Environment

## 6. Capital-Weighted Byzantine Consensus

The results we have obtained in our toy environment and world-government hypothetical must be replicated in an environment that corresponds to the real world. In this section, we begin that task. We replace the toy environment with a realistic model of resource-weighted consensus in the presence of an adversary capable of executing both book-prize attacks and Sybil attacks.

### 6.1 Resource-Weighted Sybil Resistance

In the toy environment, the adversary $\mathcal{A}$ was prohibited from engaging in pseudo-spoofing. In reality, the internet is a permissionless network, where a single agent $n$ can create an unlimited number of identities. Let the collection of all identities be $I$, indexed by $i = 1, 2, \ldots, I$. Each identity $i \in I$ is controlled by an agent $n \in N$, such that $N[i] = n$. The identity $i$ inherits the status $Y_n$ such that $Y_i = 0$ if $N[i] = n \in N_C$ and $Y_i = 1$ if $N[i] = n \in N_F$.

Let $I_S$ be the subpopulation of identities within a protocol. The protocol is vulnerable to a Sybil attack if, even though $\bar{y}N_S < k$, consensus fails because $\bar{y}I_S \geq k$. In other words, even if the requisite fraction of agents in the protocol is correct, so that consensus would be guaranteed if each agent voted only once, the adversary $\mathcal{A}$ can still rig the vote by having faulty agents create extra identities.

*Sybil Resistance*

Traditional Byzantine consensus is inherently vulnerable to Sybil attacks, because it assumes a reliable identity system. Protocols for *weighted* consensus overcome this problem. [26] Every identity $i$ is assigned a weight $\omega[i]$, where $\omega[i] \geq 0$. Even if the adversary controls a super-threshold fraction of identities — i.e., if $\bar{y}I_S \geq k$ — consensus is still guaranteed if correct agents control a sufficient fraction of the weight within the protocol.

Formally, a weighted consensus protocol's security guarantees will hold if and only if

$$\sum_{i=1}^{I_S^F} \omega[i] \Big/ \sum_{i=1}^{I_S} \omega[i] \geq k$$

,

where $\omega[i]$ is a non-negative weight assigned to an identity $i$ and $I_S^F$ is the population of identities in the protocol controlled by faulty agents. For every weight-assignment protocol $P$, there will be a collection of potential combinations of agents within $N$, the elements of which are sets of agents whose members control enough weight to break the protocol.[4] That collection is the weighted protocol's control structure, $F$. If a deterministic quorum-based consensus algorithm is employed, then $I_S$ will represent all the weight that has been issued, whereas if a chain-length algorithm is employed as required to assume that $k=0.5$, then $I_S$ will represent only weight that is contemporaneously online and contributing to consensus.

*Resource Weighting*

In most modern protocols, weights are assigned in proportion to stakes within a specified resource, such as hashing power or cryptocurrency. To preserve the generality of our model, we define an abstract staking resource, $V$. Let $e[n,v]$ be agent $n$'s endowment $v$ of resource $V$, such that the supply of $V$ is the sum of the individual endowments of $N$'s members.

The weighted voting power of every agent $n$ within $N_S$ is scaled in proportion to $e[n,v]$. This can be modeled using a "flat world" transformation, in which every agent who owns $v$ units of $V$ becomes $v$ replicas who each own one unit of $V$. When this transformation is applied to $Ns$, let the resulting population be $M_S$, within which $e[m,v]=1$ for all $m$. We assign every replica $m$ of $M_S$ a status $Y_m$ whose value matches the status $Y_n$ of the original agent $n$ within $N_S$.

Resource-weighted consensus among the members of $N_S$ can be guaranteed if and only if within $M_S$ the mean corruption status falls below the weighted-consensus algorithm's security threshold — i.e., it is necessary that $\bar{y}M_S < k$. In other words, after weights are applied to create a simulated electorate on a virtual network, that electorate must contain the honest majority or supermajority required by an unweighted consensus algorithm. Otherwise, the weighting procedure will place the adversary in control of whether the protocol's security guarantees are violated.

## 6.2 Allocation-Adaptive Adversaries

The formal proofs of security employed by today's cryptocurrency ledgers do not demonstrate that $\bar{y}M_S < k$. Instead, it is standard to specify the maximum fraction of the staking resource that the adversary can acquire before the protocol breaks — and then to "prove" that the protocol is secure by assuming that this threshold will never be exceeded. The problem is that, in real life, an adversary can change how it allocates its capital in response to the introduction of an anti-Sybil resource-weighting scheme. A static adversary can have the agents it controls exchange their shares of non-staking resources for shares of the staking resource; an adaptive adversary has the further ability to change which agents it has corrupted, so that it can obtain control of whichever combination of resource endowments will maximize the probability of successfully trading for a super-$k$ fraction of the staking resource. The axioms employed in existing proofs are therefore unreliable, because they are not guaranteed to remain invariant with respect to changes in how voting weight is assigned. [27]

---

[4] The sole, trivial exception is the null protocol, $\mathcal{P}_\emptyset$, wherein $\omega[n] = 0$ for all $n$, because $\omega[i] = 0$ for all $i$. We exclude it from our analysis, because it is nonfunctional.

*Bitcoin: A Case Study*

Bitcoin offers a cautionary example. When it was designed, Nakamoto chose hashing power as the protocol's staking resource in order to approximate a "1 person, 1 vote" weighting, based on the assumption that every person owned roughly one CPU. [5] [28] [29] Yet, as soon as the success of Bitcoin increased the market value of hashing power, agents adapted by acquiring extra hardware to inflate their weights in the protocol. Today, a standard CPU receives effectively zero weight in Bitcoin, because participants in the consensus protocol exaggerate their hashing power using distributed supercomputers comprised of ASICs. This undermines Bitcoin's core value proposition — its claimed ability to act as a store of value superior to gold, because its absolute cryptographic scarcity guarantees a higher stock-to-flow ratio. [30]

In reality, Bitcoin's reliability as a store of value depends on two stock-to-flow ratios, not one. The familiar ratio is the quantity of BTC held by protocol participants relative to the BTC that miners earn from block rewards. If one equates Bitcoin with a physical commodity, then this represents the most impressive stock-to-flow ratio in history, because the total future flow is cryptographically limited. However, Bitcoin is different from a physical commodity, in that the value of the existing stock can be destroyed even if the future supply is not inflated. If an adversary acquires a super-threshold fraction of voting power on the network, then it can begin executing double-spending attacks, and the willingness of agents in the economy to accept Bitcoin as money will thereby be diminished or destroyed. Even a large but sub-$k$ stake may be sufficient for the adversary to force transaction confirmation delays that undermine the usability of the network. It is therefore crucial to consider not only the stock-to-flow ratio of BTC itself, but also the stock-to-flow ratio of the staking-resource $v$ — which, in the case of Bitcoin, is algorithm-specific hashing power.

Viewed through that lens, BTC's properties are less impressive. The success of the Bitcoin protocol caused demand for hashing power to explode, and the flow of new $v$ onto the market increased at a rate akin to hyperinflation. If the efficiency of the equipment for mining gold were ever to increase at a comparable rate, then the flow of gold onto the world market would have sped up exponentially, and the price of gold would immediately have begun dropping. In Bitcoin, that outcome is avoided because the difficulty of the underlying cryptographic puzzles automatically increases, so that the flow of BTC remains stable no matter how quickly the quantity and efficiency of the hardware being operated to mine Bitcoin grows. However, the fact that mining equipment cannot be directly used to flood the market with new Bitcoin does not alter the fact that control of Nakamoto consensus will inevitably shift to the agents who have acquired the newest, fastest mining equipment. This limits the scalability of the protocol, because its maximum user base $N_U$ encompasses only agents who have reason to believe that whoever has the newest, fastest mining equipment is trustworthy.

---

[5] In our notation, Nakamoto's analysis of Poof-of-Work starts with the axiom that $\overline{Y} < k$ within $N$, assumes that this inequality will hold within $N_S$, and then reasons that $\bar{y}M_S < k$ because $N \approx M$. [26]

Worse, the agents with the latest mining equipment can use their ability to degrade network performance or execute double-spending attacks to hold the protocol hostage unless other stakeholders (Bitcoin users, exchanges, etc.) accede to modifying the original supply schedule, so that more than 21-million BTC can be mined.[6] It is therefore incorrect to portray Bitcoin as a "synthetic commodity," which possesses innate scarcity equivalent to physical resources like gold, merely because a tentative agreement has been reached about the future supply schedule. [31] Indeed, the ability to adhere to a predefined supply schedule was the original benefit promised by unbacked fiat money as a replacement for the gold standard. [32] That theoretical benefit never materialized because, when placed in charge of administering a currency ledger, governments eventually succumb to the temptation to deviate from the supply schedule to which they are nominally bound.

As one 19th century advocate of private currency observed, "The idea that governments must necessarily control the currency arises . . . from the erroneous but prevalent idea that governments are more trustworthy than their subjects." [33] Today, the idea that the Bitcoin protocol should control the money supply arises from the erroneous but prevalent idea that it is more trustworthy than the government. If Milton Friedman's vision of "replacing the Fed with a computer" is ever going to be meaningfully realized, control of that computer must be distributed among a set of agents whom the owners of money can verifiably trust. [34] Bitcoin was designed to accomplish that, by distributing voting power in proportion to a resource that everyone already owned. Unfortunately, it failed because agents were able to dynamically adjust their resource allocations.

---

[6] Switching to a different hashing algorithm can temporarily fix this problem by making the ASICs employed by existing mining pools ineffective, but the structural factors that determine which entities dominate Proof-of-Work — including expertise in creating and operating supercomputers, access to cheap electricity, and acquiescence of local government — will ultimately reemerge.

## GLC: An Eternal Gordian Knot

The underlying problem is not unique to Bitcoin or cryptocurrencies. It is an example of Goodhart's Law: when a measure becomes a target, it ceases to be accurate. [35] A familiar example is the problem of "teaching to the test" in education: when student performance is measured by standardized tests, teachers devote disproportionate time to the material covered by the exams, artificially boosting their pupils' scores. Variants of the same phenomenon are known as "Campbell's Law" and "the Cobra Effect." [36] These are all names for the core, recurring problem in fields that attempt causal interventions in the presence of volitional agents, whose ability to alter their choices in response to interventions creates a reflexive "feedback loop" between cause and effect. [37] [38]

Proof-of-Balance is based on a simple solution to this problem, which is nonetheless apparently novel. [39] The solution can be phrased informally as a rejoinder to Goodhart's Law: "if whatever you measure will become a target, measure whatever is already a target." In the present section, we will begin defining the realistic environment needed to model an adversary's ability to circumvent a Sybil-resistance scheme by modifying its resource allocations. This will lay the groundwork for a full formalization of our solution in Section 8.

## Endowments and Allocations

Let there be $G$ resources, indexed by $g = 1, 2, …, G$. One of these resources is the staking resource, $V$. We adopt the standard assumptions of general equilibrium in an Arrow-Debreu environment, including that supply and demand will converge on the market-clearing prices $p_1, p_2, …p_G$ for all resources. [40] Let $e[n,g]$ be agent $n$'s initial endowment of resource $g$, such that the supply of each resource is the sum of the endowments of that resource among all members of $N$. Let $x[n,g]$ be agent's $n$'s ultimate allocation of resource $g$ after all exchanges, if any, have been completed.

For general equilibrium, it is standard to assume that every agent $n$ has Cobb-Douglas utility, such that $\alpha[n, g] > 0$ for all $n$ and $g$. The utility $\alpha[n]$ of an agent $n$ is therefore

$$\alpha[n] = \prod_{g=1}^{G} x\,[n,g]^{\alpha[n,g]}$$

,

which represents the amount of each resource it controls, raised to the exponent defined by its utility function for that resource.

*Adversarial Utility Function*

Formalizing the Byzantine adversary $\mathcal{A}$ in this economic framework is a non-trivial challenge. $\mathcal{A}$ will always prefer breaking the protocol to not breaking the protocol, so it has a lexicographic preference for the staking resource in quantities sufficient to prevent consensus. Formally,

$$\left\{ x[\mathcal{A}, v \colon \overline{y}_{M_s} \geq k]^{\alpha[\mathcal{A}, V \colon \overline{y}_{M_s} \geq k]} \right\} > \forall \left\{ x[\mathcal{A}, g \colon \overline{y}_{M_s} < k]^{\alpha[\mathcal{A}, g \colon \overline{y}_{M_s} < k]} \right\}$$

Unfortunately, like all lexicographic preferences, this ordering is strictly discontinuous, so it cannot be replicated by any von Neumann-Morgenstern utility function. We will employ a variant of the Cobb-Douglas utility function that is continuous up to an asymptotic limit, and which approximates a Byzantine adversary's preferences with arbitrary precision.

Let there be two forms of utility, standard utility, $\alpha_N$, and faulty utility, $\alpha_F$. Both correct and faulty agents receive $\alpha_N$ from all resources. However, $\alpha_F$ represents the utility of a successful attack on the protocol, so it is only available to faulty agents, and those agents only receive it from controlling units of the staking resource, $V$. The aggregate utility of the adversary $\mathcal{A}$ is therefore

$$\alpha[\mathcal{A}] = \alpha_N[\mathcal{A}] + \alpha_F[\mathcal{A}] = \prod_{g=1}^{G} x\,[\mathcal{A}, g]^{\alpha_N[\mathcal{A}, g]} + x[\mathcal{A}, v]^{\alpha_F[\mathcal{A}, v]}$$

where

$$\alpha_F[\mathcal{A}, v] = \left( \lim_{\alpha_F \to \infty} \right) \left( \rho[\overline{y}_{M_s} \geq k] \right)$$

The strength of the adversary's preference for a successful attack is asymptotically close to infinity, but the value the adversary attaches to the staking resource is discounted by the probability of acquiring the necessary fraction of voting weight to successfully attack, $\rho[\overline{y}_{M_s} \geq k]$.[7]

If this probability is 0, then the adversary's demand for the staking resource is based exclusively on the standard utility $\alpha_N$ derived from the resource's non-faulty uses, because no faulty utility $\alpha_F$ is available. This reflects the fact that, if amassing the staking resource is not a viable means to attack the protocol, the adversary $\mathcal{A}$ will prefer to allocate its capital towards other avenues of attack, like attempting to destroy or compromise critical network infrastructure.

---

[7] Technically, this utility function is tailored to Proof-of-Stake and Proof-of-Balance. $\mathcal{A}$ will rationally continue acquiring the staking resource after it knows that it can violate a protocol's safety guarantees, because the staking resource is also the monetary resource that must be transferred and then reclaimed in the course of a double-spending attack.

## 6.3 Honest Majority of Capital

The simple economic model we have just defined is sufficient to obtain a structural axiom for permissionless consensus. The axiom must represent a feature of the world that cannot be volitionally altered by any agent, including the adversary, $\mathcal{A}$. The axiom that a majority of an arbitrary resource, such as computing power, is owned by honest agents before the introduction of the protocol does not satisfy that standard. As Chen and Micali observe, "since computing power can be bought with money, if malicious users own most of the money, then they can obtain most of the computing power." [41]

Indeed, since every resource can be bought or sold for money, an adversary of sufficient wealth can acquire the majority of *any* resource. Our structural axiom must therefore be the maximum overall wealth of $\mathcal{A}$, as measured by the combined value of all the resources it controls. This axiom is intrinsically reliable, because even if $\mathcal{A}$ trades its endowment for a different allocation of resources, it does not have the power to volitionally increase its wealth.

In formal terms, the individual wealth of an agent $n$ is the sum of the exchange value of all its resource endowments:

$$w[n] = \sum_{g=1}^{G} p_g(e[n,g]).$$

The total wealth $w[N]$ of all agents with access to the network is the sum of the individual wealth $w[n]$ of every agent $n \in N$:

$$w[N] = \sum_{n=1}^{N} w[n]$$

The wealth available to the adversary, $w[\mathcal{A}]$, is a subset of $w[N]$ comprised of the wealth of faulty agents:

$$w[\mathcal{A}] = \sum_{n=1}^{N_F} w[n]$$

For simplicity, all wealth in our model is sufficiently liquid to be employed as capital in a consensus protocol. The terms "wealth" and "capital" can therefore be employed interchangeably. However, when we input real-world data, only liquid wealth should be included.

The structural axiom we introduce is the existence of an *Honest Majority of Capital*. The term "majority" is used broadly to encompass either a simple majority or a supermajority, depending on the security threshold of the underlying consensus algorithm. Formally, the axiom specifies that

$$\frac{w[\mathcal{A}]}{w[N]} < k$$

This assumption is implicit in every protocol whose security guarantees are derived from the axiom that the majority of its designated resource, $V$, will remain controlled by correct agents. The implicit dependence is expressed in formal terms by the following theorem.

**Theorem 2**: *Resource-weighted permissionless Byzantine consensus is impossible if $\frac{w[\mathcal{A}]}{w[N]} \geq k$, i.e., if the wealth of the adversary exceeds the wealth of all network agents by a ratio greater than the security threshold of the protocol's consensus algorithm.*

**Proof:** An agent whose total endowment has value $w$ can exchange that endowment for any possible allocation of value $w$ or less. This is a standard assumption in Arrow-Debreu environments, which we have adopted in our model. It follows that any endowment whose value represents a $k$ fraction of all $w$ can be exchanged for an allocation that contains a $k$ fraction of every resource $g \in G$. Therefore, if $w[\mathcal{A}] / w[N] \geq k$, then $\mathcal{A}$ can obtain a super-threshold fraction of every resource, such that $\frac{\sum_{n=1}^{N_F} x[n,g]}{\sum_{n=1}^{N} x[n,g]} \geq k$ for all $g \in G$. If $\mathcal{A}$ employs this strategy, then $\frac{x(\mathcal{A},V)}{x(N,V)} \geq k$, because $V = g \in G$. This guarantees that $\bar{y}M_S \geq k$ for every possible choice of $V$. $\square$

## 6.4 Minimum Capital Threshold

In the realistic environment, the Honest Majority of Capital axiom serves the same role that was filled in the toy environment by the axiom that $\bar{Y} < k$ — i.e., that a majority of all agents with access to the network were honest. As we demonstrated in Section 3, that axiom is not inherently sufficient to ensure liveness or safety; valid security guarantees can be derived from knowledge of the population's mean corruption status if and only if the number of agents participating in the protocol ensures a representative sample — i.e., it is also necessary that $|N_S| \geq |N_S|_{MIN} : \bar{y}N_S \geq k$. Similarly, the Honest Majority of Capital axiom does not ensure liveness or safety in a weighted consensus protocol; valid security guarantees can be derived if and only if the quantity of capital participating in the protocol ensures a representative sample of the economy as a whole. *Formulas*

To formalize this concept, we employ the same flat-world transformation technique that we previously applied to $N_S$ in order to obtain the staking-resource normalized population $M_S$. However, this time the transformation is applied to the full population, $N$, which is normalized by wealth to define the population $W$, such that $w[w]$=1 for every replica $w \in W$. Note that the same technique cannot be applied to $N_S$ in order to derive the sample size $|W_S|$. The result of that transformation would be the total wealth of all the agents who are participating in the protocol, when what we require is the quantity of capital needed to obtain $\omega[N_S]$, all the weight that the protocol has assigned. The correct formula for $|W_S|$ is therefore

$$|W_S| = |M_S|(p_V)$$

i.e., the quantity of capital participating in the protocol is the cardinality of the normalized population of replicas multiplied by the price of one unit of the staking resource.

Once we have obtained these definitions of $W$ and $W_S$, it is trivial to apply the deterministic and probabilistic participation thresholds from the toy environment to the realistic environment. The minimum deterministic sample $|W_S|_{MIN(D)}$ is given by the formula

$$|W_S|_{MIN(D)} = |W_S^F|_{MAX} + |W_S^C|_{MIN} : \frac{|W_S^F|_{MAX}}{|W_S|} < k$$

where

$$|W_S^F|_{MAX} = (\bar{y} W_{MAX})(|W|_{MAX})$$

The minimum probabilistic sample $|W_S|_{MIN(P)}$ is given by the formula

$$|W_S|_{MIN(P)} = E(|W_S^F|) + |W_S^C|_{MIN} : \frac{E(|W_S^F|)}{|W_S|} < k$$

where

$$E(|W_S^F|) = (\bar{y} W_{MAX} + Bias)(|W_S|)$$

Other than the substitution of $W$ for $N$, these equations are identical to the versions we applied to calculate the minimum-participation threshold in the toy environment, and we can therefore apply them using the techniques introduced in Section 3.

*Application*

For the probabilistic threshold, the requirement of a mean response propensity of at least 0.2 is as relevant to capital-normalized samples as it was to agents in the toy environment. Thus, $W_S$ must include at least 1/5 of $W$. However, as we previously cautioned, our model's definition of wealth is equivalent to liquidity in the real world, so to input real-world data into the formula we must define $W$ based on the quantity of liquid wealth in the global economy. There are edge cases that make it difficult to objectively define what wealth is insufficiently liquid to be included in $W$, much like it was difficult to define an objectively tolerable quantity of bias in the toy environment. Fortunately, we can once again employ apophatic reasoning to sidestep the difficulty.

It is indisputably necessary for any definition of liquid wealth to include the global economy's most liquid asset, fiat money. Given that the combined value of all fiat money is approximately 80 Trillion USD, we can therefore conclude that $|W| \geq 80 \times 10^{12}$. If $k=0.5$, then verifiably secure permissionless resource-weighted Byzantine consensus requires a staking resource whose value exceeds \$16 Trillion USD. This figure is orders of magnitude larger than the sample sizes achieved by existing protocols. The largest Proof-of-Stake ICO in history purportedly attracted \$4 Billion in investment, which is 0.00025 of the minimum sample size we have just calculated.

*Stake Laundering*

Why "purportedly?" Even when investments in an ICO are made publicly on a platform such as Ethereum, it is difficult or impossible to verify how much capital agents actually expended to acquire voting power. Before the SEC was created to regulate securities offerings in the U.S., it was common for parties to engage in "wash trading" by selling securities to themselves or their associates to create an illusion of demand. It is plausible, if not likely, that similar forms of market manipulation have been engaged in by some parties administering unregulated ICOs for Proof-of-Stake protocols.

The administrator simply needs to make a side-agreement with one or more high-wealth "whales" to eventually return the funds that those parties contribute to the ICO. An audit of the ICO itself will not reveal this fraud, because no double counting occurs; every unit of money "invested" is transferred into the ICO account once and only once. Proof that capital was actually expended would require a full audit of the disposition of the funds received in the ICO. In the absence of such an audit, the objectively verifiable quantity of $|W_S|$ is limited to the price of "renting" money for the period of the ICO.

Even agents who subjectively trust a particular ICO administrator not to engage in this form of fraud should expect the objective risk of such fraud to decrease the ultimate value of the protocol's native asset, because the protocol's maximum user base will be limited to the set of agents with subjective faith in the administrator's past honesty. In effect, ICOs are an extremely pernicious form of trusted setup. They cannot be superseded in the same manner as e.g. signing ceremonies, because they irreversibly dictate the substantive state of the protocol, not merely the composition of the cryptographic primitives used to maintain that state.

# 7. Inter-Temporal Staking Costs

## 7.1 The Automatic Increase Fallacy

The designers of today's Proof-of-Stake protocols believe that it is not necessary for the amount of capital invested in an ICO to provide robust security guarantees, because as the value of the protocol's native asset grows, the strength of its security automatically increases. According to the position paper for the Tezos protocol, Proof-of-Stake "automatically scales up the cost of an attack as the currency appreciates." [42] This is also the theory that underpins the Cardano protocol, according to the CEO of its corporate developer, IOHK. [14] Logically, the theory is the underpinning of all ICO-launched Proof-of-Stake protocols, which inherently aspire to grow in security so that in the future they will be able to securely store and transmit greater amounts of value than was possible in the past or is possible in the present.

If the theory of automatically increasing security were true, then our criticisms of Proof-of-Stake would be overly harsh. Although the ICO-based sample size $|W_S^{\langle s_{GEN} \rangle}|$ is inadequate, and although the present sample size $|W_S^{\langle s \rangle}|$ is inadequate, a protocol could achieve verifiable security by a future slot $s'$ if $|W_S^{\langle s' \rangle}| \geq |W_S|_{MIN}$. Because its control structure $F$ would verifiably evolve, there would be no limit to an ICO-launched protocol's maximum user base. The world's future global monetary standard could be bootstrapped from a Proof-of-Stake protocol with an arbitrarily small market capitalization. In an extreme case, an ICO that attracted an arbitrarily small quantity of initial investment, such as a mere $1,000 USD, could theoretically launch a protocol that would later be used to securely transfer billions or trillions of dollars.

Unfortunately, the theory is false. It overlooks the risk of pseudo-transfer attacks, which we demonstrated in Section 5. We will first make our point with a simple hypothetical, then present a more rigorous formalization. The hypothetical we offer concerns Tezos, but we are not attempting to pick on that protocol or its founders. The underlying criticism applies to all ICO-launched Proof-of-Stake protocols.

*156 Agents vs. The World*

Our hypothetical starts with the publicly documented facts that the Tezos protocol raised $250 million in an ICO, and that half of the protocol's voting power was thereby issued to 156 accounts. [43] A ½ share of all voting power exceeds the security threshold $k$, so whoever controlled these 156 accounts received the power to successfully attack the protocol. In the best-case scenario that the accounts were owned by 156 different people, the control structure of the Tezos protocol during its genesis slot therefore included a control set composed of just 156 agents. Formally, $F^{\langle s_{GEN} \rangle} \ni f^{\langle s_{GEN} \rangle} : \left| f^{\langle s_{GEN} \rangle} \right| = 156$.

This means that, when the protocol was initiated, its maximum user base $N_U^{\langle s_{GEN} \rangle}$ was limited to agents whose trust set included one of the 156 agents. However, the 156 agents were anonymous, so their trustworthiness could not be evaluated on the basis of their identities. The control set was effectively an abstract group of 156 anonymous agents who paid $125 million for the collective power to control the protocol. It is possible that all 156 agents were taking part in a book-prize attack. It is also possible that they were 156 disconnected individuals who shared nothing but an enthusiasm for acquiring large quantities of voting power. In either event, the objective ability of the 156 agents to successfully execute an attack made the Tezos protocol initially unsuitable as a global trust-minimized computation platform.

Hence the need for the theory of automatically increasing security. If growth in the market capitalization of the protocol made its security guarantees more and more reliable, then the initial concentration of power in the hands of a small group of anonymous strangers could eventually be forgotten. The problem is the existence of pseudo-transfer attacks, which invalidate the theory's assumption that the migration of cryptographic stake to new addresses proves that voting power is becoming decentralized.

To retain control of the protocol, each of the 156 agents can create an arbitrarily large number of new cryptographic addresses which it secretly controls, and it can then restrict future transfers of stake from its account to the set of addresses that it individually controls. The 156 agents are thereby guaranteed the ability to attack the protocol in any future slot, subject only to the requirements that their stake is not diluted to a sub-threshold level by block rewards or transaction fees. If those conditions are satisfied, then in every slot, it is possible that the protocol's control structure contains the 156 agents who were in control of it during the genesis block. Formally, $\diamondsuit F^{\langle \forall s \rangle} \ni f^{\langle \forall s \rangle} : (\left| f^{\langle \forall s \rangle} \right| = 156) \wedge (f^{\langle \forall s \rangle} \equiv f^{\langle s_{GEN} \rangle})$.

Growth in the apparent market value of the protocol's native asset does not offer any protection against pseudo-transfer attacks. There are two reasons for this. First, the 156 agents are not required to relinquish control of their stake in order for the protocol's market capitalization to grow, because the transactions that are used as the basis for calculating the price of the asset can be executed with the ½ of total stake that is owned by other agents. Second, and more dangerous, the 156 agents can effectively "spoof" a large market capitalization as part of a pseudo-transfer attack by engaging in wash trading. By purchasing stakes from one another, they can "spend" money that is actually returned to their own pockets, creating the illusion that the protocol's control structure $F$ is evolving based on large expenditures of capital, when in reality the same 156 agents remain in power.

Because the return available from executing such an attack grows as the currency appreciates, it would be rational for an agent who has gained the option to attack at the moment of its choosing to delay the exercise of that option until the slot that delivers the greatest time-discounted payoff. Thus, in any given slot, the fact that the protocol has not yet been successfully attacked is not reliable evidence that the 156 agents will not execute an attack in the future. The initial centralization of power in their hands acts as a permanent obstacle to reliable security.

Tezos is not the only protocol whose security guarantees are undermined by pseudo-transfer attacks. Any protocol bootstrapped from an ICO will have the same security problem. What appears to be decentralization of power can be entirely spoofed by the adversary.

If we adopt the definition of Proof-of-Stake from the Ouroboros paper, then "[w]hat distinguishes a PoS-based blockchain from those which assume static authorities is that stake changes over time and hence the trust assumption evolves with the system…" [12] Pseudo-transfer attacks mean that the latter deduction is invalid. A Proof-of-Stake protocol's trust set does *not* evolve over time, because an adversary who corrupts one control set in the protocol's initial control structure can successfully launch an attack in any subsequent slot. This is an inherent shortcoming of attempting to bootstrap a global, decentralized monetary standard from an ICO.

## 7.2 Staking Asynchrony

Agents risk finding themselves in the position of the turkey in the famous parable about the problem of induction if they assume that the ongoing absence of an attack provides increasingly strong security guarantees on a Proof-of-Stake ledger, so that larger and large sums of money can be transferred with confidence. Indeed, the risk of pseudo-transfers is more severe than the abstract problem of induction, because the known incentives of the adversary provide an affirmative game-theoretic reason to distrust observations that the adversary has the power to rig. An agent does not face merely the abstract possibility that the future will not conform to the past; it has an affirmative reason to distrust evidence that a rational adversary would fabricate, much like an agent trying to ascertain whether it has been targeted in a Ponzi scheme should not be reassured by its past ability to withdraw money,

In the prior literature on Byzantine consensus, the distinction between these forms of risk has been formalized in terms of the powers available to an adversary in a betting game about the results of a coin flip. [44] Treating a coin flip that the adversary has the power to bias as fair is objectively irrational, because it will cause an agent to systematically lose money. Similarly, in an asynchronous environment, the betting game may be rigged even if all of the individual coin flips are fair, because the adversary has the power to control when the outcome of the game is determined. Agents are effectively playing a game in which they must guess the probability that the outcome of the *last* coin flip was favorable to the adversary, and in which the adversary decides which flips are included *after* it knows their results.[8]

The same principle informs our analysis of pseudo-transfer attacks, which present a temporal vector for injecting adversarial bias that is similar to the one in the asynchronous coin-flip game. All that has changed are the details of what is being guessed. Here, a game is being played in which the recipient must infer the wealth of the signaler based on proof that the signaler controls a quantity $v$ of the staking resource $V$. As in the asynchronous coin-flip game, the odds of the game depend on whether the adversary can control the slot that is used to determine the state of one of the relevant parameters. In the signaling game, the parameter that varies over time is the market price $p_V$ of the staking resource.

It is impossible to know exactly which of the historic prices a real-world attacker paid, so even though it may be unlikely that any agent had perfect timing, that is the appropriate power to confer on the adversary if our goal is to establish robust security guarantees. The point is not to prepare for a real-world agent with such overwhelming powers, any more than the purpose of the basic Byzantine adversary formalism is to prepare for omnipresent coordinated attacks. Guaranteeing that the protocol's security properties will remain intact even in the worst-case scenario is a means of deriving objective answers to questions that would otherwise depend on unreliable judgments about the likelihood of relevant events.

---

[8] Treating the outcome of those flips as unbiased is equivalent to making the wrong choice in the Monty Hall Problem, the infamous puzzle about whether it is rational to accept an invitation from the "house" to switch guesses in a three-option guessing game given rules that bias which of the three options was withdrawn from play. [72]

Historic purchases of the staking resource by the adversary are an example of such events. We therefore extend the technique employed in the existing literature for dealing with asynchrony by defining a new class of adversary.

In the context of resource-weighted consensus, let a *price-adaptive adversary* have the power to purchase any unit of the staking resource at the lowest price ever paid by an agent for that unit of the staking resource. Formally, let the units of the staking resource $V$ be indexed $v_1, v_2, \ldots, v_V$, with the corresponding price of each unit denoted $p_{v_1}, p_{v_2}, \ldots, p_{v_V}$. The unit-specific prices of the staking resource can be divided into temporal parts, with the collection of all such temporal parts indexed $p_{v_1}{}^{\langle \forall s \rangle}, p_{v_2}{}^{\langle \forall s \rangle}, \ldots, p_{v_v}{}^{\langle \forall s \rangle}$. Let the price $\breve{p}[v_v]$ paid by a price-adaptive adversary for a given unit of the staking resource be the lowest price among all the temporal parts of the price for that unit, such that $\breve{p}_{v_v} \leq p_{v_v}{}^{\langle \forall s \rangle}$. [9] In the signaling game, that represents the appropriate formalism for playing against an adversary that has the power to control when the price used to impose verifiable costs is selected.

*Inter-Temporal Cost Estimator*

In practice, it is often impossible to track the price history of individual units of the staking resource. The minimum cost imposed on a price-adaptive adversary must therefore be estimated based on whatever data is, in fact, available. Here, we will assume that the only information available is the historic price of the staking resource in each slot, $p_V[s]$, and the volume of the staking resource transacted in each slot, $q_V[s]$. By definition, the quantity of the staking resource acquired by a price-adaptive adversary in a given slot is a subset of the total quantity of the staking resource acquired in that slot.

Let all slots of a given price $p_v$ be members of the same price era, $d$, such that $p_V[s]$ returns the same value for every slot $s \in d$ within a price era. The collection of eras, $D$, is indexed $d_1, d_2, \ldots, d_D$ in ascending order of their respective prices. For convenience, we can also abbreviate the price $p_V[d]$ for a specified era using $p_d$, such that $p_1$ represents the lowest price ever paid for the staking resource, because it is associated with the first era in the list of eras indexed in ascending order of price. The volume of an era, $q_V[d]$ may be expressed using the equivalent abbreviation, $q_d$. It represents the sum of the volume during all periods in the era, per the formula:

---

[9] Staking asynchrony may also introduce higher-order uncertainty if economies of scale for acquiring the staking resource vary across slots, because a price-adaptive adversary can choose the slots where the slope of volume-based discounting is steepest.

$$q_d = q_V[d] = \sum_{s=1}^{d} q_V[s]$$

Let $V_S$ be the quantity of the staking resource in the protocol, and $V_S^F$ be the quantity of the staking resource within the protocol controlled by faulty agents. To execute a successful Sybil attack, the adversary's allocation of the staking resource, $x[\mathcal{A}, v]$, must be greater than or equal to $V_S^F \min: \bar{y}M_S > k$, the quantity of the staking resource needed to violate the protocol's weighted consensus threshold. When the volume of all eras is progressively summed, starting at the first indexed era $d_1$ and proceeding in ascending order, let $d_k$ represent the first era in which the summed volume exceeds the cardinality of $V_S^F \min: \bar{y}M_S > k$.

We divide $d_k$ into two sections, the portion $d_{k\,(PRE)}$ before the threshold is crossed and the portion $d_{k\,(POST)}$ after the threshold is crossed. Let the same operation be performed on the indexed sequence of all eras, such that $D_{k(PRE)}$ designates the eras before the threshold is crossed, including $d_{k\,(PRE)}$, and $D_{k\,(POST)}$ designates the eras after the threshold is crossed, including $D_{k\,(POST)}$.

The minimum cost to a price-adaptive adversary of attacking a consensus protocol is given by the formula:

$$W_{S\,MIN}^F: \bar{y}M_S > k = \sum_{s=1}^{D_{k(PRE)}} q_d(p_d)$$

This represents the sum of the cost of the cost of purchasing all of the units of the staking resource during the price-eras before the security threshold is reached, plus the cost of purchasing the pre-threshold fraction of the units within the price-era in which the threshold is reached.

A visualization of the formula is provided in **Figure 7.1**. It shows a protocol with strong security: the protocol's security threshold is crossed many eras away from the first era $d_1$, where the adversary has the opportunity to pay the lowest historic price, $p_1$. The era in which the threshold is reached, $d_k$, involves a much higher price for the staking resource, because the volume of the staking resource available in the initial eras is not sufficient to mount an attack. Therefore, even a price-adaptive adversary would have to pay a mean price per unit that greatly exceeds the staking resource's historic minimum price.

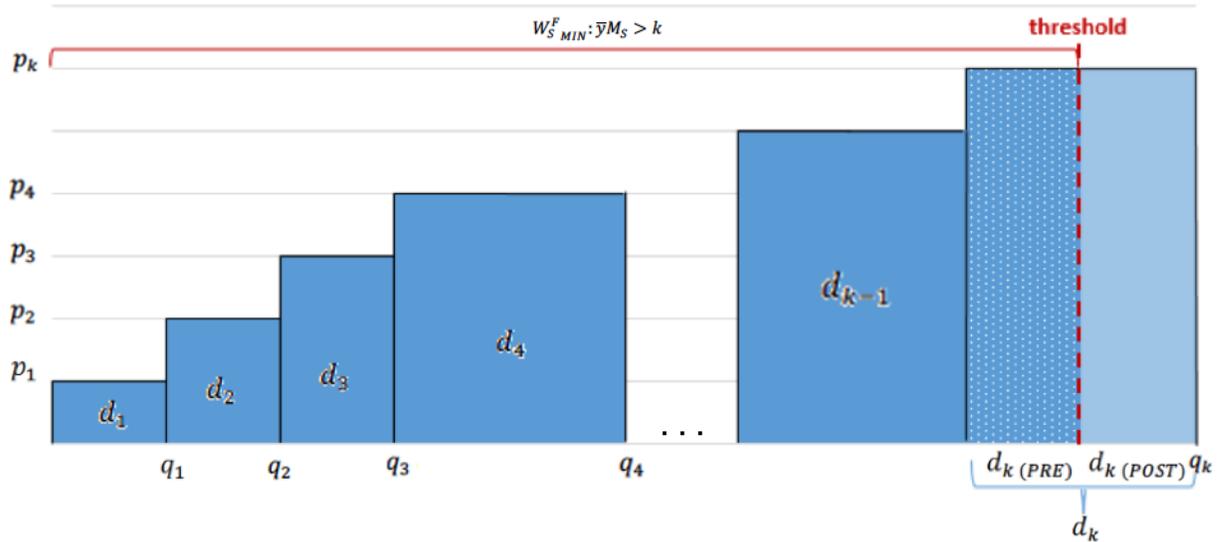

**Figure 7.1:** The verifiable costs imposed on a price-adaptive adversary depend on the past volume and price of staking-resource transactions.

In contrast, if the staking resource were auctioned for its minimum historic price in an ICO, then $d_1$ and $d_k$ would be the same era. That is, the security threshold would be reached within the era of the lowest historic price, $p_1$, because the volume $q_1$ available in that era would be sufficient for the adversary to mount an attack. The theory that security increases automatically as the currency appreciates is incorrect, because the verifiable minimum cost imposed on the adversary remains locked at the floor set in the ICO. From the perspective of signaling theory, the information conveyed by a signal is limited by the minimum cost of spoofing, so the adversary's ability to pay a "discounted" rate must be taken into account when seeking a Nash equilibrium that permits permissionless Byzantine consensus.

## 7.3 Errors of Ergodicity

It is an objective error to assume that the cost to the adversary of acquiring half the stake in a protocol is ½ of the protocol's current market capitalization, just as it is an objective error to assume that a traditional asynchronous adversary will lose ½ of the betting games it retrospectively chooses to play. [44] Naïve application of the normal rules of probability to outcomes that are under the potential control of a competing player in a zero-sum game is a losing strategy. The fact that major Proof-of-Stake protocols were designed on the basis of this false assumption is a warning sign that something has gone very wrong.

The culprit becomes easy to spot if we broaden our view. Computer scientists have fallen victim to the same error that plagued decision theorists and economists for 250 years: They have conflated "ensemble probability" and "time probability." [24] [25] This is a special case of the error involved in making decisions based on unweighted odds — i.e., without regard to the magnitude of relative consequences.

A classic example of the latter error is judging the skill of stock-market prognosticators according to who can predict the direction of the market, rather than on who can make the most money over time by placing bets on the basis of that information. In real life, winning strategies depend on the magnitude of costs and rewards. This is particularly true if, as soon as players go insolvent, they lose their opportunity to continue playing the game. Such a game is "non-ergodic" in the sense that play is not guaranteed to reach all of the time slots in the game; the mean payoff available across the ensemble of all slots is not an accurate predictor of the payoff that an agent will receive, because a sufficiently bad outcome in one slot will deprive the player of the opportunity to access the payoffs available in all other slots. [24] [25] Simply put, the consequences of losing a bet matter even more than the odds.

In Bitcoin, the consequence of an agent controlling the majority of all voting weight (i.e., the majority of the hashing power online and active in the protocol) in one slot are limited, because violation of the honest-majority axiom in a single slot does not permit the agent to execute an attack against the protocol during an arbitrary slot in the future. As we have demonstrated, pseudo-transfer attacks mean that today's Proof-of-Stake protocols cannot replicate this property. An agent can acquire the majority of all stake in an early slot, paying a severely discounted price based on the protocol's lack of adoption, and it can then employ that stake to execute an attack in a future slot of its choosing. A Proof-of-Stake protocol can effectively be "murdered" by an adversary who overpowers it in its infancy, just as an agent is especially vulnerable to "death" at the beginning of a non-ergodic game when it has not yet accumulated payoffs that provide a buffer against insolvency.

*Weightless Sleep*

Priors claims by Proof-of-Stake protocols to have replicated the security properties of Bitcoin are incorrect, because they overlook the significance of introducing path dependence into a system's mean corruption status. We will demonstrate the flaw in their reasoning through the lens of the "sleepy consensus" or "dynamic availability" properties they invoke to substantiate these claims. [12] [13]

In Bitcoin's implementation of Nakamoto consensus, new agents can come and go from the protocol whenever they want. This free-entry condition ensures that the set of protocol participants is a constantly updated subset of the population. It was the inspiration for our discussion in Section 5 of a toy protocol with a disposable sampling frame, wherein the outcome of every round of consensus depends only on which agents are present and online. Such a protocol does not involve anything analogous to trusted setup; it can be bootstrapped from nothing, because its present security guarantees do not depend on the past mean corruption status within the protocol. A group of agents that was in the majority during the early slots of the protocol will be a control set that belongs to the control structures of those early slots. The same group will not belong to the control structures of later slots if the number of agents participating in the protocol grows large enough that the group becomes a minority. The group itself does not shrink; all of its agents remain in the protocol. However, the influx of new agents who each possess one unit of voting power allows for verifiable dilution to occur: a quantity of voting power that previously exceeded the security threshold $k$ becomes a fraction of voting power that cannot be used to perform an attack.

To convert our example from the toy environment to the realistic one, we simply specify that when new agents join the protocol their voting weight depends on their allocation of the voting resource, $v$. Nakamoto chose to use computing power as the voting resource because he knew that most agents *already* owned roughly one unit of CPU power, and he hoped that the cost of acquiring additional CPU power would be sufficient to prevent effective Sybil attacks. [28] In other words, in the protocol's initial slot $s_{GEN}$ every agent $n$ effectively receives a weight assignment $\omega[n]$, regardless of whether it is asleep or participating, merely by virtue of already owning a CPU. This avoided the need for trust to be placed in the initial set of protocol participants; the overwhelming majority of agents are initially asleep, and so the honest majority axiom has almost no reliability in the early slots. However, as more agents decide to participate in the protocol, the axiom becomes more and more reliable, because security is verifiably based on larger samples of the population.

The fact that sleeping agents effectively receive weight assignments in Nakamoto consensus is the basic reason that the algorithm was suitable for bootstrapping a ledger. No trust had to be placed in the protocol's initial control sets during later slots, because their weight was verifiably diluted as new agents joined the protocol with the weight that they already controlled. If we assume that agents do not alter their allocations of the staking resource, $v$, then the mean corruption status $\bar{y}M_S$ within the protocol will verifiably converge towards the mean corruption status $\overline{Y}$ on the network as larger samples of agents from the network choose to participate, because on average each of them will bring in a preexisting weight $\omega$ that corresponds to the hashing power of one CPU, and there is no reason to expect an initial correlation between corruption status and hashing power.

Unlike Bitcoin, existing Proof-of-Stake protocols award voting weight exclusively to the set of agents who are awake during the initial slot, $s_{GEN}$. Agents who are asleep during that slot do not receive any weight, so they do not have the right to participate in consensus. This forces trust to be placed in the protocol's earliest control structure, since the agents that comprise it start out with a monopoly on the present ownership of and future ability to acquire all the weight ω that will ever exist. If a single one of the protocol's initial control sets is corrupted by the adversary, then the adversary can attack the protocol in any future slot, because its fraction of the voting weight inside the protocol cannot be diluted. The agents that were asleep when the protocol was initiated did not receive any weight, so they cannot join the protocol and increase the total amount of weight that is active in consensus. As we showed in Section 5, there is no guarantee that the mean corruption status within the protocol will converge towards the mean corruption status in the population. The closed sampling frame means that most of the population is effectively "locked out" out of the protocol. If the initial set of participants choose to abuse that fact by retaining their monopoly on power, voting weight can remain centralized indefinitely.

*Gambler's Ruin Revisited*

The process of wealth-normalized replicas within the population $W$ "waking up" and joining the protocol can be modeled with a variant of the gambler's ruin formalism that Nakamoto used in the original Bitcoin whitepaper. [29] Let us assume that a coin exists, with one side labeled "faulty" and one-side labeled "correct," whose ergodic value is known to be biased in favor of the correct side because a significant majority of all wealth is controlled by correct agents. Because faulty replicas executing a book-prize attack can all join the protocol together, the ordering of results from flipping the coin may deviate from the ergodic average.

During the pendency of an ICO, it is possible for the mean corruption status of agents who own stake to decrease if correct agents invest additional wealth. Therefore, every flip of the coin represents one replica $w \in W$ joining the protocol, and the initial distribution of agents in the protocol is based on a number of coin flips equal to the total quantity of wealth invested in the ICO. If the fraction of times the coin has landed "faulty" exceeds the security threshold $k$ during the ICO period, the game does not end, because the adversary is not guaranteed to ultimately win. The remaining flips in the ICO may dilute the mean corruption status below the security threshold. However, if the threshold is still violated when the sequence of ICO flips is complete, then the adversary wins and the game ends.

In order to predict the results of this coin-flipping game, it is not enough to know the ratio of faulty to correct flips that will emerge if the coin is flipped an infinite number of times. If the adversary can arrange the order of the flips, so that the coin is likely to begin with a streak of faulty results, then it can bias the results of the sequence of flips included in the ICO period. The adversary will win the game as soon as the ICO is complete, so it will be irrelevant if the early concentration of faulty flips would have been offset by a later series of correct flips. The adversary achieves victory before the later correct flips can return the mean corruption status associated with flipping the coin to its ergodic average.

The dynamic sampling frame that we described in Section 5, and which is realized in the KRNC protocol, can be understood as a mechanism for extending the sequence of coin flips that are guaranteed to occur before the results of the game are decided. If the adversary can only re-order a limited number of coin flips, then prohibiting it from achieving victory until an extremely large but finite number of flips have occurred can eliminate the adversary's opportunity to win the game. The finite bias that the adversary can initially introduce will be diluted by the results of additional correct flips, so that once the outcome of the game is on the line, the mean corruption status is guaranteed to have converged below the threshold $k$ with overwhelming probability.

## 8. Adaptive Capital Signaling

In the last section, we translated the minimum-participation threshold from the toy environment into the realistic environment, and we proved that existing permissionless protocols have not attracted enough capital to achieve verifiable security. In this section, the focus shifts to obtaining a solution to this problem by designing a superior Sybil-resistance mechanism.

To formalize what is required to prevent book-prize attacks while also preserving Sybil-resistance, we extend our game-theoretic model of participation costs with three new layers of decision nodes, including a capital-allocation subgame. The results we obtain provide clear criteria for designing a secure permissionless protocol.

### 8.1 Weighted Extensive-Form Game

In the extensive-form game from the toy environment, the only choice that agents faced was whether to join the protocol. In the realistic environment, the protocol must choose a resource $v$ to employ for assigning weights in a consensus protocol. Each agent who joins the protocol must therefore choose a quantity of the staking resource to verify, and the protocol must respond to that choice by assigning a quantity of voting weight.

We will prove that the equilibrium cost of participation depends on which resource is adopted by the protocol as a staking resource. Identifying a resource that reduces the cost of participation to zero will enable us to design a protocol that satisfies the requirements of Theorem 1.

*Staking Resource Selection*

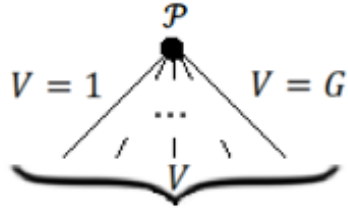

**Figure 8.1:** The protocol selects a resource that will be used to assign weights in the consensus protocol.

The first addition that we make to the extensive form game is to add a new root node. The root node is now a decision node for the protocol, at which it chooses one of the $G$ resources to be the staking resource, $V$. As we will prove in this section, this choice by the protocol determines the outcome of the game, since choosing the correct staking resource will ultimate dictate whether the resource-weighted mean corruption status $\bar{y}M_S$ crosses the security threshold $k$ in equilibrium.

*Capital Allocation Subgame*

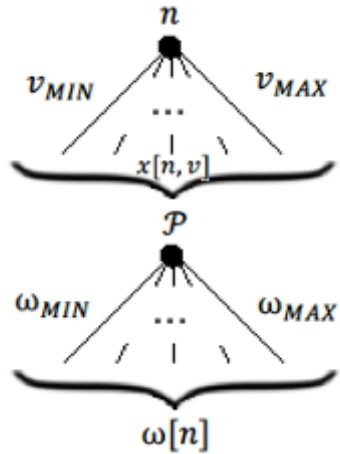

**Figure 8.2:** The protocol assigns a weight to a participating agent based on the agent's allocation of the staking resource.

When an agent $n$ joins the protocol, it now enters a capital-allocation subgame. At the subgame's root node, $n$ must choose a quantity $v$ of the staking resource to distribute among its identities in the protocol from within the closed interval $[v_{min}, v_{max}]$, where $v_{min} \leq v \leq v_{max}$. The protocol $\mathcal{P}$ assigns a weight $\omega[i]$ to every identity in $I_S$ based on the identity's share of the staking resource, $x[i,v]$. One identity $i$ in the subset of $I_S$ controlled by $n$ receives $x[i,v] \geq v_{min}$, and the others each receive $x[i,v] \geq 0$, with the sum totaling $x[n,v]$. Thus, no matter how $v$ is divided among $n$'s identities, the total weight they receive is still equal to the quantity of weight $\mathcal{P}$ assigns in response to $x[n,v]$.

In the extensive-form representation, $n$'s choice of $v$ is followed by a decision node where $\mathcal{P}$ responds by assigning a weight $\omega[n]$. This is an abstraction that represents the sum of every individual weight $\omega[i]$ that $\mathcal{P}$ assigns to an identity $i$ within $I_S$ for whom $N[i]=n$. When those assignments are made, $\mathcal{P}$ has no information about each $i$ except for $x[i,v]$. However, $\mathcal{P}$ is able to adopt a strategy for assigning $\omega[i]$ that is guaranteed to assign a weight $\omega[n]$ to any agent $n$ proportionate to $n$'s share $v$ of the staking resource. The difference between employing such a strategy and directly choosing $\omega[n]$ is that the former does not involve $\mathcal{P}$ acquiring information about $n$'s actual choice of $v$. We account for this difference by specifying that $\mathcal{P}$ knows it has followed its chosen strategy for mapping $x[n,v]$ to $\omega[n]$, but $\mathcal{P}$ forgets $n$'s specific choice of $v$ and $\mathcal{P}$'s corresponding choice of $\omega[n]$. If the subgame were extended with an additional layer of decision nodes for $\mathcal{P}$, those nodes would all belong to the same information set due to the memory-erasure property we have just specified.

*Wealth Signaling and Estimation*

Let the strategy $\mathcal{P}$ employs for mapping $x[n,v]$ to $\omega[n]$ be $\mathcal{P}_\omega$. The strategy can be understood as a rule of inference when we recall that weights must be assigned in proportion to capital. It is common knowledge that the protocol $\mathcal{P}$ will attempt to guarantee that $\bar{y}_{M_s} < k$ by ensuring that $\omega[n] = w[n]$ for all $n$, because $\frac{w[\mathcal{A}]}{w[N]} < k$. As a corollary, every weight assignment $\omega[n]$ by $\mathcal{P}$ based on $x[n,V]$ represents an estimate $\mathcal{P}[v]$ that an agent $n$ controlling $v$ units of the staking resource has total capital of $w[n]$. Given that wealthier agents are able to acquire more of any resource, we further assume that an agent $n$'s endowment $e[n,v]$ of the staking resource is a function of its wealth $w[n]$, and that $e[n,v]$ increases linearly with wealth, such that

$$e[n,v] = v_{MIN} + (w[n] - w_{MIN})\beta$$

where $v_{MIN}$ is the minimum endowment of the staking resource for an agent $n \in N$, $w_{MIN}$ is the minimum wealth for an agent $n \in N$, and $\beta$ is the slope of the relationship between each agent's wealth and its endowment of the staking resource. The linearity of $\beta$ is a simplifying assumption that we must take into account when considering the generality of our results.

The probability-distribution function for $w$ among the members of $N$ is denoted by $\Omega[w]$, where the points of increase for $\Omega$ are the whole interval $[w_{min}, w_{max}]$. The probability-distribution function for $w$ among the subpopulation of protocol participants is $\Omega[w_S]$, where $\Omega$ is constrained to values of $w$ realized by $w[n] : n \in N_S$. In the special case that all agents join the protocol, $N_S \equiv N$, and as a corollary $\Omega[w] \equiv \Omega[w_S]$.

*Costs and Rewards*

In our model, the amount of the staking resource an agent can distribute among its identities is limited by the total exchange value of all the resource it controls. Formally, $\mathcal{P}$'s attempt to estimate $w$ from knowledge of $v$ presupposes that the equilibrium quantity of $v$ an agent controls is constrained by $w$, because if $w$ did not constrain $v$, then $v$ would not be informative of $w$. However, this constraint will not necessarily apply with equal force to faulty and correct agents, because faulty agents may receive more utility from ownership of the staking resource than correct ones, per Lemma 1.

The total utility of each agent $n$ is therefore a function of four factors: its corruption status, its allocation of the voting resource, the weight its identities are assigned by the protocol in response to that allocation, and its wealth as measured by the market value of its initial resource endowment. Formally, we introduce a total-utility function $\alpha_T$ for each agent $n$, which takes four arguments:

$$\alpha_T[n] = \alpha_T(Y[n], x[n,v], \omega[n], w[n])$$

Let the strategy employed by an agent $n$ to choose an allocation $x[n,v]$ that maximizes its total utility $\alpha_T[n]$ be $X(Y[n], w[n])$. It is sometimes convenient to specify a type-dependent strategy, so we use $X_C(w[n])$ to denote the strategy employed by a correct agent $n \in N_C$, and $X_F(w[n])$ to denote the strategy employed by a faulty agent $n \in N_F$.

The specific form of $\alpha_T$ that we employ is a modified version of the bell-shaped distribution from [45], which has been translated into the notation from our model, and extended to permit type-dependent utility differentials:

$$\alpha_T[n] = \exp\{\omega[n](\lambda_Y - \psi_Y)\} * \exp\left\{-\frac{1}{2}\left(\frac{x[n,v] - e[n,v]}{\infty - p_V}\right)^2\right\}$$

The left term captures the utility that an agent receives from being assigned weight by the protocol; the quantity of weight it is assigned is scaled by the type-dependent expected payoff.[10] The right term captures the utility that an agent loses from expending capital to acquire additional units of the staking resource; the difference between its final allocation of the staking resource and its original endowment is scaled by the price of the staking resource.

---

[10] The type-dependent term $\psi_Y$ captures potential costs to faulty agents that are unrelated to purchasing units of the staking resource, such as the risk of criminal penalties from taking part in an attack.

To find a Nash equilibrium, it is also necessary to specify a corresponding utility function for the protocol $\mathcal{P}$, to constrain its weight-assignment strategy, $\mathcal{P}_\omega$. Let $\mathcal{P}$'s utility be dictated by a loss function $D(w[n], \omega[n])$, which penalizes it in proportion to the discrepancy between an agent's true wealth and the weight it is assigned. This loss function only describes the utility penalty associated with assigning weight to the subset of identities controlled by a single agent, $n$. To determine the utility penalty across all weight assignments by $\mathcal{P}$, the loss function must be averaged over the distribution of wealth within the subpopulation $N_S$ of protocol participants:

$$\int_w D\left(w[n], \mathcal{P}_\omega\big(X(Y[n], w[n])\big)\right) d\Omega[w]$$

*Equilibrium Conditions*

All of the toy-environment Nash-equilibrium conditions from Section 4 apply in this extended version of the game, however we must also introduce new equilibrium conditions that capture the concept of stable resource allocations. The solution concept we will prove is an evolutionarily stable Nash equilibrium ("ESS"), in which neither the protocol nor the agents can increase their utility by modifying their respective strategies. Formally, let an allocation strategy $X$ that has been adopted by every agent $n$ be $X^*$, and let the protocol's corresponding weight-assignment strategy be $\mathcal{P}_\omega^*$. An ESS for $X^*$ and $\mathcal{P}_\omega^*$ must satisfy the following two conditions, which translate Grafen's definitions into our model. [6]

First, the utility that each agent $n$ receives from strategy $X^*$ must be greater than or equal to the utility available from adopting any alternative allocation strategy. Formally, for all $x[n, v]$ and $w[n]$:

$$\alpha_T\big(Y[n], X^*(Y[n], w[n]), \mathcal{P}_\omega^*\big(X^*(Y[n], w[n])\big), w[n]\big) \geq$$
$$\alpha_T\big(Y[n], x[n, v], \mathcal{P}_\omega^*(x[n, v]), w[n]\big)$$

Second, the utility penalty that $\mathcal{P}$ receives from $\mathcal{P}_\omega^*$ must be less than or equal to the utility penalty available from adopting any alternative weight-assignment strategy. Formally, for all $x[n, v]$ and $w[n]$:

$$\int_w D\left(w[n], \mathcal{P}_\omega^*\big(X^*(Y[n], w[n])\big)\right) d\Omega[w] \leq$$
$$\int_w D\left(w[n], \mathcal{P}_\omega\big(X^*(Y[n], w[n])\big)\right) d\Omega[w]$$

Consistent with the original signal-theoretic models we are extending, we assume continuity and differentiability for $\alpha_T$ and $D$, and measurability for $X^*$ and $\mathcal{P}_\omega{}^*$. [45]

## 8.2 Proving Evolutionary Stability

To find the ESS, we employ a proof technique drawn directly from the biological signaling-theory literature. [45] First, we show how the original technique can be translated into our model's terminology and notation, but defer consideration of whether the type-dependent argument $Y[n]$ prevents an ESS. Once we have demonstrated the proof technique, we consider its generality and prove that the ESS it yields will exist in our model if certain additional requirements are satisfied.

Given that an agent $n$'s initial endowment $e[n,v]$ of the staking resource reflects its freely chosen pre-protocol apportionment of capital to $V$, the agent will necessarily suffer a utility penalty from switching to a different allocation $x[n,v]$, with greater divergences from $e[n,v]$ inflicting greater utility penalties. Formally,

$$sign(\alpha_{T_2}(Y[n], x[n,v], \omega[n], w[n])) = -sign(x[n,v] - e[n,v])$$

where the subscript two in $\alpha_{T_2}$ indicates that a partial derivative is taken with respect to the second argument in the total-utility function, $x[n,v]$. If an agent's allocation is larger than its initial endowment, the derivative $\alpha_{T_2}$ is negative, which indicates that the agent would prefer to allocate less of its wealth to the staking resource. If an agent's allocation is smaller than its initial endowment, the derivative $\alpha_{T_2}$ is positive, which indicates that the agent would prefer to allocate more of its wealth to the staking resource.

To find an ESS that is consistent with this assumption, we take the first derivative of an agent $n$'s utility with respect to $x[n,v]$ at the candidate ESS, $X^*(Y[n], w[n]), \mathcal{P}_\omega{}^*(X^*(Y[n], w[n]))$. Logically, the slope of the first-order utility gradient must equal zero at the candidate ESS, because otherwise an agent $n$ could improve its utility by buying or selling the staking resource. An ESS therefore requires

$$\{\alpha_{T_2}(Y[n], x[n,v], \mathcal{P}_\omega{}^*(Y[n], x[n,v]), w[n])\}$$
$$+ \{\mathcal{P}_\omega{}^*_2(Y[n], x[n,v])\alpha_{T_3}(Y[n], x[n,v], \mathcal{P}_\omega{}^*(Y[n], x[n,v]), w[n])\} = 0$$

The first bracketed expression is the partial derivative of the agent's total-utility function, taken with respect to its allocation of the staking resource. Reducing this expression to zero ensures that the agent's intrinsic capital-allocation preferences are consistent with the ESS. The second bracketed expression represents the partial derivative of the protocol's weight-assignment strategy taken with respect to an agent's allocation of the staking resource, multiplied by the partial derivative of an agent's total-utility function taken with respect to its weight assignment. Reducing this expression to zero ensures that the reward offered by the protocol does not induce the agent to have capital-allocation preferences that are inconsistent with the ESS.

To find a version of $\mathcal{P}_\omega{}^{*\prime}(x[n,v])$ that satisfies the zero-slope requirement, we consider how the protocol assigns weight to agents of the lowest possible wealth within the distribution $\Omega[w_S]$. If we assume that $\mathcal{P}_\omega{}^*(v_{min}) = w_{min}$ – i.e., that the protocol accurately infers that an agent who controls the minimum quantity of the staking resource also controls the minimum quantity of wealth — then the zero-slope requirement is satisfied by

$$\mathcal{P}_\omega{}^{*\prime}(x[n,v]) = -\frac{\alpha_{T_2}(Y[n],x[n,v],\mathcal{P}_\omega{}^*(Y[n],x[n,v]),\mathcal{P}_\omega{}^*(Y[n],x[n,v]))}{\alpha_{T_3}(Y[n],x[n,v],\mathcal{P}_\omega{}^*(Y[n],x[n,v]),\mathcal{P}_\omega{}^*(Y[n],x[n,v]))}$$

.

Note that the fourth argument in the total-utility function has been modified. Ordinarily, it is $w[n]$; here, $w[n]$ is replaced with $\mathcal{P}_\omega{}^*(Y[n],x[n,v])$, because at the ESS it is necessary for the protocol to accurately infer wealth based on control of the staking resource, such that $w[n] = \mathcal{P}_\omega{}^*(X^*(Y[n],x[n,v])$.) We will prove that this requirement is satisfied for an initial range, starting at $v_{min}$. This provides a solution for $\mathcal{P}_\omega{}^*$ within that range, which we can invert to obtain $X^*$ over the same initial range. If the range of the solution extends to $w_{max}$, we will have found an ESS, because the protocol will have assigned an accurate weight to every agent, and neither the protocol nor any agent will be able to improve its total utility by modifying its strategy.

First, we prove that the monotonic section at the beginning of $\mathcal{P}_\omega{}^*$ must be increasing, based on the conditions we have already specified. At $v_{min}$, the initial slope of $\mathcal{P}_\omega{}^*$ is zero, because

$$\alpha_{T_2}(Y[n],v_{min},\omega[n],w_{min}) = 0.$$

Given that $\mathcal{P}_\omega{}^*$ initially has zero slope, and that $v$ increases immediately above $v_{min}$, we can infer that the slope of $\mathcal{P}_\omega{}^*$ is positive immediately above $v_{min}$. The slope of the agent's staking strategy $X^*$ is therefore effectively infinite, such that $x[n,v] > e[n,v]$, and $\alpha_{T_2} < 0$. Because $\mathcal{P}_\omega{}^*$ is the inverse of $X^*$, it follows that $\mathcal{P}_\omega{}^*$ has a positive initial slope, and that its slope remains positive for at least some interval above $v_{min}$. By inverting $\mathcal{P}_\omega{}^*$ over that interval, we can obtain an agent $n$'s partial staking strategy,

$$X^* : \mathcal{P}_\omega{}^*(X^*(Y[n],w[n])) = w[n]$$

,

whose properties include that $X^*(Y[n], w_{min}]) = x[n: w[n] = w_{min}, v]$, that the initial slope at $v_{min}$ is infinite, and that the slope is positive for all values within the range of the solution.

*Generality and Type-Dependence*

Is this sufficient to prove an ESS in our model? Yes, but only if additional conditions are satisfied.

Even without the potential for a type-dependent payoff differential, if the expected utility of receiving a weight assignment from the protocol is too small, such that $\alpha_{T_3}$ assumes a near-zero positive value, then the protocol's weight-assignment strategy $\mathcal{P}_\omega{}^*$ may issue infinite weight in response to a finite quantity of the staking resource. We overcome this and similar failure modes by adopting the viability axiom that an ESS exists if: (1) the existence of the ESS is consistent with our model; and, (2) the existence of the ESS is necessary for permissionless Byzantine consensus. This is a much narrower axiom than assuming that permissionless Byzantine consensus is, in fact, possible; it forces us to identify a viable ESS in a model where the adversary is powerful enough to defeat all existing permissionless Byzantine consensus protocols.

The challenge arises from the first argument $Y[n]$ in the total-utility function $\alpha_T$. Because an agent's type $Y[n]$ is private information, the protocol $\mathcal{P}$ cannot observe the agent's true type when deciding how much weight to assign, so its strategy $\mathcal{P}_\omega$ will represent a probability-weighted average of the appropriate weight assignment for a correct agent and the appropriate assignment for a faulty agent. If the partial derivative $\alpha_{T_1}$ of total utility with respect to corruption status is positive, then faulty agents receive greater utility than correct agents from allocating capital to the staking resource. The result is a type-dependent "wealth gap" for an observed quantity of the staking resource, between the true wealth of a faulty agent with that allocation $w[n: n \in N_F]$ and a higher wealth of a correct agent, $w[n: n \in N_C]$. The mean weight assignment from $\mathcal{P}_\omega$ will therefore be systematically biased in favor of faulty agents, which prevents the pooling equilibrium mandated by Theorem 1. To obtain an ESS that enables permissionless Byzantine consensus, we will need to find a way to verifiably reduce the partial derivative $\alpha_{T_1}$ to 0, so that faulty and correct agents receive the same utility from the staking resource.

*Bayesian Inference Rules*

We start by applying the ESS proof technique to the specific bell-shaped distribution of $\alpha_T$ we previously formulated. The protocol's candidate weight-assignment strategy $\mathcal{P}_\omega{}^*$ is given by the formula

$$\mathcal{P}_\omega{}^* = w_{min} + \frac{x[n,v] - v_{min}}{\beta} - \frac{(\lambda_Y - \psi_Y)(\infty - p_V)^2}{\beta^2}$$

$$\left\{ 1 - \exp\left( -(v - v_{min}) \frac{\beta}{(\lambda_Y - \psi_Y)(\infty - p_V)^2} \right) \right\}$$

The first two term describe the wealth $w[n]$ of an agent for whom $e[n,v] = x[n,v]$, i.e., whose original endowment of the staking resource is equal to the allocation of the staking resource observed by the protocol, $\mathcal{P}$. The rest of the equation accounts for the fact that the introduction of the protocol may induce agents to exaggerate their holdings of the staking resource, such that $e[n,v] < x[n,v]$, which forces the protocol to downgrade its wealth estimates. The result is a type-independent wealth gap, which we refer to as the "Sybil gap" to distinguish it from the type-dependent gap we have already defined.

At $v_{min}$, the Sybil gap is zero, because the protocol accurately infers that $v_{min} = w_{min}$. However, as $v$ increases the Sybil gap expands; it assumes progressively larger negative values, indicating that the wealth associated with a given quantity of the staking resource is diminishing. Formally, the Sybil gap converges exponentially towards an asymptote of

$$\left. (\lambda_Y - \psi_Y)(\infty - p_V)^2 \middle/ \beta^2 \right.$$

with the distance along the $v$ axis for the gap to halve (i.e., the function's half-life) given by

$$\left( \left. (\lambda_Y - \psi_Y)(\infty - p_V)^2 \middle/ \beta \right. \right) \ln 2$$

For each agent $n$, there is a corresponding "staking gap" between the pre-protocol endowment $e[n,v]$ and post-protocol allocation $x[n,v]$ for an agent $n$ of a given wealth $w[n]$. Unlike the Sybil gap, the staking gap assumes a positive value, reflecting the post-protocol increase in an agent's quantity of the staking resource. After beginning at 0 for $w_{min}$, the staking gap converges asymptotically to

$$\left. (\lambda_Y - \psi_Y)(\infty - p_V)^2 \middle/ \beta \right.$$

with a half-life, of

$$\left( {(\lambda_Y - \psi_Y)(\infty - p_V)^2} \Big/ {\beta^2} \right) \ln 2$$

To satisfy the requirements of Lemma 1 and achieve permissionless Byzantine consensus, we will need to find an ESS that eliminates all of the potential gaps we have identified. As Lemma 2 demonstrates, the only way to accomplish that is to reduce the cost of participation in the protocol to zero. In the next sub-section, we move closer to that goal by formalizing the factors that determine the equilibrium costs of participation.

*Equilibrium Staking Costs*

To identify the equilibrium cost of participation in Sybil-resistant consensus, we scale the difference between an agent's post-protocol allocation and pre-protocol endowment by the market price of the staking resource. Formally, we obtain the following type-dependent cost function $c[n]$ for agents of a given wealth:

$$c[n] = c(Y[n], w[n]) = \frac{X^*(Y[n], w[n]) - e[n, v]}{\infty - p_V}$$

$$= \frac{(\lambda_Y - \psi_Y)(\infty - p_V)}{\beta} \exp\left( -(X^*(Y[n], w[n]) - v_{min}) \frac{\beta}{(\lambda_Y - \psi_Y)(\infty - p_V)^2} \right)$$

The equation makes it clear that the equilibrium cost of protocol participation depends on $\beta$, the slope of the pre-protocol association between wealth and the staking resource.

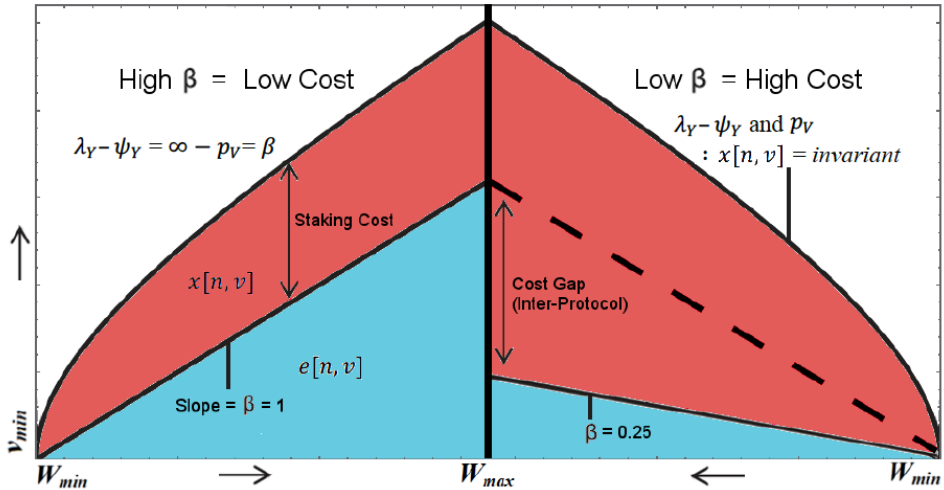

**Figure 8.3:** Equilibrium costs depend on $\beta$, the slope of the pre-protocol relationship between wealth and the staking resource.

As shown in **Figure 8.3**, a higher value of $\beta$ indicate that the slope of the association is steeper. A steeper slope means that there is a greater difference between the low-wealth and high-wealth agents' initial endowments of the staking resource — i.e., that the pre-protocol quantity of the staking resource an agent owns grows more quickly as a function of its wealth. This increases the cost that a lower-wealth agent would have to pay to imitate a higher-wealth agent, because it forces the lower-wealth agent to purchase a larger amount of the staking resource in order for its post-protocol allocation to match the higher-wealth agent's pre-protocol endowment. The greater the cost of exaggeration, the less that low-wealth agents' post-protocol allocations will deviate from their pre-protocol endowments, and therefore the less that high-wealth agents will be forced to increase their own allocations of the staking resource in order to obtain accurate weight assignments.

A simple way of understanding this result is to consider the difference between sunk costs and prospective costs. High values of $\beta$ indicate that high-wealth agents have already incurred significant sunk costs to acquire large pre-protocol endowments of the staking resource, while low-wealth agents have not. This creates a natural cost asymmetry: high-wealth agents can acquire weight "for free" based on the costs they have already incurred, while low-wealth agents attempting to acquire the same quantity of weight would be forced to incur new costs, because they do not already possess the necessary quantity of the staking resource. The higher the value of $\beta$, the greater this natural asymmetry, and the less that high-wealth agents are forced to incur new costs to distinguish themselves from low-wealth agents.

In contrast, low values of $\beta$ are associated with high participation costs at equilibrium, because there are fewer sunk costs to provide asymmetry. In the absence of sunk costs, the only asymmetry available to guarantee honest signaling is the prospective ability of high-wealth agents to expend more capital than their low-wealth counterparts. Such high-cost signaling is employed as the Sybil-resistance mechanism in Proof-of-Work and Proof-of-Stake protocols, which force agents to verifiably expend computational or financial resources. If the only factor affecting cost tolerance were access to capital, these methods would be reliable. However, per Lemma 2, faulty agents have an inherently higher cost tolerance than correct agents in permissionless protocols that have not cleared the minimum participation threshold required to prevent the adversary from successfully attacking. Existing Sybil-resistance methods are therefore unsuitable for permissionless Byzantine consensus.

## 8.3 Zero-Cost Staking

To overcome the problem we have just formalized, we need a Sybil-resistance mechanism that relies on an asymmetry in pre-protocol sunk costs, rather than an asymmetry in present cost tolerance. This concept becomes intuitive if we recall the biological mechanisms that inspired the underlying signal-theoretic models

*Tigers and Peacocks*

It is already well known that Sybil-resistance methods like Proof-of-Work are based on the "handicap principle" that an untrusted signaler can prove that it is an agent with a high cost tolerance by sending a signal that is verifiably expensive. [4] In biology, this is the signaling mechanism famously employed by male peacocks to communicate their fitness to females: Wasting finite biological resources to grow an oversized tail imposes a verifiable fitness penalty that only the fittest peacocks can survive.

As the example of the peacock illustrates, handicap-authenticated signaling is inherently wasteful. The waste itself is what makes the signal reliable, "because by wasting one proves conclusively that one has enough assets to waste." [3] In Proof-of-Work, by wasting computing power on arbitrary puzzles, agents conclusively prove that they are wealthy enough to acquire the necessary computing power. In Proof-of-Stake, by transferring funds to an ICO administrator, agents conclusively prove that they are wealthy enough to tolerate the loss of the funds. In both instances, security is based on the assumption that the combined cost tolerance of all the correct agents who join the protocol will exceed the combined cost tolerance of all the faulty agents who join the protocol.

The problem is that, unlike peacocks, human agents can choose to opt out of handicap-authenticated signaling. Most of the agents on the internet have exercised that option; they have declined to join permissionless consensus protocols that require all users to sacrifice computing power or money. Thus, even if Proof-of-Work and Proof-of-Stake were effective as Sybil-resistance methods, they still would be impractical for building a verifiably secure permissionless protocol, because the costs of the handicaps they impose deter agents from participating in consensus. Assigning capital-weighted voting power equally among the set of all consensus participants is meaningless if the set of protocol participants is so small that the adversary can simply execute a book-prize attack.

Fortunately, there are alternatives to handicap-authenticated signaling that do not impose the same costs. One of them is identity-authenticated signaling, which biologists formalized in the context of group selection. This has already become one of the pillars of modern computer networking, as evidenced by the numerous permissioned protocols that rely on identity. However, there is a third form of biological signaling that computer scientists have overlooked: cue-authenticated signaling. [45]

Like a handicap, a cue is a trait whose verifiable cost makes it impossible for low-fitness organisms to spoof. The difference is that cues represent traits that developed for an initial non-signaling purpose; they allow authentication to be performed based exclusively on sunk costs, with zero added costs. For example, male tigers compete to grow as large as possible in order to kill one another and their prey, so female tigers are able to identify high-quality males based on their larger body size (or, indirectly, from the elevation of scratch marks left on trees by male tigers to mark their territory). [46] This is a far more efficient signaling protocol than the method employed by peacocks, because it does not force male tigers to handicap themselves in any way.

High-quality male tigers allocate their finite biological resources in the same manner that they would have in the absence of any signaling protocol, because the signaling protocol is based on their non-signaling preferences. A low-quality male who was unable to compete against high-quality males in the original body-size competition has no ability to "spoof" a larger body size once that trait is adopted by females as an informative cue. The cue is informative precisely because it is based on the results of a competition in which all male tigers are already guaranteed to be fervent competitors.

*The Measure of All Things*

A cue-authenticated Sybil-resistance scheme would overcome the shortcomings of today's handicap-authenticated methods, because it would decrease the cost of participation to zero. The key is to identify a staking resource that everyone is already competing to acquire, because then everyone will already own the resource needed to participate in consensus. Agents will not need to incur any new costs, because the protocol will be able to measure their sunk costs in society's preexisting resource competition.

There is one and only one resource that can fill this role: Money itself. It is the one resource that everyone is already competing to acquire, because it is not really a resource at all. [47] Rather, it is an abstraction that records every economic agent's individual balance of trade with the rest of society. [48] Those balances can then be used to obtain whatever resource an agent requires in the future, so there is no limit to the quantity of money that agents desire. [49] This accounts for why money is inherently the most valuable resource. It also explains why the slope of the relationship between money and wealth is orders of magnitude steeper than for other resources.

Formally, those are the two critical attributes for achieving a zero-cost ESS. Because receiving weight in a successful permissionless protocol will inherently be valuable, we must anticipate a significant positive value for $\lambda_Y$. This value cannot reliably be reduced by $\psi_Y$ for two reasons. First, the incomplete-information conditions specified in Section 3 mean that the value of $\psi_Y$ is uncertain. Second, even if the value of $\psi_Y$ were significant, it would still only offset $\lambda_Y$ for faulty agents in the event of an attack by the adversary. To verifiably reduce the cost of participation to zero for all agents in all subgames, it is necessary for the values of $\beta$ and $p_V$ to be so high that agents cannot practically imitate the endowments of their higher-quality counterparts. The unique characteristics of money guarantee that using it as the staking resources will ordinarily maximize the value of these parameters, so that the equilibrium cost of protocol participation will be lower than is possible with any other resource.

An exception to this general principle occurs in the event of hyperinflation, during which the supply of money increases so much that it ceases to be the most saleable resource in the economy. We exclude this scenario from our analysis by adopting an axiomatic limit on the size of the incumbent monetary supply, defined using constants that avoid the need to make a specific scale-setting assumption.

**Monetary Scarcity Axiom:** The total supply of the monetary voting resource $V$ is constrained : $p_V - p_g < u$ for $\forall g \in G$ where $u$ is the threshold at which $\exists g \in G : c[n]$ would increase if $g$ were substituted for $V$.

**Theorem 3:** *If a protocol's staking resource is the incumbent monetary resource, the evolutionarily stable equilibrium cost of participation in permissionless Byzantine consensus is zero.*

**Proof**: We obtain this result through a proof by contradiction. Negative costs do not exist in our model. Therefore, if the equilibrium cost of participation is not zero, then it must be positive. If the equilibrium cost is positive when the staking resource $V$ is money, then the equilibrium cost is positive for $\forall g \in G$, because money has the optimal available combination of $\beta$ and $p_V$ given the monetary scarcity axiom. If the equilibrium cost is positive for $\forall g \in G$, then Lemma 2 is not satisfied, and permissionless Byzantine consensus is impossible per Theorem 1. The result is a contradiction: permissionless Byzantine consensus would be verifiably impossible, but our model would include an ESS that is consistent with permissionless Byzantine consensus, which contravenes the viability axiom. □

The fact that the equilibrium cost of participation drops to zero if money is used as the staking resource enables us to tie up all the remaining loose ends in our game-theoretic model. Because it costs agents nothing to participate in consensus if the staking resource is money, Lemma 2 establishes that the dominant strategy for all agents is to join the protocol. When all agents join the protocol, $\Omega[w_S] \equiv \Omega[w]$, so the wealth-weighted mean corruption status of protocol participants is guaranteed to be below the security threshold, $k$. Theorem 3 guarantees that weight assignments made in proportion to money will accurately reflect wealth, so all agents can verify in advance that $\bar{y} M_S < k$. The possibility for a type-dependent payoff differential therefore vanishes: $\lambda_F = \lambda_N$, because play is guaranteed to reach the left subgame where faulty and correct agents receive identical payouts. In economic terms, there is no faulty utility $\alpha_F$ available, so $\mathcal{A}$ is motivated exclusively by $\alpha_N$, and its behavior is therefore indistinguishable from that of a correct agent.

## 8.4 The "New Money" Trap

The security and performance improvements that a permissionless consensus protocol can achieve by using the incumbent monetary resource as its staking resource $v$ are such an improvement over current technology that it is natural to suspect trickery or gimmickry. Those fears subside when one understands the deeper principles that underpin our approach.

Whenever a new resource is adopted as a monetary store of value, agents respond by producing more of the resource, which then destroys the resource's value by expanding its supply. This has been dubbed the "easy money trap." [30] A well-known solution to this problem is to continue employing the incumbent monetary resource as money. Because the incumbent monetary resource has already achieved an equilibrium price that reflects its value as money, its rate of production has already been virtually maximized, and the available supply cannot suddenly expand.

Gold enjoys this as a key advantage over other metals: it has been treated as money for so long that most of the gold available on Earth has already been mined. [30] The existing "stock" of gold vastly exceeds the "flow" of new gold onto the market, since all of the easily mineable gold has already been obtained. In contrast, whenever a different metal starts to become accepted as money, the rate of production for that metal increases, inflating the circulating stock and diminishing the metal's per-unit value. Gold has the highest stock-to-flow ratio of all physical commodities, which is what makes it a reliable store of value.

There is an inherent circularity to what we have just described: gold has the highest stock-to-flow ratio because it is a form of money, and it is a form of money because it has the highest stock-to-flow ratio. As it turns out, this circular pattern extends beyond gold. For anything to function as money in society, it is necessary that "everybody wants it because everybody knows that all the others want it." [49] Money is thus the ultimate Keynesian beauty contest. [50] Agents insist on receiving payment in whatever resource they expect other agents to treat as money in the future, and the best predictor of what will be treated as money in the future is what has attained the status of money in the present. [51]

So far, we have described this process in terms of physical money, but there is no formal difference between using ownership of tokens to track monetary balances and recording those balances on a ledger. [52] In game-theoretic terms, the choice of which metal to accept as money when used in coins can be directly compared to which ledger to accept as a record of monetary balances. In both instances, an agent is forced to choose between competing versions of reality, since the canonical definition of money dictates how much money everyone owns. [53]

*Fractal Salience*

There is an obvious parallel to Nakamoto consensus, in which the canonical branch of the blockchain determines how much Bitcoin everyone owns. Miners all want to receive their rewards on the branch that will become the canonical version of reality in the future, so the equilibrium strategy is for them to mine on the longest branch in the present. This "incumbent" branch represents the obvious Schelling point, and selecting it enables miners to reach a Nash equilibrium. [54]

In the Tezos shell, the same concept is extended to a higher level of abstraction: instead of using the "follow the crowd" mechanism merely to define a canonical ordering of transactions for a given consensus algorithm, a meta-protocol allows participants to express preferences about which implementation of the ledger protocol should be treated as canonical. [42] The logical Nash equilibrium is for all agents to accept only the version of the protocol that has received the required endorsement of the crowd. Thus, at both the base level of transaction ordering and the higher level of protocol design, the principle of "follow the crowd" enables groups of strangers to effectively coordinate their decisions.

When a protocol stacks coordination games on top of one another, it is important for the Schelling points at both levels to be consistent. By definition, a Schelling point is an option that players in a coordinate game expect one another to pick. [55] This requires: (1) that there is a characteristic of the option that distinguishes it from the others; and (2) that each player expects the other players to treat that characteristic as salient. The latter requirement depends on the decision rules that players follow for identifying salient features. The choice of properties upon which salience is evaluated can be modeled as a meta-game, in which rational players exclude properties that lack *meta*-salience given the structure of the coordination space. Players converge on the focal property precisely *because* it yields the focal point. [56]

In a two-level coordination game, if there is one potential decision rule that designates a Schelling point at two levels, and other potential decision rules that each designate a Schelling point at only one of the two levels, then the "meta-Schelling point" is for players to follow the rule that delivers Nash equilibria at both levels. In a slight abuse of terminology, we will refer to this variant of subgame perfection as *fractal salience.* [56]

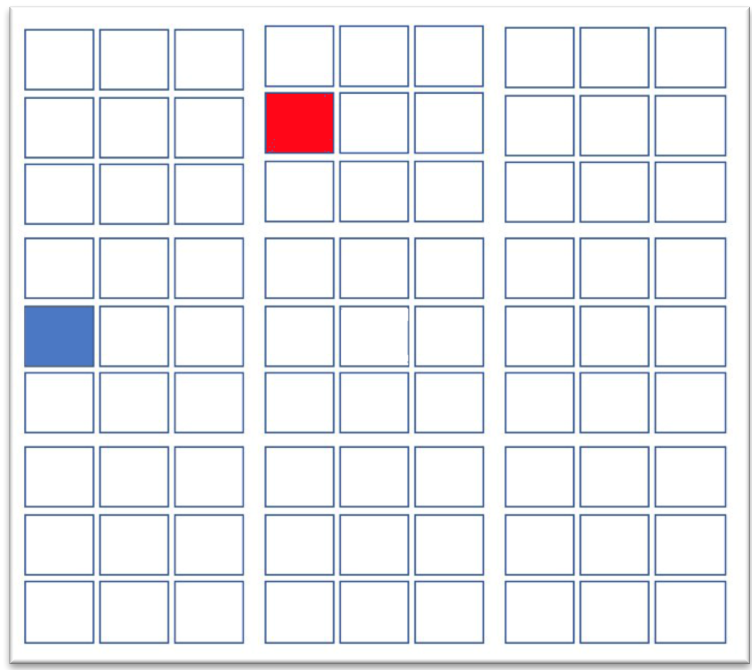

**Figure 8.4:** The blue cell has fractal salience, because it can be reached by a consistent decision rule. The red cell does not.

A simple example of fractal salience is offered in **Figure 8.4**, which depicts a two-level coordination game. At the first level, the players must select one cell within the 3x3 grid that spans the entire figure. At the second level, whatever cell the player selected becomes another 3x3 grid, and the player can either choose to select one of the cells within that grid or return to the first level of the game. In the absence of any colored cells, the Schelling point would be to choose the center cell at both levels of the game. However, the introduction of colored cells provides two stronger Schelling points at level two, and the rational strategy at level one is therefore to choose one of the two cells whose internal sub-grids contain a colored cell — i.e., to choose one of the two subgames that contains its own Schelling point. Despite the fact that there are facially two competing Schelling points available in level 2, rational agents can still achieve Nash equilibrium by adhering to the principle of fractal salience: the blue cell is the superior Schelling point, because it permits agents to choose the same cell at both levels of the game.

The Schelling points available in Bitcoin, Tezos, and other cryptocurrency protocols are equivalent to the red cell. At the intra-protocol level of choosing a canonical ledger version, they invoke the "follow the crowd" decision rule. Yet at the inter-protocol level of choosing the canonical version of money, they force agents to *deviate* from the crowd by accepting an arbitrary set of bootstrapped balances rather than the balances recorded in the incumbent fiat monetary system.

*The OLG Prisoner's Dilemma*

The overwhelming network effects that favor incumbent currencies mean that a new, bootstrapped currency like Bitcoin will only achieve adoption as a global monetary standard if the existing system crumbles in, e.g., a severe economic crisis. [57] In contrast, when balances on a new cryptographic ledger are assigned in proportion to the data recorded in the world's incumbent monetary ledger, then the congruence between the two versions of monetary reality means that the new ledger is "at no disadvantage whatsoever against the established ledger." [53]

|  | Option 1: Fiat | Option 2: Cryptocurrency | **Option 3: Crypto-Fiat** |
|---|---|---|---|
| Protocol | Keep | Change | **Change** |
| Data | Keep | Change | **Keep** |

**Figure 8.5:** The third option broadens the search space by allowing agents to express a new preference combination.

When the only way for a market agent to express a desire for cryptographic inflation protection is to switch to using a cryptocurrency, then agents can only express that preference if it is so strong that it overrides their desire to maintain the stability of the incumbent monetary order. Even people who individually stand to profit from the replacement of fiat money by cryptocurrency do not wish for their loved ones and friends to have their savings destroyed in a monetary-protocol transition. As **Figure 8.5** illustrates, assigning balances in proportion to existing money allows the market to converge on agents' true preferences more accurately. It broadens the search space, so that a new preference combination can be expressed — one that favors keeping the world's existing monetary balances while upgrading the protocol used to track those balances.

For the sake of argument, let us assume that the network effects favoring fiat money are not, in fact, strong enough to fend off the challenge from cryptocurrency. This is the view held by many in the Bitcoin community, who have publicly cheered perceived dips in the value of fiat money, which they believe presage the death of the dollar and today's other national currencies. Is this an outcome that the public should embrace? Is it an outcome that their elected officials should tolerate? Only if they wish to crown a new crypto-oligarchy.

It is clear why the early adopters of Bitcoin are seeking a million- or billion-fold return on their initial "investments" of computing power, but it remains a mystery why the rest of humanity should acquiesce to their plan. There is nothing inherent in the creation of a nongovernmental monetary protocol that requires it to function like a pyramid scheme. That was never the intention of early advocates of private money, who were at pains to design systems that would empower the people — not impoverish them. [33] [58] [59] [60] [61]Some of these visionaries have been appropriated by today's cryptocurrency advocates, but they would not have supported concentrating the world's wealth in a handful of early protocol adopters.

The transition to a new monetary protocol can be accomplished smoothly only if it inherits the balances stored in the existing system. Those balances represent the public's savings — value earned for past work, which has not yet been redeemed for goods or services. [48] As Nobel Laureate Paul Samuelson warned, it has always been in the myopic self-interest of the young to "unilaterally repudiate the money upon which the aged hope to live in retirement." [62] To prevent this injustice, "a continuing social compact is required," one in which each new generation resists the temptation to betray its elders by repudiating their money. [62]

This same game-theoretic mechanism explains why the reign of any bootstrapped successor to the incumbent monetary system would inherently be short-lived. Having defected against the prior incumbent, Bitcoin has forfeited access to the Nash equilibrium in which every generation honors its predecessor's money. The alternative equilibrium is one in which every succeeding generation defects against its predecessor by repudiating the incumbent money. [63] This is an equilibrium in which everyone loses, because long-term value storage is impossible. Yet it is the equilibrium that has been mandated by the limits of prior technologies.

Ironically, this problem was recognized by the creators of the Tezos protocol, but their empathy extended only to the holders of *cryptocurrencies* whose savings would be destroyed if their crypto-balances were not included in new monetary ledgers. [42] No concern was given to the billions of owners of fiat money who would suffer the same fate if Tezos or Bitcoin became a new universal medium of exchange and value storage. Or, more realistically, no concern was given to the fact that Tezos and Bitcoin will never reach that status, due to the overwhelming inertia in favor of fiat money as the "longest chain of value" in society.

At the intra-protocol level, Nakamoto consensus embodies the same basic game-theoretic equilibrium that has underpinned money for thousands of years: When everyone puts skin in the game, the game itself can become 'too big to fail'– which makes it the game that everyone wants to play. Trying to bootstrap a new game is a critical mistake. The security of money has always come from the protection of the herd. [49]

*A Perfect Circle*

The lessons of this sub-section may sound as if they relate to economics rather than protocol security. Yet, as we have demonstrated with our formal model, these concepts cannot be separated from one another: the value of the staking resource is ultimately what dictates the security of any resource-weighted protocol.

This is not a wholly new insight. It has long been recognized that a "chicken-and-egg" problem exists in Proof-of-Stake: a protocol needs to have a valuable native asset in order to be secure, but it also needs to be secure in order for its native asset to be valuable. [64] In the conventional view, the flaw in Proof-of-Stake is therefore that it is circular. [65] We have proved that the opposite is true: The flaw in Proof-of-Stake is that it is not circular enough.

Money itself is circular. It "is accepted as money by everybody merely because it is accepted as money by everybody else." [51] By assigning cryptographic weight in proportion to money, a consensus protocol can be created that everyone will join merely because they know that everyone else will join. That consensus protocol will represent a meta-agent more trustworthy than any institution or actor in human history.

# Part IV. Protocol Design

## 9. Key Retroactivity Network Consensus

To harness our theoretical results in the real world, we need a way to employ the incumbent monetary resource as the staking resource in a permissionless Byzantine consensus protocol. Our solution is the Key Retroactivity Network Consensus ("KRNC") protocol. For expositive clarity, we introduce it in three steps.

First, we provide an informal summary of the KRNC protocol's intended architecture. We start by drawing an analogy to the gold standard, then explain the differences and advantages of KRNC.

Second, we formalize the intended architecture in an idealized environment, where a PKI exists and all commercial banks operate a public, sharded ACID database. A toy Proof-of-Balance protocol is defined, which surpasses the verifiable security and performance of Proof-of-Work and Proof-of-Stake systems by a factor of ~40,000.

Third, we turn the idealized environment into a realistic environment by removing the assumptions that allowed the toy protocol to function. We iteratively extend the toy protocol until we have obtained the full KRNC protocol, which delivers the same security and performance advantages in the real world.

## 9.1 A New Gold Standard

The basic goal of the KRNC protocol is to distribute something inherently scarce to the present owners of government-issued digital money. There is thus an obvious parallel to the historic use of precious metals like gold to provide commodity backing for government-issued paper money. In the simplest version of the gold standard, every paper banknote is effectively a ticket, which can be redeemed for gold bullion of a specified weight. In KRNC, "weight" is cryptographic, but it serves the same purpose as gold bullion by protecting the value of the public's money.

The inherent flaw in the gold standard is that, if a central bank is entrusted with maintaining adequate gold reserves, then it can simply renege on its promise. In that event, the owners of gold-backed paper money can suddenly find themselves owners of unbacked paper money. KRNC corrects this problem by issuing cryptographic backing directly to the owners of government-issued money, rather than having physical backing stored by a central bank. In each transaction, both the original fiat money and its cryptographic backing are transferred from the purchaser to the seller. There is accordingly no "single point of failure" that can eliminate inflation protection.

Sellers can confidently ask buyers to pay in backed money rather than unbacked money, because obtaining backing is free for owners of fiat money. In KRNC, using a monetary "ticket" to claim cryptographic weight does not mean losing the original government-issued money. Instead, a record of which tickets have already been redeemed is maintained, which ensures that each ticket can be used once and only once to unlock its corresponding cryptographic backing. Because protocol participants are not forced to part with their money in order to gain access to its backing, the cost of participation is zero. There is nothing to buy, and no need to waste computing power. Users simply verify their unbacked fiat holdings, and they receive the weight needed to back that quantity of money.

*A Meta-Dollar*

For simplicity, let us temporarily imagine that the U.S. Dollar is the world's only currency, and that all agent who own dollars choose to join the KRNC protocol and unlock their free cryptographic weight. The result is similar to a "hard fork": for every dollar that previously existed, there is now one dollar plus one unit of cryptographic weight, both of which are the property of the dollar's owner.[11] However, when a hard fork is performed on a cryptocurrency, the two resulting assets are generally in competition with one another. In KRNC, they are complimentary: the unbacked fiat dollar can be used to pay taxes, while the cryptographic weight offers inflation protection and smart-contract functionality.

Each of these two assets exists on a different layer-one ledger, so they can be transferred independently. However, when used in combination, one dollar and one unit of cryptographic weight can be treated as a single unit of account. We refer to this meta-asset as a cryptographically backed dollar. It is a modern twist on the "symmetallic standard" proposed by Alfred Marshall, one of the founders of neoclassical economics, in the late 1800s. [66] Under Marshall's proposal, the base money in society would have been a meta-commodity composed of gold and silver in a specified ratio. The virtue of this concept was the protection that it provided against supply or demand shocks in either of the meta-commodity's constituent elements. In the KRNC protocol, cryptographically backed dollars deliver the same diversification of risk. The only difference is that the commodities in question are digital, rather than physical.

Under the symmetallic standard, if an unexpected technological innovation destroyed the value of either gold or silver (a fate that befell aluminum, the 19[th] century's most precious metal) the price of the meta-commodity would have remained relatively stable; demand for the other metal would have proportionately increased, given that gold and silver were the only two credible base-money candidates. This was a precursor to "handcuffing" in modern fantasy sports, a strategy in which a fan drafts both a starter (generally, an American football running back) and that starter's backup. The backup player acts as an insurance policy, of sorts; if the starter is injured during the season, then the backup will receive the bulk of the point-scoring opportunities that previously went to the starter, ensuring stability in the weekly point-scoring value of the fantasy-sports team. The same principle holds true in KRNC: cryptographic weight is effectively the "backup" to unbacked dollars, so that in the event of a hyperinflationary crisis the value originally assigned to government-issued dollars will naturally shift to their scarce cryptographic backing. This change in the relative value of the components of the meta-resource will not affect the value of crypto-backed dollars, just as shifts in the relative value of gold and silver would not affect the value of base money under Marshall's symmetallic standard.

---

[11] Technically, cryptographic weight in KRNC is a form of "forked fiat currency." [14] The term "weight" has been employed to make Proof-of-Balance technology more intuitive.

The diversification of control over the monetary supply that this architecture enables may be compared to the advantage of a bicameral legislature. From an engineering standpoint, splitting responsibility for maintaining a working form of base-level money is a prudent redundancy. If political instability or economic miscalculation causes traditional sovereign money to lose its effectiveness as a store of value, the danger of a global economic crisis can be reduced by the ready availability of a parallel monetary system that is not exposed to the same risks. The converse is also true: should anything go wrong with the cryptography used to maintain the protocol's blockchain, the traditional fiat monetary system would remain intact and ready to take over.

*An International Standard*

The entire world does not, of course, use the U.S. Dollar. If we wish to maximize the security and performance of the KRNC protocol, it will also be necessary to assign weight to the owners of other currencies, and to do so in a manner that makes those weight assignments compatible with one another.

This is where the analogy to the gold standard once again becomes useful. In the contemporary monetary system, the prices of different currencies float against one another; under a gold standard, each national currency is backed by a specified quantity of gold, which creates fixed exchange rates. If the gold standard were still in place, we could then accomplish our goal merely by assigning cryptographic weight in proportion to the weight of the gold bullion specified as the backing for each national currency. Can we apply the same approach in the absence of a gold standard? Yes. First, based on the value of the U.S. Dollar, we define a weight necessary to achieve full backing of U.S. Dollars. Second, we take a snapshot of every foreign currency's exchange rate against the U.S. Dollar, and use those ratios to determine the weight necessary to achieve full backing of each of those currencies. An abstract reconstruction of the gold standard is thereby created, which can be used to assign the appropriate quantity of cryptographic weight to every unit of electronic fiat currency on Earth.

The credibility of cryptographic weight as an alternative to sovereign currency may be a powerful mechanism to dissuade central banks from taking excessive inflationary risks — a role that has been suggested for Bitcoin, but which it does not seriously fill because of fiat money's incumbency advantage. [67] If excessive quantities of sovereign currency are introduced, then the supply of sovereign money will exceed the supply of cryptographic weight available to provide backing to the sovereign money. In the event of a global hyperinflationary crisis, rapid growth in the supply of sovereign money would make it trivially inexpensive to obtain the national currencies that, when matched with their corresponding cryptographic weight, would represent one unit of cryptographically backed fiat money. As the limiting factor in the set of two inputs needed to produce the standard unit of account, cryptographic weight would organically evolve into the world's de facto monetary standard.

## 9.2 Toy Protocol

Now that the reader has an intuitive understanding of the protocol, it is time to begin formalizing its architecture.

*Definitions*

We model the world monetary system as a pair of abstract arrays — one for settlement, the other for custody and clearance. Central banks exercise their authority to create new units of base money on the settlement array, which the public cannot access directly. The accounts that members of $N$ maintain at financial institutions display the data recorded on the custodial array, where commercial banks introduce new digital money into circulation when they make loans. There is more fiat money in the custodial array than the settlement array due to the "money multiplier," which determines the fraction of the balances shown on the custodial array that must actually be backed by reserves on the settlement array.

Because the power to issue new currency is exercised by modifying the balances on the settlement array, the base version of a fiat currency is necessarily recorded and updated on the shard of the settlement array controlled by the central bank with authority to issue that currency. Similarly, every cryptocurrency is recorded and updated on the shard of the settlement array administered by its consensus protocol. A decentralized Proof-of-Balance protocol also administers a shard on the settlement array, except the balances on that shard – which also serve as weights in the consensus algorithm that controls the shard – are "forked" by importing data from past states of the custodial array. This enables the users of the custodial array to collectively trust the consensus algorithm, since its security guarantees are verifiably derived from their own aggregate trustworthiness.

Formally, we index the custodial array by the 7-tuple (*o, i, a, m, p, b, s*), where *o* is a custodial institution such as a commercial bank or credit union, *a* is an account number, *i* is the identity of the accountholders, *m* is a specified currency, *p* is the price of that currency in the numeraire, *b* is a balance, and *s* is a slot. We denote the column of values to which an element belongs using the capitalized version of the indexing character; thus, *o* represents a financial institution while *O* denotes the collection of all financial institutions, *i* represents an identity while *I* denotes the collection of all identities, and so on for all elements of the 7-tuple. Values are associated with accounts only in slots after their creation and before their termination, but the array is otherwise dense — i.e., all elements have defined values. Thus, at all times that an account exists at a financial institution, it is associated with one or more accountholders, and it contains a specified quantity of currency whose objective market value is a function of the exchange-rate expressed by the numeraire. We assume that the participants in the protocol have access to accurate records of past exchange rates and that these records are consistent with one another. In a real-world execution, the records do not actually have to be consistent or even perfectly accurate; they simply have to be accurate enough that agents are willing to tolerate the protocol's canonical version.

In our model, the custodial array is divided into $|O|$ shards. Each financial institution $o$ administers the shard that contains the elements of the array associated with $o$. This corresponds to the fact that commercial banks are responsible for individually managing the accounts opened by their customers.

*Simplifying Assumptions*

We now introduce three unrealistic axioms that make a fork of the fiat custodial array possible in our idealized model. Later, in the non-idealize model, the axioms must be replicated or approximated by extending the protocol's capabilities.

First, in the idealized model, we assume that all transactions are performed through real-time gross settlement, such that a global state-transition function instantly updates the ledger in every slot. If in slot $s$ an accountholder $i$ transfers a quantity $q$ of currency $m$ from account $a$ to account $a'$, then in slot $s+1$ the balance $b$ associated with $a$ will be $q$ less than in slot $s$, while the balance $b'$ associated with $a'$ will be $q$ greater. However, the transaction will be executed if and only if it is valid: i.e., the summed $q$ of all transfers authorized from $a$ for slot $s+1$ must be less than or equal to the balance $b$ associated with $a$ in slot $s$. The combination of synchrony and validity ensures that our idealized custodial array maintains a perfect global state $GS$ in all slots.

Second, we assume that the global state $GS$ is recorded publicly, so that it can be verified directly by all members of $N$. Any agent can download the latest set of balances for all accounts at all custodial institutions and use them to populate balances in a consensus protocol.

Third, we assume that every identity $i$ is already associated with a public key $pk_i$, corresponding to the secret key $sk_i$ controlled by the agent behind the identity. This makes it possible for a balance in the new protocol to be placed under the control of its rightful owner, $n : n=N[i]$, merely by associating the balance with the key $pk_i$.

These three assumptions make it trivial to assign weights in a new consensus protocol based on balances in the incumbent monetary resource. Let $s_{fork}$ be the slot in which balances are imported from the incumbent monetary ledger. The weight ω[$pk_i$] assigned to each public key in the new protocol is given by the formula

$$\omega[pk_i] = \sum_{a=1}^{A_i} (b[a, s_{fork}])(p[m, s_{fork}])$$

where $A_i$ is the subset of all accounts within $S$ associated with $i$, $b[a, s_{fork}]$ is the balance recorded for account $a$ at the slot of the fork, and $p[m, s_{fork}]$ is the price of the account's currency, $m$, at the slot of the fork.

Every unit of voting weight in the new protocol is transferrable, because it is also the native asset of the new monetary ledger. For convenience, we assume that the numeraire is the U.S. Dollar, and we therefore refer to units of voting weight in the new protocol as "forked dollars." A cryptographically backed dollar consists of both a government-issued dollar and a forked dollar. By definition, the existing monetary system is already capable of executing arbitrary transfers of government-issued dollars. In order to demonstrate the viability of transferring cryptographically backed dollars, it will therefore suffice to define a protocol that is capable of transferring the corresponding forked dollars. That is the approach that we take here, because it greatly reduces the complexity of the protocol that we must define.

The constituent elements of crypto-backed dollars always remain separate on the settlement array, but for clearance purposes they can either be transferred in parallel or merged into a token that represents a claim on both a government-issued dollar and a forked dollar. To become the owner of one forked dollar in the genesis block of the new protocol, at the time of the fork an agent must own a quantity of fiat currency equal in value to one U.S. Dollar. This can be accomplished either by directly holding a U.S. Dollar or by holding a quantity of other fiat currency equal in value to one U.S. Dollar, as measured by the exchange rates in effect at the time of the fork. This enables all fiat currencies to receive cryptographic backing.

For this simple implementation, the initial distribution of forked dollars is specified by a list of public keys and corresponding balances included in the new protocol's genesis block. Proof-of-Balance acts as a verifiably secure mechanism for assigning those balances, but in our idealized model its role ends there. Once the genesis block has been finalized, the new ledger can be maintained using an off-the-shelf consensus algorithm, such as Tendermint or Ouroboros. [68] [12] The initial set of balances can also be used to initiate a meta-protocol, like the Tezos shell, so that the ledger can become transparently self-governing. [69] We anticipate that weight owners will vote to determine the initial consensus algorithm, so we leave that decision to their discretion.

*Protocol Characteristics*

If a ledger could be constructed with the properties of the new protocol in our idealized model, it would provide unprecedented capabilities. Given that the total value of fiat currency is estimated to be ~80 Trillion USD, if we assume that $k$=0.5, the wealth required to unlock a super-$k$ fraction of all voting weight in the idealized model is ~40 Trillion USD. It has often been estimated that the cost of a successful attack on Bitcoin is \$1 Billion, which is 80,000-times less than the attack cost we have just derived. To our knowledge, the highest estimated cost is \$4 Billion, and even this liberal figure is 20,000-times less than the attack cost we have just calculated for Proof-of-Balance. If we make the conservative assumption that the true cost is \$2 Billion, then the cost of an attack on the Proof-of-Balance system in our idealized model is 40,000-times greater than the cost to compromise Bitcoin's Proof-of-Work system. If we treat $w[\mathcal{A}]$ as a constant, the maximum fraction of adversarial voting power is $1/40,000^{th}$ of Bitcoin. Because the adversary's maximum fraction of voting power is so severely limited, Proof-of-Balance can support lower confirmation times and more aggressive sharding than any proposed Proof-of-Work or Proof-of-Stake system.

A potential objection to this figure is that it is derived from an estimate of the total value of fiat currency that includes coins and banknotes, which are not actually recorded in the electronic fiat ledger. If we adopt an estimated value of ~5 Trillion USD for all physical monetary media and exclude them from our calculation, then the value of the electronic balances recorded on the fiat ledger is ~75 Trillion USD. Given our assumption that $k$=0.5, the adversary would require control of more than ~37.5 Trillion USD to execute a successful attack. [12]

A second potential objection is that these figures represents the price of the original fiat currencies, not the price of their cryptographic weight, i.e., if the weight assignment were performed at slot $s$, then the adversary $\mathcal{A}$ would be guaranteed not to initially control a super-$k$ fraction of voting weight, but it would have the opportunity to purchase the additional weight required at slot $s+1$, $s+2$, etc. The security of the protocol therefore depends, not merely on the price to the adversary of acquiring a super-$k$ fraction at slot $s$, but also in later slots. This is tautologically true for any protocol facing a price-adaptive adversary: the initial cost of acquiring a staking resource sets the ceiling on verifiable security, but that ceiling is lowered if the price of the staking resource drops. In effect, we have defined what is the equivalent of a Proof-of-Stake protocol bootstrapped from a Proof-of-Work distribution in which $75 Trillion was expended by miners. Only the market can decide how much of that security will be preserved by the future exchange value of the asset that was mined.

---

[12] This calculation is conservative in that it treats the ratio of electronic to physical fiat currency as a structural parameter that cannot be altered by the introduction of the protocol. It is possible that the protocol will induce agents to collectively store more of their fiat money at commercial banks.

The question is how to actually build the ledger required to make forked-fiat money a reality, once we remove the convenient assumptions of our idealized model.

## 9.2 KRNC: Private Non-ACID Ledger

We now define a real-world model of the fiat ledger. We will replace the unrealistic assumptions from our idealized model with accurate descriptions of the obstacles that must be overcome to fork the commercial banking system.

*Synthetic Consistency*

First, in the real-world model we can no longer assume strict consistency or validity. Settlement is not performed in real time for most transactions, overdrafts sometimes permit unearned funds to be spent, and a balance may be recorded incorrectly in a given slot due to myriad other factors, from human error to electrical faults. A balance $b[a,s]$ recorded in the real world may therefore differ from the version recorded in the idealized model.

We adopt the notation $c[a,s]$ to denote the latter value, which is not affected by the faults found in the real-world model. Based on this distinction, the set of slots is partitioned into $S_C$ correct slots and $S_F$ faulty slots. If the balance of the account is accurately recorded at slot $s$, such that $b[a,s] = c[a,s]$, let $s$ be a correct slot, $s_c$, where $s_C \in S_C$. If not — i.e., if $b([a,s] \neq c[a,s]$ — let $s$ be a faulty slot, $s_F$, where $s_F \in S_F$.

We refer to a sequence of one or more slots with identical balances as a period, $l$, and the set $L$ of all periods is indexed $l_1, l_2, \ldots, l_L$. $L$ is partitioned into $L_C$ correct periods and $L_F$ faulty periods, such that every correct slot belongs to a correct period ($s_C \in l_C \in L_C$) and every faulty slot belongs to a faulty period ($s_F \in l_F \in L_F$).

In Proof-of-Balance, rather than assigning a weight $\omega$ to an account based on its value at a single slot $s$, the value is measured across periods and averaged.[13] We refer to the sequence of periods measured by the protocol as $l_P$, where $P \geq 1 \leq L$. Our task is to determine how quickly Proof-of-Balances causes the effect of faults to converge to zero.

For indexing purposes, we assume that a fault arises at the initial slot ($s_1 = s_F \in S_F$), so that the initial period is faulty ($l_1 = l_F \in L_F$). We adopt the scale-setting assumptions that: (1) the initial faulty period contains the same number of slots as subsequent correct periods, such that $|l_1|=|l_C|$ and $|l_1|P=L_P$; and (2) the absolute difference between the correct value $c[a,s]$ and the faulty value $\neg c[a,s]$ equals the numeraire, such that $|(b[a, s_C]) - (b[a, s_F])| = 1$. The mean balance of slots within all protocol periods is therefore

$$\bar{b}L_P = \frac{1}{|L_P|}\sum_{s=1}^{L_P} b[a,s]$$

---

[13] When multi-currency staking is being performed, for maximum accuracy the balance $b[a,s]$ from each slot should be multiplied by $p[m,s]$ before being averaged.

This value is derived from the mean balance within the initial faulty period, and the mean balance within the ensuing correct periods:

$$\left\{\left(\frac{1}{|l_1|}\sum_{s=1}^{l_1} |c[a,s] - 1|\right) + \left(\frac{1}{|L_P| - |l_1|}\sum_{s=1}^{l_{2-P}} c[a,s]\right)\right\}\frac{1}{2}$$

Through a series of straightforward reductions, we obtain

$$\frac{\bar{b}l_1 + \bar{b}l_{2-P}}{P} = \frac{|c[a,s] - 1| + c[a,s] + \cdots + c[a,s]}{P}$$

$$= \frac{|c[a,s] - 1| + c[a,s](P-1)}{P}$$

$$= \frac{|c[a,s]P - 1|}{P} = \left|c[a,s] - \frac{1}{P}\right|$$

i.e., as the number of periods captured by the protocol is extended, Proof-of-Balance exponentially converges on the correct value.

The use of staking periods not only allows Proof-of-Balance to asymptotically replicate the security guarantees from the idealized ledger, it actually provides stronger security guarantees than are available in the idealized model. The key is that the honest majority of capital axiom does not assign a privileged position to a specific slot: i.e., our knowledge that $\frac{w[\mathcal{A}]}{w[N]} < k$ is equally reliable in $s$ or $s'$, so taking the mean value of the slots provides at least as much assurance as measuring any single slot.[14] The

---

[14] Technically, a slot $s$ can enjoy a privileged position indirectly if it corresponds to a pre-intervention state of the world in which $\frac{w[\mathcal{A}]}{w[N]} < k$ is axiomatic, while $s'$ corresponds to a post-intervention state in which the axiom's reliability is diminished. This is not a problem in KRNC, but it is

reason it provides greater assurance is that, as our earlier signal-theoretic model demonstrated, an agent may attempt to exaggerate its holdings of the staking resource in order to artificially increase its voting power, e.g., by borrowing money to boost its bank balances when they are measured. Sampling the value of those balances over a longer period of time increases the duration of the necessary loan, thereby raising the costs paid by agents who attempt to exaggerate their wealth.

*Balance Verification*

Second, in the real-world model, we can no longer assume that every agent $N$ has read access to the entire fiat ledger. Indeed, no single agent or institution has access to all the values that comprise $S$. Instead, an institution $o$ has access to all elements that are associated with $o$, because those elements are properties of the accounts that it maintains on its internal database, which functions like a shard of the full abstract array. An agent $n$ has access to all of the accounts it controls across financial institutions, even if it assumes different identities to do so. Formally, let $o[a]$ designated the institution that administers an account, such that $o[a]$ has both read access and write access to the value of elements associated with $a$, and let $i[a]$ designate the identity associated with the account, such that $n$ has read access to the value of the elements associated with $a$ if and only if $i[a]=i$ : $N[i]=n$.

Third, although an institution $o$ can effectively associate itself with a public key $pk_o$ using official channels (e.g., its website or verified social-media accounts), we can no longer assume that every identity $i$ is already associated with a public key $pk_i$. This reflects the fact that accountholders at commercial banks do not, in fact, rely on asymmetric cryptography to access their funds. Instead, banks rely on many different authentication factors, including but not limited to control of a registered email address or phone number, knowledge of a password or alphanumeric code, possession of a hardware authentication device, and/or in-person presentation of physical credentials, such as a driver's license or passport.

To simplify our formal notation, we define any combination of authentication factors sufficient to control account $a$ within the ledger shard administered by institution $o$ as an *account key,* denoted by $ok_a$. An account key is not an actual cryptographic key; it is simply a useful abstraction for representing the credentials required to access the specified account. However, an account key can be employed by an agent $n$ to authorize the execution of a transaction by $o$, much like a private key can be used to sign

---

relevant to cryptocurrencies, e.g., global adoption of Bitcoin could theoretically deliver a sufficient ROI for the protocol's early adopters to acquire the majority of all wealth, undermining the reliability of the "honest majority of capital" axiom needed for permissionless consensus.

a message that authorizes a transaction on a blockchain. We capture this similarity by using the notation $SIG_{pk_i}$ for asymmetric cryptographic signatures and $SIG_{ok_a}$ for conventional transaction authorizations. We use $ENC_{pk_i}$ to denote encrypting a message with public key $pk_i$, such that it can be decrypted with $sk_i$. An account key cannot be used to encrypt a message, since it is not actually a member of a cryptographic key pair.

The challenge in the real-world model is how to specify a protocol that will enable every agent $n$ to trust the accuracy of both its own weight assignment and also the weight assignments to other agents, whose balances it cannot directly observe. Our solution relies on two related axioms. First, we adopt the axiom that an agent $n$ who has deposited fiat money at an institution $o$ necessarily also trusts that institution to act as a custodian of $n$'s forked-fiat money. Formally, $o \in T[n]$ where $o[a]=o$ and $i[a]=i : N[i]=n$. However, we restrict the scope of the axiom by specifying that it holds if and only if $n$ retains the ability to withdraw its money from $o$ on demand. This reflects the empirical relationship between demand deposits and bank runs, in which trust in a custodian vanishes as soon as it freezes withdrawals.

Second, we assume a weak version of transitivity, in the sense that an agent $n$ who trusts an institution $o$ necessarily trusts balances whose validity $o$ accepts based on attestation from $o'$, where $o' \in O : o \neq o'$. This reflects the fact that the customers of one bank are inherently placing trust in the institutions their bank partners with to execute, e.g., SWIFT transfers and other permissioned transactions.

*Synchronous Weight Assignment*

In this sub-section, we will define a simple implementation of the protocol that requires all members of $O$ to cooperate at its inception. This is not technologically impossible, but it may be bureaucratically impractical.

A cryptographic weight $\omega[o]$ is hardcoded into the protocol's genesis block for every institution $o \in O$. An institution $o$ calculates its own weight, $\omega[o]$, according to the formula

$$\omega[o] = \sum_{a=1}^{A_o} \omega[a]$$

,

where

$$A_o = \{a \in A : o[a] = o\}$$

,

and

$$\omega[a] = \frac{1}{|L_P \cup|} \sum_{l=1}^{L_P} (b[a,s])(p[m,s])$$

The custodial weight represents the mean quantity of wealth controlled by the institution during the protocol's staking periods. It is calculated by normalizing each past fiat balance according to its contemporaneous exchange value, which yields the mean quantity of wealth held in each account during the staking periods, then summing that quantity for every account within an institution's shard of the fiat ledger. The denominator in the fraction to the left of the sigma expression is the cardinality of the transitive closure of the protocol periods, i.e., the number of slots within those periods.

Each institution $o$ generates an asymmetric key pair and discloses its public key, $pk_o$. It then employs the corresponding secret key, $sk_o$, to generate and broadcast a message of the form $SIG_{pk_o}(o, CLM, \omega[o])$, where $CLM$ is the opcode for claiming cryptographic weight. The genesis block contains a valid weight-assignment signature from every member of $O$. The weight-assignment signatures define the quantity of forked currency that each custodial institution $o$ receives on behalf of its customers. Once the protocol commences, a custodial institution's voting power changes whenever the quantity of forked currency it controls increases or decreases. Customers may transfer their forked-fiat balances to other custodial institutions or to their own private keys for direct access to the settlement layer. Voting power is thus "pre-delegated" to each user's existing financial institution(s) of choice, but there is no limit to how decentralized the protocol can become as users become comfortable managing their own keys.

We refer to this system as *liquid custody*, because it complements and extends existing stake-delegation mechanisms, which are based on proposals for "liquid democracy." In that framework, which is also known as delegative democracy, individuals can choose to participate in the consensus process either actively by casting votes directly or passively by delegating their vote-casting authority. KRNC realizes that ideal more faithfully than today's Proof-of-Stake systems, because it makes it possible for agents to exercise their voting power without ever interacting with cryptographic keys, and because all agents who own fiat money receive voting power by default instead of authority being concentrated in the small pool of agents who opt to purchase stakes in a new cryptocurrency.

To place the forked-fiat balances under the initial control of their customers, the members of $O$ modify their respective shards of the custodial array by creating new accounts containing forked fiat currency. For an original fiat account $a$ that receives a weight assignment of $\omega[a]$, an institution $o$ creates a forked-fiat account $\hat{a}$ whose initial balance $b[\hat{a}, s]$ is equal to $\omega[a]$ and whose initial account key $ok_{\hat{a}}$ is identical to the account key $ok_a$ associated with $a$. The result is that every agent $n$ is placed in control of the appropriate weight $\omega[n]$ by virtue of the account keys that it possesses. Specifically, the initial weight $\omega[n]$ controlled by $n$ is given by the formula

$$\omega[n] = \sum_{a=1}^{A_n} \omega[a]$$

where

$$A_n = \{a \in A : i[a] = i : N[i] = n\},$$

i.e., $n$ controls a weight equal to the summed weight of all accounts whose associated identities are controlled by $n$.

Agent $n$ has the ability to withdraw its forked-fiat funds from $\hat{a}$ at its convenience. It simply generates an asymmetric key pair, such that it controls the secret key $sk_n$, then privately transmits to institution $o$ a message of the form $SIG_{ok_{\hat{a}}}(i', o, \hat{a}, pk_n, q, m)$, where $i'$ is the identity associated with $\hat{a}$ and $pk_n$ is the public key that corresponds to $sk_n$. If the requested transfer is valid, then institution $o$ publicly broadcasts a message of the form $SIG_{pk_o}(pk_o, q, m, pk_n)$, causing the forked-fiat funds to be transferred to $n$'s control on the settlement array, based on $n$'s knowledge of $sk_n$.

An agent $n$ can employ a similar procedure to transfer forked-fiat funds to a third-party. The third-party, $n'$, generates an asymmetric key pair, such that it controls the secret key $sk_{n'}$, then it discloses the corresponding public key, $pk_{n'}$ to agent $n$. Agent $n$ then privately transmits a message to institution $o$ of the form $SIG_{ok_{a'}}(i', o, \hat{a}, pk_{n'}, q, m)$. If the requested transfer is valid, then institution $o$ publicly broadcasts a message of the form $SIG_{pk_o}(pk_o, q, m, pk_{n'})$, causing the forked-fiat funds to be transferred on the settlement array to the control of the third party, $n'$, based on its knowledge of $sk_{n'}$.

As we have just demonstrated, although the forked-dollar ledger is initially controlled by custodial institutions, normal users have the ability to take direct responsibility for managing their own asymmetric keys. However, one advantage of the protocol is that users are never forced to do so against their will, because their forked currency is automatically available on the custodial platforms where they chose to store their original fiat currency. This is a matter of not only convenience, but also security: forked balances represent weights in the consensus algorithm, and forcing billions of bank customers to suddenly start managing cryptographic keys for the first time in their lives would represent a critical vulnerability that the adversary could exploit.

To authorize a transfer of forked-fiat money from account $\hat{a}$ without using an asymmetric key pair, an agent $n$ transmits a message to institution $o : o = o[\hat{a}]$ of the form $SIG_{ok_{\hat{a}}}(i', \hat{a}, q, m, o, i'', \hat{a}')$, where $\hat{a}'$ is the forked-fiat account receiving the transfer, $o$ is the institution $o[\hat{a}]$ administering the account receiving the transfer, and $i'$ is the identity associated with the receiving account. For simplicity, we assume that the quantity $q$ is low enough for the transfer to be valid. If $o[\hat{a}']=o$, then the transfer is executed internally by $o$ without affecting the settlement layer of the decentralized ledger, simply by debiting $q$ from $\hat{a}$ and crediting it to $\hat{a}'$. If $o[\hat{a}']=o'$, such that $o[\hat{a}']\neq o$, then $o$ debits $q$ from $\hat{a}$, privately transmits a message to $o'$ of the form $ENC_{pk_{o'}}(SIG_{pk_o}(o,\hat{a},q,m,o',\hat{a}',i,z))$, and publicly broadcasts a message of the form $SIG_{pk_o}(pk_o,q,m,pk_{o'},z)$, where $z$ represents an arbitrary transaction identifier for conveniently associating the public and private messages with one another. The public message causes the transfer of forked dollars to be executed on the decentralized settlement layer, so that the funds involved in the transaction leave the control of $o$ and become controlled by $o'$; the private message allows the institution receiving the transfer, $o'$, to associate the incoming funds with the accounts and identities involved in the transaction, which span its shard of the custodial array and the shard administered by $o$. It modifies its shard of the custodial array by crediting $q$ units of $m$ to account $\hat{a}'$ of $i''$, and it associates that operation with the account and identity of the transaction initiator, $\hat{a}$ and $i'$. This enables the ultimate recipient of the funds, the agent $n : n=N[i'']$, to verify that it has been paid without the need to interact with asymmetric keys or addresses; it simply uses $ok_{\hat{a}'}$ to gain read access to the relevant portion of the custodial shard administered by $o'$.

There are, of course, many improvements in efficiency and functionality that can be obtained by modifying the fields just described or specifying procedures for batching cleared payments into larger, less-frequent transfers on the settlement array. This paper intentionally omits those and other obvious refinements in favor of highlighting the novel aspects of the KRNC protocol. In the next sub-section, we will remove the need for an initial, synchronized distribution of voting weight.

*Retroactive Weight Assignment*

In the prior specification of the protocol, all members of $O$ must simultaneously participate in the creation of the protocol by assigning themselves synchronized weights, because the initial set of weight distributions is directly encoded in the genesis block. The need for synchronization is a byproduct of the genesis block's immutability: if a custodial institution is excluded from the initial distribution of weights but later chooses to join the protocol there is no mechanism to expand the supply of forked dollars so that the late-joining institution's customers can receive their fair allotment.

One solution to this problem would be to encode the public keys of all $O$ institutions in the genesis block, but to leave the accompanying weight assignments undefined. Each public key would instead act as a *minting authorization* memorializing the right of the key owner $o$ to claim its custodial weight $\omega[o]$ if and when it joins the protocol. If such a scheme were implemented, then when an institution $o$ chose to join the protocol, it would calculate its weight assignment $\omega[o]$ and then use its secret key $sk_o$ to sign and broadcast a message of the form $SIG_{pk_o}(o, pk_o, \omega[o])$. In other words, upon joining the protocol each institution $o \in O$ would have a one-time opportunity to *retroactively* issue itself the quantity of forked dollars that must be given to its customers in order to provide them with their rightful weights on the new ledger.

The advantage of retroactive weight assignment is that the fork of the fiat ledger can be initiated by a single founding institution, $\dot{o}$, then joined by other institutions as the protocol and its native asset grow in popularity. For now, we assume that the founder is one of the existing institutions within $O$, such that $\dot{o} \in O$.

Without retroactive weight assignment, no single institution $\dot{o}$ would be able to launch a credible forked-fiat ledger, because everyone except that bank's customers would be excluded from the protocol's finite weight distribution. That would prevent its consensus algorithm from providing reliable security guarantees, and it would make it impossible for the ledger's native asset to piggyback on the network effects of fiat money to achieve mass adoption. Those problems are resolved by retroactive weight assignment, which enables the quantity of forked dollars authorized by the genesis block to be revised and expanded. This increases the expected equilibrium price of the protocol's native asset, since the market capitalization of the ledger is a time-discounted function of the protocol's anticipated future user base. [70]

The potential user base of a protocol that can accommodate arbitrarily many late-joining institutions is limited only by the size of population, $|N|$. Critically, users whose institutions join later are not placed in an inferior position to those whose institutions join early, because all users receive their proportionate weight assignments. The structure of the custodial array is employed as a statistical sampling frame to ensure that voting weights are properly assigned, regardless of the order in which the members of the population initially join the protocol.

However, there are two immediate drawbacks to the retroactive-weighting scheme we have just described. First, although it reduces the work that every individual institution $o$ must perform at the outset of the protocol, it does not fully eliminate the need for all members of $O$ to opt in before the genesis slot, since each must generate and broadcast its public key $pk_o$ before the genesis block can be created. Second, although the scheme is secure if all the members of $O$ remain correct indefinitely, the corruption of even a single institution $o$ by the adversary can trigger a catastrophic failure. If the institution is corrupted before it has generated and broadcast its retroactive-weighting message, then the adversary receives an opportunity to issue itself an unlimited number of forked dollars. This is a more dangerous failure mode than a standard 51% attack, in which only the safety guarantee of the consensus algorithm is violated. Here, the adversary also controls a nominally unlimited quantity of money on the ledger, which magnifies the harm it can inflict by reverting its own transactions.

*Precommitment Scheme*

To deliver robust real-world security guarantees, our proposed implementation of the KRNC protocol must not only minimize the probability of institutional Byzantine faults, but also mitigate the severity of the failures that would result from their occurrence. For practical commercial adoption, KRNC will also need to eliminate the requirement that institutions lay claim to their public keys prior to the creation of the genesis block. Fortunately, these goals can all be accomplished using a cryptographic precommitment scheme.

The founding institution $\dot{o}$ embeds the precommitment message $SIG_{pk_{\dot{o}}}(\dot{o}, GEN, \acute{z}, terms)$ in the genesis block, where $GEN$ is the opcode indicating the commencement of the protocol, $\acute{z}$ is a nonce, and *terms* are the terms of ledger administration, which we will define iteratively below.

First, the terms include a pledge from $\dot{o}$ that it will award institutional authority to any member of $O$ that chooses to join the protocol before a designated slot number, $\acute{s}$. Possession of institutional authority replaces control of a public key listed in the genesis block as the prerequisite to validly laying claim to a custodial weight $\omega[o]$. As of the genesis block, $\dot{o}$ is the only member of $O$ with institutional authority, but in subsequent blocks it can award institutional authority by broadcasting a message of the form $SIG_{pk_{\dot{o}}}(\dot{o}, INS, \grave{o}, pk_{\grave{o}}, \acute{z})$, where $INS$ is the opcode indicating an award of institutional authority, $\grave{o}$ is the institution within $O$ receiving institutional authority, and $pk_{\grave{o}}$ is its public key. To reduce centralization of power in the founding institution, $\dot{o}$, the terms also specify that any recipient of institutional authority can award such authority to other institutions. This makes it rapidly infeasible for any coalition within $O$ to withhold authority from all remaining members once that authority begins to disperse.[15]

The ability of institutions to lay claim to their public keys through official channels ensures that all protocol participants can easily verify that $\dot{o}$ is fulfilling its obligations. Any institution $o$ that wishes to be awarded institutional status can broadcast a message of the form $SIG_{pk_o}(o, REQ, \dot{z})$ where $REQ$ is the opcode indicating a request for institutional status. If the signature provided matches the public key that $o$ has claimed through its official channels, and $o$ has not previously been awarded institutional status, then the message containing the $REQ$ code is valid, and all institutions that have received institutional status are mandated to award institutional status to the requester. If they fail to broadcast the mandatory $INS$ message, then they are faulty; the ongoing failure of the requestor to receive institutional status would be evidence that an adversary has either broken the protocol's liveness guarantees or corrupted all the institutions that have received institutional status.

---

[15] In practice, once an initial degree of institutional decentralization has been achieved, it may be prudent to mandate that awards of institutional authority be co-signed by e.g. 3 other authorized institutions.

Second, the terms include a global shut-off date for all minting authority derived from the genesis block, specified by the shut-off date's slot number, $\bar{s}$. This ensures that the risk of an adversary abusing the retroactive-weighting system introduced in the genesis block verifiably drops to zero if the risk has not manifested itself after $\bar{s}$ slots. The protocol's potential user base $N_U$ is thereby expanded: prior to $\bar{s}$, $N_U$ includes all agents who are willing to tolerate the risk of the attack until no later than $\bar{s}$, then once $\bar{s}$ has passed $N_U$ grows to encompass all agents whose aversion to the risk of attack foreclosed participation pre-$\bar{s}$.

Third, the terms minimize the consequences of institutional Byzantine faults by specifying cryptographic limits on weight assignment. For example, the genesis block may specify $\omega_{MAX}[o]$ to cap the maximum quantity of cryptographic weight that a specified institution $o$ can claim or $\omega_{MAX}[m]$ to cap the maximum quantity of weight that all institutions can collectively claim based on deposits of the specified fiat currency. These are cryptographic restrictions that must be complied with by future weight assignments in order to maintain their validity.

The scope of the limitations depends on what information is axiomatically available about the maximum weights that could validly be associated with a given institution or national currency within the protocol's staking periods. The more specific the common knowledge of the maximum quantity of forked dollars that could validly be issued, the more efficiently the potential harm from institutional Byzantine faults can be mitigated. Where axiomatic knowledge is available about the quantities of specific fiat currencies held by specific institutions during the protocol' staking periods, restrictions may be inserted of the type $\omega_{MAX}[m \cap o]$. The consequences of the adversary corrupting an institution can thereby be mitigated, because the maximum quantity of cryptographic money and voting weight it can obtain from each institution $o$ is pre-restricted.

*Balance Verification* (*Pre-Bank Participation*)

A downside of the implementation we have just specified is that it forces members of $N$ to rely on their pre-selected institutions of choice in order to obtain their voting weight, $\omega[n]$. This is fine for agents whose institutions choose to participate in the protocol from its inception. But an agent whose bank or credit union delays in joining the protocol may be forced to wait an unreasonably long time to access the forked-fiat money that he or she rightfully owns. In this sub-section, we introduce a solution to that problem.

The solution relies on an additional institution, the *remote verifier*, which takes responsibility for issuing weights to agents whose original custodial institutions have not yet joined the protocol. For simplicity, we will assume that the found institution is the remote verifier. If the founding institution is $\dot{o}$, one of the financial institutions that the public already relies on as a custodian of fiat money, then agents can trust the remote verifier based on the weak transitivity assumption we have already introduced. In this sub-section, we will consider the more difficult case in which the remote verifier is $\ddot{o}$ – a brand new institution. Our goal is to achieve *retroactively untrusted setup*: once all institutions join the KRNC protocol, its security guarantees should no longer depend on whether the remote verifier was correct.

The first step is for the remote verifier to create its shard of the custodial array. Initially, the shard is empty, because the remote verifier does not have custody over any traditional fiat money. Its custodial shard will only be used to store and move forked-fiat money.

An agent $\dot{n}$ who wishes to unlock forked-fiat money controlled by an institution that has not yet joined the protocol must register an account with one of its identities on the remote verifier's custodial shard. During the registration process, $\dot{n}$ demonstrates control of authentication factors (e.g., email address + phone number) required for $\ddot{o}$ to define the account key, $\ddot{o}k_{\hat{a}}$: $\hat{a} = i[\hat{a}] : N[i] = \dot{n}$.

We assume that $\dot{n}$ wishes to unlock a forked-fiat balance earned from storing money at bank $\ddot{o}$, which has not yet joined the KRNC protocol. Agent $\dot{n}$ uses a private channel to send $\ddot{o}$ a message of the form $SIG_{\ddot{o}k_{\hat{a}}}(\ddot{o}, REV, \ddot{o}, \ddot{a})$, where $REV$ is the opcode for requesting remote verification, and $\ddot{a}$ is the account at $\ddot{o}$ where the funds were stored during the protocol's staking periods, such that $o[\ddot{a}] = \ddot{o}$.

When $\ddot{o}$ receives a remote-verification request for $\ddot{a}$, it first checks whether the institution $\ddot{o}$ has joined the protocol — i.e., whether $\ddot{o}$ has requested institutional status. If so, then the remote-verification request is invalid, and $\ddot{o}$ will not attempt to execute it. If not, then $\ddot{o}$ checks whether this is the first remote-verification request for an account at $\ddot{o}$. If so, then $\ddot{o}$ generates a new key pair that $\ddot{o}$ will control and exercise on behalf of $\ddot{o}$ in advance of $\ddot{o}$ joining the protocol. We refer to this as the *provisional* key pair for $\ddot{o}$, where the provisional secret key is $sk_{\underline{\ddot{o}}}$ and the provisional public key is $pk_{\ddot{o}}$. If $\ddot{o}$ has already created provisional keys for $\ddot{o}$ then it simply selects retrieves the existing public key, $pk_{\ddot{o}}$.

Next, to determine the correct value of $\omega[\ddot{a}]$, $\ddot{o}$ must query the values associated with the account during the protocols' staking periods, which are recorded on the shard administered by $\ddot{o}$ within the custodial array. There are three methods that can be employed to access the data:

- The optimal method is delegated access through a modern standard like OAuth 2.0, which enables $\dot{n}$ to employ its existing credential, $\ddot{o}k_{\ddot{a}}$, to authorize that $\ddot{o}$ issue a cryptographic token to $\ddot{o}$ permitting $\ddot{o}$ to download $\sum_{l=1}^{L_P} b[\ddot{a}, s, m]$, i.e., the quantity of a specific fiat currency held in account $\ddot{a}$ during all staking periods. This functionality is already mandatory for institutions in the European Union due to PSD2 open-banking directive, which requires banks to let customers share their personal financial data with third-party services.

- The second-best method is intermediated access, where $\dot{n}$ authorizes a trusted intermediary $\tilde{o} : \tilde{o} \neq \ddot{o}$ to employ $\ddot{o}k_{\ddot{a}}$ to download $\sum_{l=1}^{L_P} b[\ddot{a}, s, m]$ from $\ddot{o}$ and to immediately forward that data to $\ddot{o}$. This functionality is available today through third-party services like the Plaid API, which platforms like Coinbase and Robinhood have relied on to verify bank balances. Krnc Inc. has also successfully employed the Plaid API to test forking units of fiat currency from major U.S. banks onto a permissionless ledger. The downside compared to native delegated access is the addition of another institution, $\tilde{o}$, and the attendant expansion of necessary trust. Formally, $n \notin N_U$ if $\tilde{o} \notin T[n]$.

- The final method is direct access, where $\dot{n}$ entrusts $\ddot{o}$ with $\ddot{o}k_{\ddot{a}}$ so that $\ddot{o}$ can download $\sum_{l=1}^{L_P} b[\ddot{a}, s, m]$ directly from $\ddot{o}$. Formally, this method is identical to intermediated access, except that $\tilde{o}=\ddot{o}$. It is employed by some accounting software, which requires users to input the credentials necessary to access and download their banking records. In general, we oppose this form of validation. It should only be considered in the special case where a significant number of agents trust the remote-verifier $\ddot{o}$ but not any available intermediary, $\tilde{o}$.

Upon receiving the imported data, $\ddot{o}$ is able to calculate $\omega[\ddot{a}]$ according to the formula we have already defined for $\omega[a]$. All agents and institutions, including $\ddot{o}$, axiomatically have access to the historic exchange-rate data that determines $p[m, s]$. The imported data discloses the quantity of a specified currency held in account $\ddot{a}$ during each slot within the protocol's staking periods. The remote-verifier calculates $\omega[\ddot{a}]$ by multiplying the balance in each slot by the market-value set by the exchange rate, then dividing by the number of slots to obtain the average value of the fiat balance held in the account during the staking periods.

To issue the weight on the forked-fiat shard of the settlement array, $\ddot{o}$ broadcasts a message of the type $SIG_{pk_{\ddot{o}}}(\ddot{o}, IRV, \omega, pk_{\ddot{o}}, ENC_{pk_{\ddot{o}}}(\ddot{o}, \ddot{a}))$, where $IRV$ is the opcode for issuing weight based on remote verification, $\omega$ is the weight equal to $\omega[\ddot{a}]$, $pk_{\ddot{o}}$ indicates that the weight should be issued to the remote verifier's own public key, and $ENC_{pk_{\ddot{o}}}(\ddot{o}, \ddot{a})$ represents the coordinates of the historic fiat balances on the custodial array, encrypted with the provisional public key for $\ddot{o}$ so that they can be decrypted only by an agent who possesses the provisional secret key, $sk_{\underline{\ddot{o}}}$.

To issue the corresponding funds on its shard of the custodial array, $\ddot{o}$ simply credits the money to the forked-fiat account that $\dot{n}$ created when it registered the account key $\ddot{o}k_{\hat{a}}$. If $\dot{n}$ wishes to take control of its funds directly on the settlement array, it can generate a public key $pk_{\dot{n}}$ and initiate the standard withdrawal procedure we have already specified. However, the security guarantees of the KRNC protocol would be damaged if the weight $\omega[\ddot{a}]$ could be re-issued as part of $\omega[\ddot{o}]$ once $\ddot{o}$ finally joins. To complete our specification of the remote-verification procedure, we must define a mechanism that guarantees no double-counting will occur.

That is the purpose of the provisional keys. When $\ddot{o}$ joins the protocol, it broadcasts a request for institutional status that specifies its true public key, $pk_{\ddot{o}}$, and one or more correct members of $O$ approve the request, which they know to be valid since it is signed with the public key that $pk_{\ddot{o}}$ has claimed through its official channels. However, let us assume that $\ddot{o}$'s request was made in a slot after remote verified weight has already been issued. To counter the risk of double-counting that we just described, a different procedure for obtaining $\omega[\ddot{o}]$ must therefore be followed, rather than each institution $o$ having the authority to issue itself the appropriate weight $\omega[o]$ as soon as its request for institutional status is approved.

In the new procedure, for $\omega[\ddot{o}]$ to receive authority to issue itself weight, it must receive a message from $\ddot{o}$ of the form $SIGpk_{\ddot{o}}(\ddot{o}, RFA, \ddot{o}, ENCpk_{\ddot{o}}(sk_{\underline{\ddot{o}}}))$, where the second term is the opcode for requesting attestation of remotely verified weight, and the last term is a message encrypted with the public key of $\ddot{o}$, containing the provisional secret key created by $\ddot{o}$ for use on behalf of $\ddot{o}$ to encrypt the coordinates of remotely verified weight assignments based on past fiat balances from the late-joining institution $\ddot{o}$'s shard of the custodial array. Once $\ddot{o}$ receives $sk_{\underline{\ddot{o}}}$ from ö it gains the ability to identify all of the past weight assignments that have been made on its behalf, simply by decrypting the relevant portion of their signed messages.

To confirm that the weight assignments were made in conformity with the values recorded on its shard of the custodial array, ö uses the decrypted coordinates to query its own copy of the relevant data and based on that data re-calculates the correct value $\omega[a]$ for every account listed in a signed message field decrypted by $sk_{\underline{\ddot{o}}}$. For every such decrypted field, the late-joining institution ö confirms that its independent calculation of the correct weight for the account listed in the decrypted field matches the actual quantity of weight $\omega$ issued by the remote verifier ö in the corresponding non-encrypted field of the same messaged signed by ö.

Upon confirming that its calculation of the appropriate weights matches the assignments made on its behalf by the remote verifier, ö calculates the quantity of weight that its customers must be issued. Let the *eligibility status* of an account be an indicator $ES_a$ associated with each account $a$, which takes the value 0 if the account has been remotely verified and 1 if it has not. A late-joining institution ö calculates the quantity of weight that it must be issued on behalf of its customers according to the following formula:

$$\omega[\ddot{o}] = \sum_{a=1}^{A_{\ddot{o}}} (\omega[a]) \ ES_a$$

where $A_{\ddot{o}}$ is the collection of all accounts associated with ö on the custodial array.

It issues itself the necessary weight by broadcasting a message of the form $SIGpk_{\ddot{o}}(\ddot{o}, ARA, SIGpk_{\underline{\ddot{o}}}(ALL))$, where $ARA$ is the opcode for approving a request for attestation, and the ensuing term is a message signed with the provisional secret key that ö received from ö, and $ALL$ is the opcode for approval of all specified transactions. Note that an internal signature from $pk_{\underline{\ddot{o}}}$ is included by ö because this publicly identifies ö as the specific member of $O$ that received the provisional secret key $sk_{\underline{\ddot{o}}}$ from ö. Requiring the signature is a safeguard against adversarial corruption of the members of $O$, because it prevents an institution $o$ that has not actually been transmitted a provisional secret key by ö from damaging confidence in the protocol by spoofing an attestation for weight assignments unrelated to its shard of the custodial array.

For simplicity, the sample message indicates that ö is attesting to the validity of all the transactions signed on its behalf by ö using the provisional key created by ö for that purpose. If ö instead wished to dispute the validity of certain transactions, it would need to employ an alternate opcode, *DIS*, followed by a list of all the transactions it disputes. In this paper, we assume that institution *o* reveals the specific weight assignments it is endorsing when it transmits a message with the *ALL* opcode. However, the KRNC protocol can also be implemented with zero-knowledge primitives, so that the members of *O* collectively attest to every remotely verified transaction without any individual member *o* associating itself with any specific weight assignment.

In either implementation, once all the constituent members of *O* join the protocol, the weight assignments performed by the remote verifier are entirely superseded by attestations signed by the members of *O* using public keys that they have each claimed through official channels. The protocols potential user base $N_U$ therefore ultimately encompasses the set of agents who trust the institutions comprising *O*, rather than the smaller set of agents who trust the original remote verifier, *ö*. Relying on the protocol before attestation from late-joining institutions requires temporary faith in the remote verifier, but the fact that the remote verifier will ultimately be exposed if it engages in fraud ensures that it has an incentive to execute the protocol correctly.

There are two additional refinements that can enhance the security of the protocol beyond the specification set forth above. We omit them from this initial technical paper, due to their complexity, but flag them as potential additions to the final KRNC implementation. First, instead of responsibility for acting as the remote verifier *ô* being handled by the founding institution, it may be desirable for other institutions that join the protocol to jointly take part, so that the responsibilities are effectively handled by a meta-agent who belongs to a greater fraction of agents' trust sets. Second, to refute false accusations that the remote verifier *ô* is assigning cryptographic weight to addresses that were not actually provided by balance owners, it may be important to require that individuals receiving weight assignments verifiably claim their public addresses (e.g., through social media) or that the remote verifier employ a secure enclave such as Intel SGX to create an auditable record of the weight-assignment procedures it executes.

*Pre-Verification Weight Issuance*

Many agents will be able to trust that weight assignments remotely verified by *ö* are correct before the underlying data has received external attestation, based only on the knowledge that defecting from the protocol would foreseeably harm *ö*'s own interests. In this sub-section, we demonstrate that a variation of the same principle permits weight assignments to be trusted before they are even remotely verified. Such weight assignments are *provisional*, in the sense that they are not yet cryptographically immutable. In the KRNC protocol, weight is also a new form of cryptographic money, so we employ the terms provisional weight assignment and provisional balance interchangeably.

Because provisional balances are not immutable, $ö$ cannot issue them on the forked-fiat shard of the settlement array. Instead, to acquire a provisional balance based on a fiat balance held in account $ă$ at institution $ŏ$ during the protocol's staking periods, an agent $n$ that has already registered a forked-fiat account $â$ on the remote verifier's shard of the custodial array uses the corresponding account key to transmit a message on a private channel to $ö$ of the form $SIG_{ök_â}(i, PBR, ŏ, ă, \bar{b}, m)$, where $i$ is the identity that $n$ associated with the account during registration, $PBR$ is the opcode indicating a provisional-balance request, $ŏ$ is the institution where the fiat balance was maintained during the protocol's staking periods, $ă$ is the number of the account at that institution, and $m$ is the balance's monetary unit of account.

For the provisional request to ultimately be valid, it is necessary not only for a single identity $i$ to have been used to register both account $ă$ at $ŏ$ and account $â$ at $ö$, but also for one or more of the authentication factors used to generate $SIG_{ök_â}$ in the signature to overlap with one or more of the authentication factors that comprise the account key $ök_ă$ associated with account $â$ at institution $ŏ$. The overlap requirement is necessary to ensure both (1) that an agent $n' : n' \neq N[i]$ cannot spoof a provisional-balance request for the account that $n$ registered as $i$, and (2) that $n$ cannot send a fraudulent provisional-balance request as $i$ and later blame some agent $n'$. Technically, as long as the overlap requirement is satisfied, a collection of one or more authentication factors that are collectively insufficient to qualify as $ök_â$ because they cannot authorize a transfer of funds can still be used to generate a valid signature $SIG_{ök_â}$. For example, even if two-factor authentication is necessary to spend forked-fiat money stored in $â$, verifying control of the email address associated with the account would still be enough to confirm the identity $i$ used to register the account – and, in turn, to later establish authentication-factor overlap with other accounts that agent $n$ registered as $i$ at other institutions using the same email address.

To establish verifiable incentive compatibility, it is further necessary to specify that once a valid provisional-balance request has been submitted for an account, no weight based on that account can be assigned on the forked-fiat shard of the settlement array until verification that the correct weight was less than or equal to the amount provisionally requested. To enforce this restriction, we must extend the mechanism already defined to prevent double counting.

If remote verification shows that the balance held in $ă$ was less than represented, $ö$ will broadcast a message of the form $SIGpk_ö(ö, NBE,$ $ENCpk_{\underline{\ddot{o}}}(\ddot{o}\,\check{a}, \overline{b}, \overline{b}')$), where $FPR$ is the opcode for a notice of balance exaggeration, and the next term is a message encrypted using the provisional public key created for $\ddot{o}$, which when decrypted with the corresponding provisional secret key reveals the institution and account number of the false request, followed by the claimed balance $\overline{b}$ and the actual balance, $\overline{b}'$. For the message to be valid, it is necessary that $\overline{b} > \overline{b}'$, so that an attempt to exaggerate the true balance is documented.

Upon joining the protocol and receiving its provisional secret key, $sk_{\underline{\ddot{o}}}$, $\ddot{o}$ decrypts the notice of balance exaggeration and discovers that $\check{a}$ was associated with an authenticated, false request for a provisional balance. It switches the eligibility status $ES_{\check{a}}$ of account $\check{a}$ to 0, which will prevent any weight from being assigned based on that account when $\ddot{o}$ calculates its own weight assignment, $\omega[\ddot{o}]$.

If no remote verification of account $\check{a}$ occurs before $\ddot{o}$ joins the protocol and broadcasts its request for institutional status, then $\ddot{o}$ will respond to that request by broadcasting a message of the form $SIGpk_{\ddot{o}}\left(\ddot{o}, RPV, ENCpk_{\underline{\ddot{o}}}(\ddot{o}, \check{a}, \overline{b}, i, \ddot{o}k_{\hat{a}} \cap \ddot{o}k_{\check{a}})\right)$ where $RPV$ is the opcode for requesting provisional-balance verification, and the next term is a message encrypted with the provisional public key created for $\ddot{o}$, which specifies the institution associated with the account, the account number, the claimed balance, the identity used to submit the provisional request, and the claimed intersection of one or more authentication factors comprising the two account keys.

The result is another message that $\ddot{o}$ must decrypt and include in its computation of how much weight its customers are owed. First, $\ddot{o}$ checks whether the claimed intersection between authentication factor exists; if not, the original provisional-balance request is considered invalid, and the data it contains is discarded. If the intersection is confirmed, then $\ddot{o}$ queries its records of the balance data for account $\check{a}$ and computes the average balance during the protocol's staking periods. If the mean balance returned is less than or equal to the value listed in the decrypted message, then $\ddot{o}$ has verified that the request was truthful; if not, $\ddot{o}$ determines that the request was false. If the request was false, then $\ddot{o}$ switches the eligibility status $ES_{\check{a}}$ of account $\check{a}$ to 0.

When applied to all provisional-balance requests, the procedures we have just specified guarantee that an agent $n$ who uses an identity $i$ to submit a false provisional-balance request for an account $\check{a}$ will receive a resulting weight assignment $\omega[\check{a}]$ of 0. This makes it trivial to prove that, given only the choices of submitting a true or false request for an account with a non-zero mean balance, honesty is the strongly dominant choice: any honest request for an account with a non-zero mean balance is eligible for a positive weight, $\omega[\breve{a}]$: $\omega[\breve{a}] > 0$, which is strictly greater than the payoff from a false request. This result holds for all accounts with a positive mean balance during the protocol's staking periods.

We can further prove that the strongly dominant choice is disclosure of the specific mean balance recorded on the custodial array.[16] As we have just demonstrated, all choices greater than the true value are ruled out because they guarantee a payoff of zero; the largest value within the collection of non-excluded values will necessarily guarantee the largest weight assignment — and thus, the optimal payout. That unique, strongly dominant choice is the true value — the largest value not excluded by virtue of being false. The result is a unique Nash equilibrium: the recipient of the signal expects the signaler to maximize utility by disclosing the true value; when that expectation is realized, the recipient's payout is maximized, so both parties have simultaneously maximized their payoffs.

This result establishes that, in the absence of any reward for false provisional-balance requests, the accuracy of data submitted in a non-spoofed provisional balance can be validly derived from the axiom that all agents maximize their individual utility. However, it does not establish that agents would actually submit such requests in the absence of a reward, nor does it show that the signaling equilibrium for provisional-balance requests will remain honest if a reward mechanism is introduced. Fortunately, an honest signaling equilibrium is still possible if agents have a preference for the success of the protocol over the failure of the protocol, since they will rationally take zero-cost actions that help bootstrap its adoption.

In the absence of explicit economic rewards, preliminary-balance requests from the public have the potential to act as a form of coordinated signaling, similar to voting or participating in an online petition. As the quantity of fiat money participating in the "signaling campaign" grows, the protocol enters the social consciousness, and how widespread support for its adoption truly is becomes an open empirical question. If the remote verifier $\ddot{o}$ discloses the provisional-balance requests to the members of $O$, each institution $o$ will be able to verify the authenticity of non-spoofed requests associated with its custodial shard. In effect, their customers are given a bullhorn to demand the cryptographic money that they have rightfully earned.

If a significant number of institutions have the foresight to join the protocol early in its existence, then a critical mass in favor of adoption might be reached without the need to complicate the protocol by extending the functionality available during the provisional-balance period. However, consistent with a "defense in depth" approach to the risk of insufficient inertia, we present other bootstrapping mechanisms.

---

[16] To rule out "trembling hand" errors, in this specific calculation we assume perfect execution of the protocol's calculation requirements. This eliminates the need to consider the optimal margin against such errors.

First, to minimize early vulnerability to staking-asynchrony attacks, it may be strategic to accelerate price discovery by introducing small quantities of cryptographic weight into advanced circulation. Before the periods sampled during the fork, agents submit weekly (or monthly, etc.) provisional-balance requests for the preceding week, and those who are later verified to have sent one or more honest requests become eligible to have weight assigned in proportion to the "interest rate" in effect at the time each request was submitted. Technically, the term "interest" is incorrect — no risk is being assumed, and units of account spanning multiple ledgers are involved — but there is no equally natural word for earning a monetary reward in proportion to a fiat bank balance.

One downside to such a reward scheme is the potential to violate the immutability of the forked-fiat shard of the settlement array if proper safeguards are not taken. In particular, consistency can be violated if weight assignments are based on mean balances that can be lowered by overdrafts. If a negative balance in a slot $s$ causes the mean balance in a period to be lower than 0, the weight assigned pre-$s$ based on balances verified in other slots of that period may be invalid.

Our preferred solution is to assign a non-negative balance $b_{nn}$ to each slot using the formula

$$b_{nn} = \frac{b[a,s] \; + \; |b[a,s]|}{2}$$

which leaves positive and zero balances unchanged, but converts negative balances to zero. This allows weight to be assigned safely for periods with one or more slots whose values are unknown, because the correct weight assignment for those (and all other) periods will monotonically increase if additional slot values are verified in the future. If those values reveal that the weight previously assigned for a period was insufficient, then the shortfall can be corrected in a supplementary assignment of additional weight — which is verifiably safe, because it preserves the validity of every forked dollar issued in the original assignment.

An alternative is to define which collections of slots are eligible to be aggregated in a weight assignment (e.g., only slots comprising a full calendar month) and to prohibit weight from being issued for any collection if a single one of its constituent slots has not been verified.

Second, price discovery may also be facilitated through the use of identity-based issuance of future claims on forked-fiat currency. This is one of the protocol's most elegant features: because the set of agents and institutions that will initially control forked-fiat currency on KRNC is the same set of agents and institutions that will control fiat money, it is obvious which actors can be trusted to make good on forked-fiat denominated debts issued in the present. A wealthy individual or institution that has decided to participate in the KRNC protocol in the future can thereby issue forked-fiat denominated tokens on other platforms in the present, even before the protocol has been implemented. An equivalent procedure can be used to issue tokens denominated in cryptographically backed fiat money; the only difference is that such tokens represent a claim to one traditional dollar and one forked-fiat dollar, rather than simply the forked-fiat dollar.

## Conclusion

KRNC replicates the advantages of Bitcoin and Ethereum, without the problems inherent in introducing new currencies. It gives the people of the world the cryptographic money they deserve — the money they already own.

> My faith is in the honesty of the majority of people. Actually, if you want, technically speaking, that the majority of money belongs to honest people.
>
> – Silvio Micali
> *Turing Laureate* (2012), *Gödel-Prize Winner* (1993)

Just so.